\documentclass[aps,prd,onecolumn,floatfix,nofootinbib,showpacs]{revtex4-1}

\pdfoutput=1
\usepackage{graphicx,color,float}
\usepackage{hyperref}
\usepackage{tabularx}
\usepackage{multirow}
\usepackage{verbatim}
\usepackage{longtable}
 \usepackage{amsmath}
\usepackage{amsfonts}
\usepackage{amssymb}
\usepackage{rotating}
\usepackage[dvipsnames]{xcolor}

\usepackage{setspace}

\usepackage{ulem}

\usepackage{adjustbox}

\usepackage{dcolumn}  
\newcolumntype{.}[1]{D{.}{.}{#1}} 

\newcommand{\real}{\Re {\rm e}}
\newcommand{\lsim}{\raisebox{-0.13cm}{~\shortstack{$<$ \\[-0.07cm] $\sim$}}~}

\def\slash#1{#1\!\!\!/}

\begin{document}

{\small
\begin{flushright}
IUEP-HEP-24-01
\end{flushright} }

\title{
Higgs Boson Precision Analysis of the Full LHC Run 1 and Run 2 Data
}

\def\slash#1{#1\!\!/}

\renewcommand{\thefootnote}{\arabic{footnote}}

\author{
Yongtae Heo,$^{1,2}$\footnote{yongtae1heo@gmail.com}~
Dong-Won Jung,$^{1,3,4}$\footnote{dongwon.jung@ibs.re.kr}~
and
Jae Sik Lee$^{1,2}$\footnote{jslee@jnu.ac.kr}
}

\affiliation{
$^1$ Department of Physics, Chonnam National University, 
Gwangju 61186, Korea\\
$^2$ IUEP, Chonnam National University, Gwangju 61186, Korea \\
$^3$ Department of Physics, 
Yonsei University, Seoul 03722, Korea\\
$^4$ Cosmology, Gravity, and Astroparticle Physics Group, \\
Center for Theoretical Physics of the Universe, Institute for Basic Science, 
Daejeon 34126, Korea
}
\date{May 31, 2024}

\begin{abstract}
\begin{spacing}{1.30}
We perform global fits of the Higgs boson couplings to
the full Higgs datasets collected  at the LHC
with the integrated luminosities per experiment of approximately
5/fb at 7 TeV, 20/fb at 8 TeV, and up to 139/fb at 13 TeV.
Our combined analysis
based on the experimental signal strengths used in this work
and the theoretical ones elaborated for our analysis
reliably reproduce the results in the literature.
We reveal that 
the LHC Higgs precision data are no longer
best described by the SM Higgs boson
taking account of extensive and comprehensive
CP-conserving and CP-violating 
scenarios found in  several well-motivated
models beyond the SM.
Especially, in most of the fits considered in this work, 
we observe that the best-fitted values of
the normalized Yukawa couplings are about $2\sigma$ below
the corresponding SM ones
with the $1\sigma$ errors of 3\%-5\%.
On the other hand, the gauge-Higgs couplings are consistent 
with the SM with the $1\sigma$ errors of 2\%-3\%.
Incidentally,
the reduced Yukawa couplings help to explain the excess of
the $H\to Z\gamma$ signal strength of $2.2\pm 0.7$ recently 
reported by the ATLAS and CMS collaborations.
\end{spacing}
\end{abstract}

\maketitle

\section{Introduction}
\label{sec:introduction}
Since the ATLAS and CMS collaborations have independently reported
the observation of a new scalar particle in 
the search for the Standard Model (SM) Higgs boson
in 2012~\cite{ATLAS:2012yve,CMS:2012qbp},
\footnote{The ATLAS discovery was based on 
approximately $4.8$/fb collected at $\sqrt{s} = 7$ TeV and 
$5.8$/fb at $\sqrt{s} = 8$ TeV while the CMS discovery on
up to $5.1$/fb at 7 TeV and $5.3$/fb at 8 TeV.}
more than 30 times larger number of Higgs bosons have been recorded
by the both collaborations 
at the CERN Large Hadron Collider (LHC).
Recently, ten years after the discovery,
the two collaborations have presented
the two legacy papers portraying the Higgs boson and 
revealing a detailed map of its interactions
based on the precision Higgs data collected during the Run 2 
data-taking period between 2015 and 2018
\cite{ATLAS:2022vkf,CMS:2022dwd}.
During the Run 2 period at a center-of-mass energy of 13 TeV,
the ATLAS and CMS collaborations have accumulated 
the integrated luminosities of 139/fb and 138/fb, respectively,
which exceed those accumulated during the full Run 1 period
by the factor of more than 5.
\footnote{During the full Run 1 period, 
each of the collaborations accumulated
approximately 5/fb at 7 TeV and  20/fb at 8 TeV.}

\medskip

Though there was a quite room for the new scalar boson
weighing 125 GeV
\footnote{Recently, 
exploiting the decay channels of $H\to ZZ^*\to 4\ell$ and
$H\to\gamma\gamma$,
the ATLAS collaboration reports the result
of a Higgs mass measurement of $125.11 \pm 0.11$ GeV 
with a $0.09\%$ precision
which is based on 140/fb at 13 TeV
combined with the Run 1 measurement, see Ref.~\cite{ATLAS:2023oaq}.}
to be different from the SM Higgs boson 
around the discovery stage~\cite{Cheung:2013kla} but
the increasing Higgs datasets soon revealed that
they were best described by
the SM Higgs boson~\cite{Cheung:2014noa}.
Around the end of the Run 2 period before the 
Run 2 Higgs data are fully analyzed, 
the five productions modes of
gluon-gluon fusion (ggF), 
vector-boson fusion (VBF), 
the associated production with a $V = W/Z$ boson (WH/ZH),
and the associated production with a top-quark pair (ttH) 
had been extensively investigated and, impressively,
the Higgs decays into a pair of $b$ quarks
\cite{ATLAS:2017cen,CMS:2018nsn}
and a pair of $\tau$ leptons 
\cite{ATLAS:2018lur,CMS:2017zyp}
were observed in single measurements
leading to the firm establishment of third-generation Yukawa couplings
together with the top-quark Yukawa coupling constrained by
the ggF and ttH productions and 
the Higgs decay to two photons~\cite{Cheung:2018ave}.
Now, with the Run 2 Higgs data fully analyzed
\cite{ATLAS:2022vkf,CMS:2022dwd},
the sixth Higgs production process
associated with a single top quark (tH)  
starts to be involved and the Higgs boson decays into
a pair of muons and $Z\gamma$ are emerging.
The {\it direct} searches for so-called invisible Higgs 
boson decays into non-SM particles have been also carried out.

\medskip

In Refs.~\cite{ATLAS:2022vkf,CMS:2022dwd}, the ATLAS and CMS collaborations
scrutinize the interactions of the 125 GeV Higgs boson 
using the Higgs precision data recorded by their own experiments 
during the Run 2 period and independently yield the following
Run 2 global signal strengths assuming 
that all the production and decay processes scale with the overall 
single signal strength:
\begin{eqnarray}
\mu^{\,\rm Global}_{\,\rm Run\,2\,ATLAS} &=& 1.05 \ \pm \ 0.06 \ ; \ \ \
\mu^{\,\rm Global}_{\,\rm Run\,2\,CMS} = 1.002 \ \pm \ 0.057\,,
\end{eqnarray}
in remarkable agreement with the SM expectation.
In this work, by combining the ATLAS and CMS Run 2 data 
on the signal strengths as well as including, 
though statistically less important, 
the Run 1 LHC~\cite{ATLAS:2016neq} and Tevatron global signal strengths, 
\footnote{
See Tables~\ref{tab:tev},\ref{tab:78all}, \ref{tab:ATLAS13}, and \ref{tab:CMS13}.}
we find the following global signal strength:
\begin{eqnarray}
\mu^{\,\rm Global}_{\,\rm 76\,signal\,strengths} = 1.012 \ \pm \ 0.034\,. 
\end{eqnarray}
Upon the previous model-independent analyses
\cite{Cheung:2013kla,Cheung:2014noa,Cheung:2018ave},
we have improved our analysis
by including the tH production process to accommodate the LHC Run 2 data and
by treating the ggF production signal strength beyond leading order in QCD
to match the precision of the ever-increasing Higgs data now and after.

\medskip

We demonstrate that our combined analysis
based on the experimental signal strengths used in this work
and the theoretical ones elaborated for our analysis
reliably reproduce
the fitting results presented in Ref.~\cite{ATLAS:2016neq} (Run 1) and 
Refs.~\cite{ATLAS:2022vkf,CMS:2022dwd} (Run 2) 
within $0.5$ standard deviations.
Our extensive and comprehensive CP-conserving and CP-violating fits
taking account of various scenarios found in  several well-motivated
models beyond the SM (BSM) reveal that 
the LHC Higgs precision data are no longer  
best described by the SM Higgs boson.
Especially, in most of the fits, we observe that
the best-fitted values of
the Yukawa couplings are about $2\sigma$ below 
the corresponding SM ones
with the $1\sigma$ errors of 3-5\%.
The reduced Yukawa couplings help to explain 
the combined $H\to Z\gamma$ signal strength of $2.2\pm 0.7$ recently 
reported by the ATLAS and CMS collaborations~\cite{ATLAS:2023yqk}.
Accordingly, the SM points locate outside
the 68\% confidence level (CL) region mostly and 
even the 95\% CL region sometimes in many of 
the two-parameter planes involved Yukawa couplings.
We further note that the BSM models predicting the same 
scaling behavior of the Yukawa couplings to
the up- and down-type quarks and charged leptons are preferred.
On the other hand, the gauge-Higgs couplings are consistent 
with the SM with the $1\sigma$ errors of 2\%-3\%.
Incidentally, we note that
CP violation is largely unconstrained by the LHC Higgs data
with the CL regions appearing as a circle or an ellipse or some 
overlapping of them in the CP-violating two-parameter planes.

\medskip

This paper is organized as follows.
Section~\ref{sec:data} is devoted to reviewing
the ATLAS and CMS Run 2 data on the signal strengths as well as
the Run 1 LHC and Tevatron ones used in this work.
We compare our results on the global signal strengths and
the signal strengths for the individual Higgs production 
processes and decay modes with those in 
Refs.~\cite{ATLAS:2016neq,ATLAS:2022vkf,CMS:2022dwd}.
In Section~\ref{sec:framework},
we fix the conventions and notations of the model-independent
couplings of the 125 GeV Higgs-boson $H$ to the SM particles
assuming that $H$ is a CP-mixed scalar.
And we elaborate on the parametric dependence of the
theoretical signal strengths for the relevant Higgs production 
processes and decay modes.
In Section~\ref{sec:global_fits}, we perform global fits of the
Higgs boson couplings to the full Higgs datasets collected at the LHC
and Tevatron considering various CP-conserving and CP-violating 
scenarios extensively.
Conclusions are made in Section~\ref{sec:conclusions}.
In Appendix~\ref{sec:appendix_A}, we provide the details of how we
work out the ggF production signal strength beyond leading order in QCD.
Appendix~\ref{sec:appendix_B} is devoted to the chi-square behavior 
when only a single fitting parameter is varied
while all the other ones are fixed at their SM values of either 0 or 1.
In Appendix~\ref{sec:appendix_C},
we figure out the dependence of the $H\to\gamma\gamma$ signal strength 
on the gauge-Higgs and Yukawa couplings in the two-parameter fit frequently 
referred in the literature.
We present correlations among the fitting parameters in 
some CP-conserving fits in Appendix~\ref{sec:appendix_D}.
Finally, for global fits including the 
$H\to Z\gamma$ data recently reported, see Appendix~\ref{sec:appendix_E}.

\section{Higgs signal Strength: Data and Overall Results}
\label{sec:data}
\begin{table}[t!]
\caption{\it
\label{tab:tev}
{\bf (Tevatron: 1.96 TeV)}
The signal strengths data from Tevatron (10.0/fb at 1.96 TeV).}
\setlength{\tabcolsep}{2.5ex}
\renewcommand{\arraystretch}{1.1}
\begin{center}
\begin{tabular}{c|c|c|ccc}
\hline\hline
 &  & & \multicolumn{3}{c}{Production mode}  \\
Channel & Signal strength &  $M_H$ (GeV) & ggF & VBF & WH$\oplus$ZH  \\
\hline
Combined $H\to \gamma\gamma$~\cite{tevatron_aa_ww} & $6.14^{+3.25}_{-3.19}$   & 125 & 78\%
& 5\% & 17\%  \\
Combined $H\to WW^{(\ast)}$~\cite{tevatron_aa_ww}  & $0.85^{+0.88}_{-0.81}$        & 125 &
78\% & 5\% & 17\%  \\
VH-tagged $H\to bb~$\cite{Herner:2016woc}        & $1.59^{+0.69}_{-0.72}$        & 125 & $-$    &
$-$   & 100\%  \\
\hline\hline
\end{tabular}
\end{center}
\end{table}
\begin{table}[t!]
\caption{\it
\label{tab:78all}
{\bf (LHC: 7$\oplus$8 TeV)}
Combined ATLAS and CMS Run 1 data on signal strengths~\cite{ATLAS:2016neq}
used in this work
(Approximately 5/fb at 7 TeV and 20/fb at 8 TeV per experiment).
$M_H=125.09$ GeV is taken.
}
\begin{center}
\setlength{\tabcolsep}{2.5ex}
\renewcommand{\arraystretch}{1.1}
\begin{tabular}{c|ccccc}
\hline\hline
 & \multicolumn{5}{c}{Decay mode} \\
\hline
Production mode & $\gamma\gamma$ & $Z Z^* $ & $W W^* $
& $bb$ & $\tau\tau$  \\
\hline
ggF & $1.10^{+0.23}_{-0.22}$  & $1.13^{+0.34}_{-0.31}$ & $0.84^{+0.17}_{-0.17}$
    & $-$                       & $1.0^{+0.6}_{-0.6}$  \\
VBF & $1.3^{+0.5}_{-0.5}$     & $0.1^{+1.1}_{-0.6}$    & $1.2^{+0.4}_{-0.4}$
    & $-$                       & $1.3^{+0.4}_{-0.4}$  \\
WH  & $0.5^{+1.3}_{-1.2}$     & $-$                      & $1.6^{+1.2}_{-1.0}$
    & $1.0^{+0.5}_{-0.5}$     & $-1.4^{+1.4}_{-1.4}$   \\
ZH  & $0.5^{+3.0}_{-2.5}$     & $-$                      & $5.9^{+2.6}_{-2.2}$
    & $0.4^{+0.4}_{-0.4}$     & $2.2^{+2.2}_{-1.8}$    \\
ttH  & $2.2^{+1.6}_{-1.3}$    & $-$                      & $5.0^{+1.8}_{-1.7}$
     & $1.1^{+1.0}_{-1.0}$    & $-1.9^{+3.7}_{-3.3}$    \\
\hline\hline
\end{tabular}
\end{center}
\end{table}

For our global fits, we use all the available {\it direct} Higgs data collected at the
Tevatron and the LHC.
We use 3 signal strengths measured at the Tevatron
\cite{tevatron_aa_ww,Herner:2016woc}
as shown in Table~\ref{tab:tev}.
For the Run 1 LHC data at center-of-mass energies of
$\sqrt{s}=7$ and 8 TeV, abbreviated as $7\oplus 8$ TeV for the later use, 
we use 20 signal strengths and the correlation matrix 
obtained in the combined ATLAS and CMS analysis 
\cite{ATLAS:2016neq}, see Table~\ref{tab:78all}.
\footnote{Specifically, see Table 8 of Ref.~\cite{ATLAS:2016neq} 
for the combined signal strengths and Fig. 27 therein
for the correlation matrix.
The more detailed information could be found
in the website
{\bf https://doi.org/10.17182/hepdata.78403}.}

\medskip

The Run 2 data on signal strengths are given separately
by the ATLAS and CMS collaborations.
%
%
%
The ATLAS input for the Run 2 data on signal strengths includes
the latest results from the Higgs decays into 
$ZZ\to 4\ell$~\cite{ATLAS:2020rej},
$W^\pm W^\mp\to \ell\nu\ell\nu$~\cite{ATLAS:2022ooq,ATLAS:2019vrd},
$\gamma\gamma$~\cite{ATLAS:2022tnm},
$Z\gamma\to \ell^+\ell^-\gamma$~\cite{ATLAS:2020qcv},
$b\bar b$~\cite{ATLAS:2020fcp,ATLAS:2020jwz,ATLAS:2020bhl,ATLAS:2021qou,ATLAS:2021tbi},
$\tau^+\tau^-$~\cite{ATLAS:2022yrq},
multiple leptons ($\tau^+\tau^-$, $W^\pm W^\mp$, $ZZ$)~\cite{ATLAS:2017ztq},
$\mu^+\mu^-$~\cite{ATLAS:2020fzp},
$c\bar c$~\cite{ATLAS:2022ers}, and
invisible particles~\cite{ATLAS:2022yvh,ATLAS:2021gcn}.
For our analysis, dropping the results from the Higgs decays
into $Z\gamma$, $c\bar c$, and invisible particles which 
have not been evidenced at the current stage giving
constraints rather than measurements, we use 
25 signal strengths shown in Table~\ref{tab:ATLAS13} and
the correlation matrix of the 
production cross sections times branching fractions.
\footnote{For the ATLAS Run 2 signal strengths,
see Fig.~3 in Ref.~\cite{ATLAS:2022vkf} together with
detailed information on them provided in
the website {\bf https://doi.org/10.17182/hepdata.130266}.
For the correlation matrix, see Auxiliary Fig.~14 
presented in the website.}
%
%
On the other hand,
the CMS input for the Run 2 data on signal strengths includes
the decay channels into
$\gamma\gamma$~\cite{CMS:2021kom},
$ZZ\to 4\ell$~\cite{CMS:2021ugl},
$W^\pm W^\mp\to \ell\nu\ell\nu$~\cite{CMS:2022uhn},
$Z\gamma$~\cite{CMS:2022ahq},
$\tau^+\tau^-$~\cite{CMS:2022kdi},
$b\bar b$~\cite{CMS:2017odg,CMS:2018nsn,CMS:2018sah,CMS:2018hnq,CMS:2020zge},
$\mu^+\mu^-$~\cite{CMS:2020xwi},
multileptons with ttH/tH~\cite{CMS:2020mpn}, and
invisible particles~\cite{CMS:2022qva,CMS:2021far,CMS:2020ulv}.
Dropping the Higgs decays into $Z\gamma$ and invisible particles
as in the ATLAS case, we use 28 signal strengths shown in 
Table~\ref{tab:CMS13} and  the correlation matrix.
\footnote{For the CMS Run 2 signal strengths, 
see Fig.~B.6 in Ref.~\cite{CMS:2022dwd}.
For the correlation matrix, see the figure entitled
{\it ``Production times decay signal strength modifiers correlations"}
provided in the website
{\bf https://dx.doi.org/10.17182/hepdata.127765}.
}
For the Higgs production, commonly investigated are the six 
processes of
gluon-gluon fusion (ggF), 
vector-boson fusion (VBF), 
the associated production with a $V = W/Z$ boson (WH/ZH),
the associated production with a top-quark pair (ttH), and
the associated production with a single top quark (tH).

\medskip

We compare the ATLAS$+$CMS combined Run 1 data on signal strengths
with the either ATLAS or CMS Run 2 data on signal strengths
for various combinations of Higgs boson
production and decay processes,
see Table~\ref{tab:78all} and Tables~\ref{tab:ATLAS13} and \ref{tab:CMS13}.
We note that
the tH production process and the $H\to\mu\mu$ decay mode have 
been newly measured and the mixed-production modes such as 
ggF+bbH$\oplus$VBF, 
WH$\oplus$ZH, ttH$\oplus$tH,
ggF+bbH$\oplus$ttH$\oplus$tH, and
VBF$\oplus$WH$\oplus$ZH are involved 
especially when the Higgs boson decays into 
a pair of fermions.
\footnote{We expect that the mixed-production modes
might be resolved into 
individual ones in the next Run(s) with the higher luminosities.}
Incidentally, in Run 2, the Higgs decay into $ZZ^*$ is measured
also in the WH, ZH, and ttH production modes though 
the corresponding errors are still large.
Further we observe that
the ttH and tH production processes have been always combined 
except in the ATLAS measurement of the $H\to \gamma\gamma$ decay.
For each production-times-decay mode, we note that
the measurements of 
the $H\to\gamma\gamma$ decay in tH production, 
$H\to ZZ^*$ in WH, ZH, ttH and tH,
$H\to WW^*$ in WH and ZH, and
$H\to bb$ in ggF$+$bbH and VBF are now challenging and they
might be significantly improved in near future.

%
\begin{table}[t!]
\centering
\caption{\it
\label{tab:ATLAS13}
{\bf (LHC: 13 TeV)}
ATLAS Run 2 data on signal strengths~\cite{ATLAS:2022vkf}
used in this work (139/fb at 13 TeV).
We refer to the website
{\bf https://doi.org/10.17182/hepdata.130266}
for specific information.
$M_H=125.09$ GeV is taken.
\\ }
\setlength{\tabcolsep}{2.5ex}
\renewcommand{\arraystretch}{1.1}
\begin{tabular}{c|c|c|c|c|c|c}
\hline\hline
 & \multicolumn{6}{c}{Decay mode}  \\
\hline
Production mode
 & $\gamma\gamma$ & $ZZ^*$ & $WW^*$ & $bb$ & $\tau\tau$ & $\mu\mu$ \\
\hline
ggF+bbH &
$1.04^{+0.10}_{-0.10}$ &
$0.95^{+0.11}_{-0.10}$ &
$1.14^{+0.13}_{-0.13}$ &
$-$ &
$0.90^{+0.29}_{-0.26}$ &
$-$ \\
VBF &
$1.36^{+0.30}_{-0.26}$ &
$1.33^{+0.52}_{-0.43}$ &
$1.13^{+0.19}_{-0.18}$ &
$-$ &
$1.00^{+0.21}_{-0.18}$ &
$-$ \\
WH &
$1.53^{+0.56}_{-0.51}$ &
$-$  &
$2.26^{+1.21}_{-1.02}$ &
$1.06^{+0.28}_{-0.26}$ &
$-$ &
$-$ \\
ZH &
$-0.22^{+0.61}_{-0.54}$ &
$-$ &
$2.86^{+1.84}_{-1.33}$ &
$1.00^{+0.25}_{-0.23}$ &
$-$ &
$-$ \\
ttH &
$0.90^{+0.33}_{-0.31}$ &
$-$ &
$-$ &
$-$ &
$-$ &
$-$ \\
tH &
$2.61^{+4.23}_{-3.38}$ &
$-$ &
$-$ &
$-$ &
$-$ &
$-$ \\
\hline
ggF+bbH$\oplus$VBF &
 $-$ & $-$ & $-$ & $0.98^{+0.38}_{-0.36}$ & $-$ & $-$  \\
WH$\oplus$ZH &
 $-$ & $1.50^{+1.17}_{-0.94}$ & $-$ & $-$ & $1.00^{+0.62}_{-0.59}$ & $-$  \\
ttH$\oplus$tH &
 $-$ & 
$1.68^{+1.68}_{-1.11}$ &
$1.64^{+0.65}_{-0.61}$ &
$0.35^{+0.34}_{-0.33}$ &
$1.37^{+0.86}_{-0.75}$ &
$-$  \\
\hline
ggF+bbH$\oplus$ttH$\oplus$tH &
 $-$ & $-$ & $-$ & $-$ & $-$ & $0.54^{+0.89}_{-0.88}$  \\
VBF$\oplus$WH$\oplus$ZH &
 $-$ & $-$ & $-$ & $-$ & $-$ & $2.31^{+1.33}_{-1.26}$  \\
%
%
\hline\hline
\end{tabular}
\end{table}
%

\begin{table}[t!]
\centering
\caption{\it
\label{tab:CMS13}
{\bf (LHC: 13 TeV)}
CMS Run 2 data on signal strengths~\cite{CMS:2022dwd}
used in this work (138/fb at 13 TeV). 
$M_H=125.38$ GeV is taken. 
\\}
\setlength{\tabcolsep}{2.5ex}
\renewcommand{\arraystretch}{1.1}
\begin{tabular}{c|c|c|c|c|c|c}
\hline\hline
& \multicolumn{6}{c}{Decay mode}  \\
\hline
Production mode & $\gamma\gamma$ & $ZZ^*$ & $WW^*$ & $bb$ & $\tau\tau$ & $\mu\mu$ \\
\hline
ggF+bbH &
$1.08^{+0.12}_{-0.11}$ &
$0.93^{+0.14}_{-0.13}$ &
$0.90^{+0.11}_{-0.10}$ &
$5.31^{+2.97}_{-2.54}$ &
$0.66^{+0.21}_{-0.21}$ &
$0.33^{+0.74}_{-0.70}$ \\
VBF &
$1.00^{+0.35}_{-0.32}$ &
$0.32^{+0.48}_{-0.32}$ &
$0.73^{+0.28}_{-0.24}$ &
$-$ &
$0.86^{+0.17}_{-0.16}$ &
$1.55^{+0.86}_{-0.73}$ \\
WH &
$1.43^{+0.54}_{-0.47}$ &
$0.00^{+1.55}_{}$ &
$2.41^{+0.72}_{-0.70}$ &
$1.26^{+0.42}_{-0.41}$ &
$1.33^{+0.61}_{-0.57}$ &
$-$ \\ 
ZH      &
$1.19^{+0.71}_{-0.62}$ &
$12.24^{+6.59}_{-5.69}$ &
$1.76^{+0.75}_{-0.67}$ &
$0.90^{+0.36}_{-0.34}$ &
$1.89^{+0.65}_{-0.56}$ &
$-$ \\ 
\hline
WH$\oplus$ZH &
$-$ & $-$ & $-$ & $-$ & $-$ & $5.63^{+3.04}_{-3.36}$  \\
ttH$\oplus$tH &
$1.38^{+0.34}_{-0.29}$ &
$0.00^{+0.73}_{}$ &
$1.44^{+0.32}_{-0.32}$ &
$0.90^{+0.46}_{-0.44}$ &
$0.35^{+0.44}_{-0.37}$ &
$3.07^{+2.63}_{-2.21}$ \\
\hline\hline
\end{tabular}
\end{table}

\begin{table}[t!]
\centering
\caption{\it
\label{tab:LHCPD}
LHC signal strengths
for individual Higgs production processes and decay modes.
For the combined Run 1 signal strengths, see Tables 12 and 13 
in Ref.~\cite{ATLAS:2016neq}.
For the ATLAS Run 2 signal strengths, see Fig.~2 in Ref.~\cite{ATLAS:2022vkf} together with
detailed information on them provided in
the website {\bf https://doi.org/10.17182/hepdata.130266}.
The CMS Run 2 signal strengths are from Fig.~2 in Ref.~\cite{CMS:2022dwd}.
Note that the production signal strengths 
have been obtained assuming that the Higgs boson branching fractions
are the same as in the SM and
the decay signal strengths 
have been extracted assuming that the Higgs boson production
cross sections are the same as in the SM.
\\}
\renewcommand{\arraystretch}{1.1}
\begin{adjustbox}{width= \textwidth}
\begin{tabular}{l||c|c|c|c|c|c||c|c|c|c|c|c}
\hline\hline
\multirow{2}{*}{\bf \ \ \ \ LHC}
& \multicolumn{6}{c||}{Production process} &  \multicolumn{6}{c}{Decay mode}  \\ 
\cline{2-13}
 & ggF($\oplus$bbH) & VBF & WH & ZH & ttH & tH
 & $\gamma\gamma$ & $ZZ^*$ & $WW^*$ & $bb$ & $\tau\tau$ & $\mu\mu$ \\
\hline
Run 1~\cite{ATLAS:2016neq}
 & $1.03^{+0.16}_{-0.14}$ & $1.18^{+0.25}_{-0.23}$ & $0.89^{+0.40}_{-0.38}$ 
 & $0.79^{+0.38}_{-0.36}$ & $2.3^{+0.7}_{-0.6}$ & $-$ 
 & $1.14^{+0.19}_{-0.18}$ & $1.29^{+0.26}_{-0.23}$ & $1.09^{+0.18}_{-0.16}$ 
 & $0.70^{+0.29}_{-0.27}$ & $1.11^{+0.24}_{-0.22}$ & $-$ \\
\hline
Run 2 ATLAS~\cite{ATLAS:2022vkf}
 & $1.03^{+0.07}_{-0.07}$ & $1.10^{+0.13}_{-0.12}$ & $1.16^{+0.23}_{-0.22}$ 
 & $0.96^{+0.22}_{-0.21}$ & $0.74^{+0.24}_{-0.24}$ & $6.61^{+4.24}_{-3.76}$
 & $1.09^{+0.10}_{-0.09}$ &
   $1.04^{+0.11}_{-0.10}$ &
   $1.20^{+0.12}_{-0.11}$ &
   $0.91^{+0.15}_{-0.14}$ &
   $0.96^{+0.12}_{-0.11}$ &
   $1.21^{+0.62}_{-0.60}$ \\
\hline
Run 2 CMS~\cite{CMS:2022dwd}
 & $0.97^{+0.08}_{-0.08}$ & $0.80^{+0.12}_{-0.12}$ & $1.49^{+0.26}_{-0.25}$ 
 & $1.29^{+0.24}_{-0.23}$ & \multicolumn{2}{c||}{$1.13^{+0.18}_{-0.17}$ }
 & $1.13^{+0.09}_{-0.09}$ &
   $0.97^{+0.12}_{-0.11}$ &
   $0.97^{+0.09}_{-0.09}$ &
   $1.05^{+0.22}_{-0.21}$ &
   $0.85^{+0.10}_{-0.10}$ &
   $1.21^{+0.45}_{-0.42}$ \\
\hline \hline
\end{tabular}
\end{adjustbox}
\end{table}

\medskip

In Ref.~\cite{ATLAS:2016neq} and Refs.~\cite{ATLAS:2022vkf,CMS:2022dwd} 
for the LHC Run 1 and Run 2 data, respectively, also presented are 
the individual signal strengths for Higgs boson production processes
which have been obtained assuming that the Higgs boson branching fractions
are the same as in the SM and
the individual signal strengths for Higgs boson decay modes
which have been extracted assuming that the Higgs boson production
cross sections are the same as in the SM.
We collect them in Table~\ref{tab:LHCPD}.
Comparing the ATLAS$+$CMS combined Run 1 signal strengths
with the either ATLAS or CMS Run 2 ones,
we observe that, overall, each of the signal strengths is more precisely
measured  with the errors reduced by the factor of about 2
and the most of their central values approach
nearer to the SM value of 1.
Especially,
the error of the ttH production is reduced by the factor of about 3
and the production mode has been evidenced in Run 2.
Looking into the ATLAS and CMS Run 2 signal strengths,
we note that
the ggF production signal strength $\mu^{\rm ggF+bbH}$
\footnote{Note that
the bbH process is experimentally indistinguishable from
ggF production. Though the SM bbH cross section is much smaller
than the SM ggF one by the factor of about 100,
the two production processes are grouped together
for the more precise treatment.}
has been most precisely
measured with the errors of 7\%-8\% followed
by $\mu^{\rm VBF}$ with about 12\% error.
The signal strengths of the other production modes
of WH, ZH, ttH($\oplus$tH) are measured with about 20\% errors.
For the decays, the $\gamma\gamma$, $ZZ^*$, $WW^*$, and $\tau\tau$
modes are measured with about 10\% errors while the $bb$ mode with
$15-20$\% error.
The $\mu\mu$ mode is slightly better measured
by CMS with 45\% error.

\medskip

Lastly, also available are the global signal strengths which 
are simplest and most restrictive among all kinds of signal strengths 
since they have been yielded under the assumption that 
all the production and decay 
processes scale with the same single global signal strength 
independently of the 
production and decay processes:
\begin{equation}
\mu^{\,{\rm Global}}_{\,\rm Run\,1}  = 1.09^{+0.11}_{-0.10}\,\text{\cite{ATLAS:2016neq}}; \ \ \
\mu^{\,{\rm Global}}_{\,{\rm Run\,2\,ATLAS}}  =  1.05\pm 0.06\,\text{\cite{ATLAS:2022vkf}}\,,  \ \ \
\mu^{\,{\rm Global}}_{\,{\rm Run\,2\,CMS}}  =   1.002\pm 0.057\,\text{\cite{CMS:2022dwd}}\,.
\end{equation}
Together with the Tevatron global strength
\begin{equation}
\mu^{\,\rm Global}_{\,\rm Tevatron} =  1.44^{+0.55}_{-0.54}\,,
\end{equation}
we observe that the global signal strengths
have been always consistent with the SM value of 1 and
are converging to the SM value of $1$ with the ever-decreasing errors.
Combining the three LHC global strengths and the Tevatron one,
we have obtained
\begin{equation}
\label{eq:globalss_LHCTev}
\mu^{\,\rm Global}_{\,\rm Tevatron\oplus\text{\cite{ATLAS:2016neq}}\oplus 
\text{\cite{ATLAS:2022vkf}}\oplus\text{\cite{CMS:2022dwd}}} = 1.036 \ \pm \ 0.038\,.
\end{equation}
For the above result, we assume that
each global strength is Gaussian distributed and
correlation among them could be ignored.

\medskip

In this work, we use the following 76 Higgs signal strengths 
which have been extracted 
for different combinations of Higgs boson
production and decay processes
without imposing any assumptions:
\begin{itemize}
\item{Tevatron}: 3 signal strengths based on 10/fb at 19.6 TeV,
see Table~\ref{tab:tev}
\item{LHC Run 1 (ATLAS+CMS)}: 20 signal strengths based on
about $2\times 25$/fb at 7$\oplus$8 TeV, see Table~\ref{tab:78all}
\item{LHC Run 2 (ATLAS)}: 25 signal strengths based on 139/fb at 13  TeV,
see Table~\ref{tab:ATLAS13}
\item{LHC Run 2 (CMS)}: 28 signal strengths based on 138/fb at  13 TeV,
see Table~\ref{tab:CMS13}
\end{itemize}
From the above 76 signal strengths, we obtain
the following global signal strength
\begin{equation}
\label{eq:globalss_76}
\mu^{\,\rm Global}_{\,\rm 76\,signal\,strengths} = 1.012 \ \pm \ 0.034\,, 
\end{equation}
which is consistent with the global signal strength 
given by Eq.~(\ref{eq:globalss_LHCTev}) which is
obtained from the four LHC and Tevatron global signal strengths.

%
%
\begin{table}[!t]
\caption{\it 
\label{tab:pdss76}
LHC signal strengths
for individual Higgs production processes and decay modes obtained 
from the signal strengths in Table~\ref{tab:78all} (Run 1),
Table~\ref{tab:ATLAS13} (Run 2 ATLAS), and
Table~\ref{tab:CMS13} (Run 2 CMS).
Note that,
being different from the signal strengths in Table~\ref{tab:LHCPD},
no assumptions have been imposed 
on the Higgs boson branching ratios and/or production cross sections.
The combined signal strengths are shown in the last line including
all the 76 signal strengths available.
} \vspace{1mm}
\renewcommand{\arraystretch}{1.1}
\begin{adjustbox}{width= \textwidth}
\begin{tabular}{l||c|c|c|c|c||c|c|c|c|c|c}
\hline\hline
%
\multirow{2}{*}{\bf \ \ \ \ LHC}
& \multicolumn{5}{c||}{Production process} &  \multicolumn{6}{c}{Decay mode}  \\
\cline{2-12}
& ggF($\oplus$bbH) & VBF & WH & ZH & ttH$\oplus$tH
& $\gamma\gamma$ & $ZZ^*$ & $WW^*$ & $bb$ & $\tau\tau$ & $\mu\mu$ \\
\hline
Run 1 (Table~\ref{tab:78all})
& $0.99^{+0.11}_{-0.11}$ & $1.17^{+0.20}_{-0.20}$ & $0.79^{+0.33}_{-0.33}$
& $0.78^{+0.34}_{-0.34}$ & $1.99^{+0.61}_{-0.63}$ & 
$1.11^{+0.16}_{-0.17}$ &
$1.03^{+0.28}_{-0.29}$ &
$1.01^{+0.14}_{-0.14}$ &
$0.68^{+0.29}_{-0.29}$ &
$1.04^{+0.25}_{-0.25}$ & $-$ \\
\hline
Run 2 ATLAS (Table~\ref{tab:ATLAS13})
& $1.03^{+0.06}_{-0.06}$ & $1.12^{+0.12}_{-0.12}$ & $1.22^{+0.23}_{-0.23}$
& $0.83^{+0.20}_{-0.20}$ & $0.91^{+0.20}_{-0.21}$&
$1.04^{+0.09}_{-0.09}$ &
$1.01^{+0.10}_{-0.10}$ &
$1.20^{+0.12}_{-0.12}$ &
$0.89^{+0.15}_{-0.15}$ &
$0.95^{+0.13}_{-0.15}$ &
$1.11^{+0.62}_{-0.62}$ \\
\hline
Run 2 CMS (Table~\ref{tab:CMS13})
& $0.93^{+0.07}_{-0.07}$ & $0.81^{+0.11}_{-0.11}$ & $1.40^{+0.24}_{-0.24}$
& $1.50^{+0.26}_{-0.26}$ & $1.06^{+0.17}_{-0.17}$ &
$1.20^{+0.12}_{-0.12}$ &
$0.90^{+0.13}_{-0.13}$ &
$0.95^{+0.10}_{-0.11}$ &
$1.06^{+0.23}_{-0.24}$ &
$0.79^{+0.11}_{-0.11}$ &
$1.14^{+0.49}_{-0.49}$ \\
\hline\hline
Combined  (76 signal strengths)& 
$0.99^{+0.04}_{-0.04}$ & $0.98^{+0.08}_{-0.08}$ &
$1.20^{+0.15}_{-0.15}$ & $1.03^{+0.14}_{-0.14}$ &
$1.04^{+0.13}_{-0.13}$ &
$1.10^{+0.07}_{-0.07}$ &
$0.97^{+0.08}_{-0.08}$ &
$1.04^{+0.07}_{-0.07}$ &
$0.90^{+0.12}_{-0.12}$ &
$0.87^{+0.08}_{-0.08}$ &
$1.13^{+0.38}_{-0.38}$ \\
\hline \hline
\end{tabular}
\end{adjustbox}
\end{table}

\medskip

Observing the difference between the two global signal strengths, 
see Eqs.~(\ref{eq:globalss_LHCTev}) and (\ref{eq:globalss_76}),
we also show the LHC signal strengths
for individual Higgs production processes and decay modes
in Table~\ref{tab:pdss76},
obtained by using the LHC signal strengths 
in Table~\ref{tab:78all} (Run 1),
Table~\ref{tab:ATLAS13} (Run 2 ATLAS), and
Table~\ref{tab:CMS13} (Run 2 CMS).
For the production signal strengths, we neglect 
the tH production process in ATLAS Run 2 and
all the mixed-production processes except ttH$\oplus$tH.
Note that, being different from the 
individual signal strengths in Table~\ref{tab:LHCPD},
no assumptions have been imposed 
on the Higgs boson branching ratios and/or production cross sections.
The combined individual production and decay signal strengths obtained 
by using all the 76 signal strengths are shown in the last line
of Table~\ref{tab:pdss76}.
Since any information on correlations among Run 1,
Run 2 ATLAS and Run 2 CMS datasets
are not currently available, we ignore them accordingly.

\section{Framework}
\label{sec:framework}
For the conventions and notations of the model-independent
couplings of the 125 GeV Higgs-boson $H$ to
the SM particles,
we closely follow Ref.~\cite{Choi:2021nql}.
To be most general, we assume that $H$ is a CP-mixed scalar.
And then, in terms of the CP-violating (CPV)
Higgs couplings, we calculate the 
theoretical signal strengths used for our global fits.
Especially, we have included the production signal strength 
for the tH process to accommodate the new feature of the LHC Run 2 data 
and considered the ggF production process beyond leading order in QCD
to match the level of precision of the LHC Run 2 data.

\subsection{Higgs Couplings}
The interactions of a generic 
neutral Higgs boson $H$ with the SM charged leptons and quarks,
without loss of generality, 
could be described by the following Lagrangian:
\begin{equation}
\label{eq:hff}
{\cal L}_{H\bar{f}f}\ =\ - \sum_{f=\{\ell,q\}}\,\frac{g m_f}{2 M_W}\,
\left[ \bar{f}\,\Big( g^S_{H\bar{f}f}\, +\,
ig^P_{H\bar{f}f}\gamma_5 \Big)\, f\right]\, H
\end{equation}
where $g^S_{H\bar{f}f}$ and $g^P_{H\bar{f}f}$ stand for the $H$ coupling
to the scalar and pseudoscalar fermion bilinears, respectively,
and they are normalized as
$g^S_{H\bar{f}f}=1$ and $g^P_{H\bar{f}f}=0$ in the SM limit.
The interactions of $H$ with the massive vector bosons $V=Z,W$ are described by
\begin{equation}
\label{eq:hvv}
{\cal L}_{HVV}  =  g\,M_W \, \left(
g_{_{HWW}} W^+_\mu W^{- \mu}\ + \
g_{_{HZZ}} \frac{1}{2c_W^2}\,Z_\mu Z^\mu\right) \, H\,,
\end{equation}
in terms of the normalized couplings of $g_{_{HWW}}$ and $g_{_{HZZ}}$
with  $g=e/s_W$ the SU(2)$_L$ gauge coupling and
the weak mixing angle $\theta_W$: 
$c_W\equiv\cos\theta_W$ and $s_W\equiv\sin\theta_W$.
For the SM couplings,
we have $g_{_{HWW}}=g_{_{HZZ}}\equiv g_{_{HVV}}=1$, respecting
the custodial symmetry between the $W$ and $Z$ bosons.

\medskip

The loop-induced Higgs couplings to two photons are described 
through  the following amplitude for the radiative 
decay process $H\to\gamma\gamma$:
\begin{eqnarray} 
\label{eq:haa}
{\cal M}_{\gamma\gamma H}=-\frac{\alpha(0) M_{H}^2}{4\pi\,v}
\bigg\{S^\gamma(M_{H})\, \left(\epsilon^*_{1\perp}\cdot\epsilon^*_{2\perp}\right)
 -P^\gamma(M_{H})\frac{2}{M_{H}^2}
\langle\epsilon^*_1\epsilon^*_2 k_1k_2\rangle
\bigg\}\,,
\end{eqnarray}
by introducing the scalar and pseudoscalar form factors denoted by 
$S^\gamma$ and $P^\gamma$, respectively, with
$k_{1,2}$ and $\epsilon_{1,2}$ being the four--momenta and wave vectors
of the two photons.
\footnote{
For the detailed description and evaluation of the amplitude, 
we refer to Ref.~\cite{Choi:2021nql} while, in this work,
we concentrate on the two form factors which are most relevant
regarding this work.}
Retaining only the dominant contributions from third-generation
fermions and the charged gauge bosons $W^\pm$
and introducing two residual form factors
$\Delta S^\gamma$ and $\Delta P^\gamma$ to parametrize contributions from the
triangle loops in which non-SM charged particles are running, the scalar and
pseudoscalar form factors are given by
\begin{eqnarray}
\label{eq:spaa}
   S^\gamma(M_{H})
&=& 2\sum_{f=b,t,\tau} N_C^f\, Q_f^2\,
    g^{S}_{H\bar{f}f}\,F_{sf}(\tau_{f}) - g_{_{HWW}} F_1(\tau_{W})
    +\Delta S^\gamma \,; \nonumber \\
   P^\gamma(M_{H})
&=& 2\sum_{f=b,t,\tau} N_C^f\,Q_f^2\,
    g^{P}_{H\bar{f}f} \,F_{pf}(\tau_{f})+\Delta P^\gamma \,,
\end{eqnarray}
where $N_C^f=3$ for quarks and $N_C^f=1$ for charged leptons,
$Q_f$ the electric charge of the fermion $f$,
$\tau_f=M_{H}^2/4m_f^2$ and $\tau_{W}=M_{H}^2/4M_W^2$.
For the definitions and behavior of the loop functions 
$F_{sf,pf,1}$, see, for example, Ref.~\cite{Choi:2021nql}.
Taking $M_H=125$ GeV,
for example, one may obtain the 
following estimation of the form factors:
\begin{eqnarray}
\label{eq:spaa_numeric}
S^\gamma&=& -8.324\,g_{_{HWW}} + 1.826\,g_{H\bar{t}t}^S
+(-0.020 + 0.025\,i)\,g_{H\bar{b}b}^S
   +(-0.024 + 0.022\,i)\,g_{H\tau\tau}^S + \Delta S^{\gamma}\,;
\nonumber \\[2mm]
P^\gamma&=& 2.771\,g_{H\bar{t}t}^P
+(-0.022 + 0.025\,i)\,g_{H\bar{b}b}^P
   +(-0.025 + 0.022\,i)\,g_{H\tau\tau}^P + \Delta P^{\gamma}\,, 
\end{eqnarray}
in terms of the Higgs-fermion-fermion and $HWW$ couplings
given in Eqs.~(\ref{eq:hff}) and (\ref{eq:hvv}) 
supplemented by the two residual form factors
$\Delta S^{\gamma}$ and $\Delta P^{\gamma}$.
In the SM limit where $g_{_{HWW}}=g_{H\bar{f}f}^S=1$ and
$g_{H\bar{f}f}^P=\Delta S^{\gamma}=\Delta P^{\gamma}=0$,  we have
$S^\gamma_{\rm SM}=-6.542 + 0.046\,i$ and
$P^\gamma_{\rm SM}=0$.

\medskip

The loop-induced Higgs couplings to two gluons are
similarly described through  the amplitude 
\begin{eqnarray} 
\label{eq:hgg}
{\cal M}_{gg H}^{ab}=-\frac{\alpha_s(M_H)\,M_{H}^2\,\delta^{ab}}{4\pi\,v}
\bigg\{S^g(M_{H})
\left(\epsilon^*_{1\perp}\cdot\epsilon^*_{2\perp}\right)
 -P^g(M_{H})\frac{2}{M_{H}^2}
\langle\epsilon^*_1\epsilon^*_2 k_1k_2\rangle
\bigg\}\,,
\end{eqnarray}
where $a$ and $b$ ($a,b=1$ to 8) are indices of the eight generators in
the SU(3) adjoint representation, $k_{1,2}$ the four momenta of the
two gluons and $\epsilon_{1,2}$ the wave vectors of the corresponding gluons.
Referring to Ref.~\cite{Choi:2021nql} again
for the detailed description and evaluation of the amplitude, 
the scalar and pseudoscalar form factors are given by
\begin{eqnarray}
\label{eq:spgg}
   S^g(M_{H})
 = \sum_{f=b,t,c}
   g^{S}_{H\bar{f}f}\,F_{sf}(\tau_{f})
   + \Delta S^g\,; \ \ \
   P^g(M_{H})
 = \sum_{f=b,t,c}
   g^{P}_{H\bar{f}f}\,F_{pf}(\tau_{f})
   + \Delta P^g\,,
\end{eqnarray}
retaining only the dominant contributions from third-generation and charm quarks
and introducing $\Delta S^g$ and $\Delta P^g$ to
parametrize contributions from the triangle loops in which
non-SM colored particles are running.
Taking $M_H=125$ GeV, one might have
\begin{eqnarray}
\label{eq:spgg_numeric}
S^g&=&  0.688\,g_{H\bar{t}t}^S + (-0.043+0.063\,i)\,g_{H\bar{b}b}^S
+  (-0.009+0.008\,i)\,g_{H\bar{c}c}^S
+ \Delta S^{g}  \,; \nonumber \\[2mm]
P^g&=&  1.047\,g_{H\bar{t}t}^P + (-0.050+0.063\,i)\,g_{H\bar{b}b}^P
 + (-0.010+0.008\,i)\,g_{H\bar{c}c}^P
+ \Delta P^{g} \,, 
\end{eqnarray}
in terms of the Higgs-fermion-fermion couplings given in Eq.~(\ref{eq:hff}) 
supplemented by the two residual form factors 
of $\Delta S^{g}$ and $\Delta P^{g}$.
We have
$S^g_{\rm SM}=0.636 + 0.071\,i$ and $P^g_{\rm SM}=0$
in the SM limit.
%

\subsection{Theoretical Signal Strengths}
\label{subsec:throretical_signal_strength}
In this work, to calculate the theoretical signal strengths,
we adopt the factorization assumption under which
the production and decay processes are well separated like as in the
resonant $s$-channel Higgs production in the 
the narrow-width approximation and neglect
nonresonant and interference effects.
Under the assumption,
for a specific production-times-decay process,
each theoretical signal strength is given by
the product of the production and decay signal strengths as
\begin{equation}
\mu({\cal P},{\cal D}) \simeq
\widehat\mu({\cal P})\ \widehat\mu({\cal D})\,.
\end{equation}
Here ${\cal P}$ stands for one of the six productions processes
of ggF, VBF, WH, ZH, ttH, and tH 
while ${\cal D}$ one of 
the six decay products of
$\gamma\gamma$, $ZZ^*$, $WW^*$, $bb$,
$\tau\tau$, and $\mu\mu$ 
which are supposed to be visible at the LHC in this work.

\medskip

The productions signal strength for the production process ${\cal P}$ is
given by
\begin{equation}
\widehat\mu({\cal P)}=\frac{\sigma_{\cal P}}{\sigma^{\rm SM}_{\cal P}}\,.
\end{equation}
On the other hand,
the decay signal strength for the process $H\to {\cal D}$ is given by
\begin{equation}
\label{eq:decay_signal_strength}
\widehat\mu({\cal D}) = \frac{B(H\to {\cal D})}{B(H_{\rm SM}\to {\cal D})} =
\frac{\Gamma(H\to{\cal D})}{\Gamma(H_{\rm SM}\to{\cal D})}\times
\frac{\Gamma_{\rm tot}(H_{\rm SM})}{\Gamma_{\rm tot}(H)+\Delta\Gamma_{\rm tot}}\,,
\end{equation}
with the branching fraction of each decay mode defined by
\begin{equation}
\label{eq:dgam}
B(H\to {\cal D})=\frac{\Gamma(H\to{\cal D})}
{\Gamma_{\rm tot}(H)+\Delta\Gamma_{\rm tot}}\,, \ \ \
B(H_{\rm SM}\to {\cal D})=\frac{\Gamma(H_{\rm SM}\to{\cal D})} 
{\Gamma_{\rm tot}(H_{\rm SM})}\,. 
\end{equation}
Note that an arbitrary non-SM contribution $\Delta\Gamma_{\rm tot}$
to the total decay width is introduced to
parametrize invisible Higgs decays into 
non-SM and/or undetected particles.
\footnote{Precisely, we calculate the total Higgs decay width
by summing over the 9 partial widths of the Higgs decay modes into $b\bar b$,
$WW^*$, $gg$, $\tau^+\tau^-$, $c\bar c$, $ZZ^*$, $\gamma\gamma$, $Z\gamma$,
and $\mu^+\mu^-$. Therefore, $\Delta\Gamma_{\rm tot}$ addresses either
all the Higgs decays into particles other than these 9 final states
such as dark matter and light quarks
or decays into $gg$, $c\bar c$, and $Z\gamma$ beyond the SM.
}
We observe that the partial and total decay widths of
$\Gamma(H\to {\cal D})$ and $\Gamma_{\rm tot}(H)$ 
becomes the SM ones of
$\Gamma(H_{\rm SM}\to {\cal D})$ and $\Gamma_{\rm tot}(H_{\rm SM})$, 
respectively, 
when $g^S_{H\bar{f}f}=1$,
$g^P_{H\bar{f}f}=0$,
$g_{_{HWW,HZZ}}=1$, and
$\Delta S^{\gamma,g}= \Delta P^{\gamma,g}=0$.
To calculate the decay widths of a generic Higgs boson,
we meticulously follow the recent review on 
the decays of Higgs bosons~\cite{Choi:2021nql}
which provides explicit analytical expressions 
and supplemental materials 
for the individual partial decay widths 
as precisely as possible 
by including the state-of-the-art theoretical calculations
of QCD corrections together with the SM electroweak corrections.

\medskip

At leading order (LO), the six production signal strengths are given by
\begin{eqnarray}
\label{eq:production_signastrength}
\widehat\mu({\rm ggF})^{\rm LO} &=&
\frac{\left|S^g(M_H)\right|^2+\left|P^g(M_H)\right|^2}
{\left|S^g_{\rm SM}(M_H)\right|^2}\,; \ \ \
\widehat\mu({\rm bbH}) = \left(g^S_{H\bar{b}b}\right)^2
+\left(g^P_{H\bar{b}b}\right)^2\,,
\nonumber \\[2mm]
\widehat\mu({\rm VBF}) &=& 
0.73\,g_{_{HWW}}^2 \ + \ 0.27\,g_{_{HZZ}}^2\,, 
\nonumber \\[2mm]
\widehat\mu({\rm WH}) &=&  g_{_{HWW}}^2\,,  
\nonumber \\[2mm]
\widehat\mu({\rm ZH}) &=&  g_{_{HZZ}}^2\,,  
\nonumber \\[2mm]
\widehat\mu({\rm ttH}) &=& \left(g^S_{H\bar{t}t}\right)^2
+\left(g^P_{H\bar{t}t}\right)^2\,,
\nonumber \\[2mm]
\widehat\mu({\rm tH}) &=&2.99\,\left[(g_{H\bar t t}^S)^2+(g_{H\bar t t}^P)^2\right]
\ + \ 3.38\,g_{_{HWW}}^2 \ - \ 5.37\,g_{H\bar t t}^S\,g_{_{HWW}}\,, 
\end{eqnarray}
in terms of the relevant form factors and couplings.
For $\widehat\mu({\rm VBF})$, we consider the $W$- and
$Z$-boson fusion processes separately and the 
decomposition coefficients are given by the ratios
of the SM cross sections of
$\sigma^{\rm SM}_{WW\to H}/\sigma^{\rm SM}_{\rm VBF}\simeq 0.73$ and
$\sigma^{\rm SM}_{ZZ\to H}/\sigma^{\rm SM}_{\rm VBF}\simeq 0.27$ 
which are largely independent of $\sqrt{s}$.

\begin{table}[h]
\centering
\caption{\it
\label{tab:thx_lo}
The LO tHq and tHW cross sections in fb at $\sqrt{s}=13$ TeV
for the three values of $g^S_{H\bar t t}$ taking  $g^P_{H\bar t t}=0$.
To calculate the cross sections, we use
{\tt MG5\_aMC@NLO}~\cite{Alwall:2014hca}
with {\bf NN23LO} PDF set~\cite{Ball:2012cx} and
no generator-level cuts are applied.
$M_H=125$ GeV and $M_t=172.5$ GeV are taken. 
\\ }
\begin{tabular}{c||c|c|c}
\hline
$\sqrt{s}=13$ TeV & $g^S_{H\bar t t}=+1$ (SM) &
$g^S_{H\bar t t}=0$ & $g^S_{H\bar t t}=-1$ \\ 
\hline
tHq & 71.8 & 264 & 890 \\
tHW & 17.2 & 36.7 & 155 \\
\hline
\end{tabular}
\end{table}

\medskip 
For the tH production, we consider the two main production processes
of $q^\prime b \to t H  q$  (tHq) and $g b \to t H W$ (tHW).
In the 5-flavor scheme (5FS), the LO tHq process is mediated by
the $t$-channel exchange of the $W$ boson with $H$ radiated from $W$ or $t$.
The LO tHW process is mediated by
the $s$-channel exchange of the $b$ quark with $H$, 
again, radiated from $W$ or $t$.
Both the production processes contain the two types of diagrams which involve
the top-Yukawa and gauge-Higgs couplings and the destructive
interference between them is very significant, leading to the large
negative value for the coefficient of the term proportional to the product of
the $g^S_{H\bar t t}$ and $g_{_{HWW}}$ couplings.
Using the LO tHq and tHW cross sections for the three values 
of $g^S_{H\bar t t}=+1,0,-1$ taking $g^P_{H\bar t t}=0$, see Table~\ref{tab:thx_lo},
we derive the three decomposition coefficients of the
tHq and tHW processes as follows:
\begin{eqnarray}
\label{eq:mu_thx_13}
\widehat\mu({\rm tHq})&=& 3.02\,
\left[(g_{H\bar t t}^S)^2+(g_{H\bar t t}^P)^2\right] + 3.68\,g_{_{HWW}}^2
-5.70\,g_{H\bar t t}^S\,g_{_{HWW}}\,, \nonumber \\[2mm]
\widehat\mu({\rm tHW}) &=& 2.88\,\left[(g_{H\bar t t}^S)^2+(g_{H\bar t t}^P)^2\right]
+ 2.14\,g_{_{HWW}}^2
-4.02\,g_{H\bar t t}^S\,g_{_{HWW}}\,, 
\end{eqnarray}
and, by combining them using the SM LO cross sections given in 
Table~\ref{tab:thx_lo}, we finally have
\begin{equation}
\widehat\mu({\rm tH}) =
\frac{\sigma^{\rm SM}_{\rm tHq}\,\widehat\mu({\rm tHq})+
\sigma^{\rm SM}_{\rm tHW}\,\widehat\mu({\rm tHW})}
{\sigma^{\rm SM}_{\rm tHq}+\sigma^{\rm SM}_{\rm tHW}} = 
 2.99\, \left[(g_{H\bar t t}^S)^2+(g_{H\bar t t}^P)^2\right] 
+3.38\,g_{_{HWW}}^2
-5.37\,g_{H\bar t t}^S\,g_{_{HWW}}\,,
\end{equation}
which gives $\widehat\mu({\rm tH})$ in Eq.~\ref{eq:production_signastrength}.

\begin{table}[b]
\centering
\caption{\it
\label{tab:ggF_coeff}
The decomposition coefficients
$c^X_{qq^{(\prime)}}$, $c^X_{t\Delta}$, and
$c^X_{\Delta\Delta}$ for 
$\widehat\mu({\rm ggF}+{\rm bbH})$ beyond LO in QCD,
see Eq.~(\ref{eq:ggF_coeff}),
at $\sqrt{s}=7\oplus 8$ TeV and 13 TeV.
In the third and fourth lines, 
the LO coefficients obtained by using Eq.~(\ref{eq:spgg_numeric})
are also shown for comparisons.
The Run 1 and Run 2 LO coefficients
are slightly different because the ratio
$\sigma^{\rm SM}_{\rm bbH}/\sigma^{\rm SM}_{\rm ggF}$
depends on $\sqrt{s}$, see Eq.~(\ref{eq:ggF_coeff}).
$M_H=125$ GeV and $M_t=172.5$ GeV are  taken.
\\}
\begin{tabular}{l||c|c|c|c|c|c||c|c|c|c|c|c}
\hline
 & 
$c^S_{tt}$ & $c^S_{bb}$ & $c^S_{tb}$ & $c^S_{tc}$ &
$c^S_{t\Delta}$ & $c^S_{\Delta\Delta}$ &
$c^P_{tt}$ & $c^P_{bb}$ & $c^P_{tb}$ & $c^P_{tc}$ &
$c^P_{t\Delta}$ & $c^P_{\Delta\Delta}$ \\
\hline
$\sqrt{s}=7\oplus 8$ TeV & 
$1.050$ & $0.018$ & $-0.054$ & $-0.010$ & $1.814$ & $0.784$ &
$2.244$ & $0.018$ & $-0.104$ & $-0.019$ & $2.690$ & $0.806$ \\
$\sqrt{s}=13$ TeV & 
$1.043$ & $0.018$ & $-0.050$ & $-0.009$ & $1.778$ & $0.758$ &
$2.222$ & $0.018$ & $-0.096$ & $-0.018$ & $2.632$ & $0.779$ \\
\hline
LO ($\sqrt{s}=7\oplus 8$ TeV) & 
$1.1452$ & $0.0248$ & $-0.1433$ & $-0.0315$ & $3.3278$ & $2.4176$ &
$2.6507$ & $0.0264$ & $-0.2512$ & $-0.0508$ & $5.0629$ & $2.4176$ \\
LO ($\sqrt{s}=13$ TeV) & 
$1.1447$ & $0.0253$ & $-0.1433$ & $-0.0314$ & $3.3262$ & $2.4164$ &
$2.6494$ & $0.0269$ & $-0.2510$ & $-0.0508$ & $5.0604$ & $2.4164$ \\
\hline
\end{tabular}
\end{table}

\medskip

Note that the LO ggF production signal strength 
given in Eq.~(\ref{eq:production_signastrength})
should be reliable only when higher order corrections to 
the numerator and those to the denominator 
are largely canceled out in the ratio
$\widehat\mu({\rm ggF})=\sigma_{\rm ggF}/\sigma^{\rm SM}_{\rm  ggF}$.
But, unfortunately, it turns out that the QCD
corrections to the diagrams in which top quarks are running
and those to the diagrams in which bottom quarks are running
are significantly different~\cite{Harlander:2013qxa} and, accordingly,
the LO ggF production signal strength given in 
Eq.~(\ref{eq:production_signastrength}) is unreliable.
In this work, combing ggF and bbH beyond LO, we use
\footnote{To combine ggF and bbH, we use the SM cross sections 
for $g^S_{H\bar t t}=g^S_{H\bar b b}=g^S_{H\bar c c}=1$
given in Table~\ref{tab:ggF_cx}
obtained by using {\tt SusHi-1.7.0}.}
\begin{eqnarray}
\label{eq:ggF_coeff}
\widehat\mu({\rm ggF}+{\rm bbH}) &=&
\frac{\sigma_{\rm ggF}^{\rm SM}\,\widehat\mu({\rm ggF})
+\sigma_{\rm bbH}^{\rm SM}\,\widehat\mu({\rm bbH})}
{\sigma_{\rm ggF}^{\rm SM}+\sigma_{\rm bbH}^{\rm SM}}
\\[2mm]
&=& \sum_{X=S,P} \left[
c^X_{tt}\,(g^X_{H\bar t t})^2+
c^X_{bb}\,(g^X_{H\bar b b})^2+
c^X_{tb}\,(g^X_{H\bar t t}g^X_{H\bar b b})+
c^X_{tc}\,(g^X_{H\bar t t}g^X_{H\bar c c})+
c^X_{t\Delta}\,(g^X_{H\bar t t}\Delta X^g)+
c^X_{\Delta\Delta}\,(\Delta X^g)^2\right]\,,\nonumber 
\end{eqnarray}
with  the decomposition coefficients given in Table~\ref{tab:ggF_coeff}.
In comparison to Refs.~\cite{Bernon:2015hsa,Kraml:2019sis}, we exploit the 
N$^3$LO (NNLO) production cross sections 
obtained by using 
{\tt SusHi-1.7.0}~\cite{Harlander:2012pb,Harlander:2016hcx}
for several combinations of the CP-even (CP-odd) Yukawa couplings
having in mind the possibility that $H$ is a CP-mixed state.
We further consider the beyond-LO effects 
on the contributions from the triangle loops in which
non-SM heavy particles are running.
For the detailed derivation of the coefficients for
$\widehat\mu({\rm ggF})$ beyond LO in QCD, 
see Appendix~\ref{sec:appendix_A}.

%
\begin{table}[h]
\centering
\caption{\it
\label{tab:lhc_cxs}
The SM cross sections from
Ref.~\cite{LHCHiggsCrossSectionWorkingGroup:2016ypw}
used for Eq.~(\ref{eq:mu_mixed})
taking $M_H=125$ GeV:
ggF from Table 191, 
VBF from Tables 25 and 26, 
WH from Table 223, 
ZH from Table 225, 
ttH from Table 231, 
tHq from Table 237, 
and
bbH from Table 247. 
On the other hand,
the 13-TeV $tHW$ cross section
is from Ref.~\cite{Demartin:2016axk}.
\\ }
\begin{tabular}{c||c|c|c|c|c|c|c|c}
\hline
$\sqrt{s}$ (TeV) & ggF (pb) &
VBF (fb) & WH (fb) &
ZH (fb) & ttH (fb) & tHq (fb) & 
tHW (fb) & bbH (fb) \\
\hline
$7$ & $16.85$ & $1241.4$ & $~\,577.30$ &
$339.10$ & $88.78$ & $12.26$ & $-$ & $155.20$ \\
\hline
$8$ & $21.42$ & $1601.2$ & $~\,702.50$  &
$420.70$ & $133.0$ & $18.69$ & $-$  & $202.10$ \\
\hline
$13$ & $48.57$ & $3781.7$ & $1373.00$  &
$883.70$ & $507.2$ & $74.25$ & $15.2$ & $488.00$ \\
\hline
\end{tabular}
\end{table}
%
%

\medskip

For the mixed-production mode 
involved with two production processes or more,
we use the following production signal strength weighted by
the cross sections of all the production processes involved:
\begin{equation}
\label{eq:mu_mixed}
\widehat\mu({\cal Q}) =
\frac{\sum_{{\cal P}_i\subset {\cal Q}}\,
\sigma^{\rm SM}_{{\cal P}_i}\,\widehat\mu({\cal P}_i)}
{\sum_{{\cal P}_i\subset {\cal Q}}\,\sigma^{\rm SM}_{{\cal P}_i}}\,,
\end{equation}
where, for the SM cross sections, we adopt those given in 
Ref.~\cite{LHCHiggsCrossSectionWorkingGroup:2016ypw},
see Table~\ref{tab:lhc_cxs}.
Explicitly, for the Run 2 mixed-productions modes at $\sqrt{s}=13$ TeV, 
we use
\begin{eqnarray}
\label{eq:muP13}
\widehat\mu({\rm ggF}+{\rm bbH}\oplus{\rm VBF}) &=& 
0.928\, \widehat{\mu}({\rm ggF}+{\rm bbH})
+  0.072\, \widehat{\mu}({\rm VBF}) \nonumber \\[2mm]
&=& 0.969\,(g^S_{H\bar t t})^2 +  0.016\,(g^S_{H\bar b b})^2  
-0.046\,(g^S_{H\bar t t} g^S_{H\bar b b})
-0.009\,(g^S_{H\bar t t} g^S_{H\bar c c})   
+1.651\,(g^S_{H\bar t t} \Delta S^g) + 0.703\,(\Delta S^g)^2 \nonumber \\[2mm]
&+& 2.063\,(g^P_{H\bar t t})^2 + 0.017\,(g^P_{H\bar b b})^2  
- 0.089\,(g^P_{H\bar t t} g^P_{H\bar b b})
-0.016\,(g^P_{H\bar t t} g^P_{H\bar c c}) 
+2.444\,(g^P_{H\bar t t} \Delta P^g) + 0.724\,(\Delta P^g)^2 \nonumber \\[2mm]
&+&0.052\, g^2_{{}_{HWW}} + 0.019\, g^2_{{}_{HZZ}}\,;  \nonumber \\[2mm]
\widehat\mu({\rm WH}\oplus{\rm ZH}) &=& 
 0.608\, g_{_{HWW}}^2 + 0.392\, g_{_{HZZ}}^2\,; \nonumber \\[2mm]
\widehat\mu({\rm ttH}\oplus{\rm tH}) &=& 
1.299\,\left[(g_{H\bar t t}^S)^2+(g_{H\bar t t}^P)^2\right] + 
0.507\,g_{_{HWW}}^2
-0.806\,g_{H\bar t t}^S\,g_{_{HWW}}\,; \nonumber \\[2mm]
\widehat\mu({\rm ggF}+{\rm bbH}\oplus{\rm ttH}\oplus{\rm tH}) &=& 
0.988\, \widehat{\mu}({\rm ggF}+{\rm bbH})
+ 0.012\, \widehat{\mu}({\rm ttH}\oplus{\rm tH}) \nonumber \\[2mm]
&=& 1.047\,(g^S_{H\bar t t})^2 +  0.017\,(g^S_{H\bar b b})^2  
-0.049\,(g^S_{H\bar t t} g^S_{H\bar b b})
-0.009\,(g^S_{H\bar t t} g^S_{H\bar c c})   
+1.757\,(g^S_{H\bar t t} \Delta S^g) + 0.749\,(\Delta S^g)^2 \nonumber \\[2mm]
&+& 2.211\,(g^P_{H\bar t t})^2 + 0.018\,(g^P_{H\bar b b})^2  
- 0.095\,(g^P_{H\bar t t} g^P_{H\bar b b})
-0.017\,(g^P_{H\bar t t} g^P_{H\bar c c}) 
+2.601\,(g^P_{H\bar t t} \Delta P^g) + 0.770\,(\Delta P^g)^2 \nonumber \\[2mm]
&-&0.010\, g^S_{H\bar t t}g_{{}_{HWW}} + 0.006\, g^2_{{}_{HWW}}\,;  \nonumber \\[2mm]
\widehat\mu({\rm VBF}\oplus{\rm WH}\oplus{\rm ZH}) &=& 
0.685\,g_{_{HWW}}^2 + 0.315\,g_{_{HZZ}}^2\,.
\end{eqnarray}

\subsection{$\chi^2$ Analysis}
Once all the theoretical signal strengths 
$\mu({\cal Q},{\cal D})\simeq\widehat\mu({\cal Q})\,\widehat\mu({\cal D})$,
each of which is associated with 
the specific production process of ${\cal Q}$ and 
the decay mode $H\to{\cal D}$,
have been obtained, one may carry out a chi-square analysis.
For the Tevatron data in which observables 
are uncorrelated, each $\chi^2$ is given by
\begin{equation}
\chi^2({\cal Q},{\cal D}) =
\frac{\left[\mu({\cal Q},{\cal D})-\mu^{\rm EXP}({\cal Q},{\cal D})\right]^2}
{\left[\sigma^{\rm EXP}({\cal Q},{\cal D})\right]^2}\,,
\end{equation}
where $\mu^{\rm EXP}({\cal Q},{\cal D})$ and
$\sigma^{\rm EXP}({\cal Q},{\cal D})$ denote the experimentally measured
signal strength and the associated error, respectively.
For the LHC Run 1 and Run 2 data,
taking account of  correlation among the observables in each set of data,
we use $\chi^2$ for $n$ correlated observables:
\begin{equation}
\chi^2_n = \sum^n_{i,j=1}
\frac{(\mu_i-\mu^{\rm EXP}_i)}{\sigma^{\rm EXP}_i}\,
\left(\rho^{-1}\right)_{ij}\,
\frac{(\mu_j-\mu^{\rm EXP}_j)}{\sigma^{\rm EXP}_j}\,,
\end{equation}
where the indices $i,j$ count $n$ correlated production-times-decay modes
and $\rho$ denotes the relevant $n \times n$ correlation matrix
satisfying the relations of $\rho_{ij}=\rho_{ji}$ and $\rho_{ii}=1$.
If $\rho_{ij}=\delta_{ij}$, we note that $\chi^2_n $ reduces to
$$
\chi^2_n=\sum^n_{i=1}
\frac{(\mu_i-\mu^{\rm EXP}_i)^2}{(\sigma^{\rm EXP}_i)^2}\,,
$$
i.e., the sum of $\chi^2$ of each uncorrelated observable.

\medskip

For our chi-square analysis, we consider two statistical
measures: (i) goodness of fit (gof) quantifying the agreement with
the experimentally measured signal strengths in a given fit, 
and 
(ii) $p$-value against the SM
for a given fit hypothesis to be compatible with the SM one:
\footnote{
For the second statistical measure to test the SM null hypothesis
with $\mu=1$, we use the likelihood ratio 
$\lambda(1)=L(1)/L(\widehat\mu)$: 
see Eq.~(40.49) and below 
in the 2023 edition of the review ``{\bf 40. Statistics}" by G. Cowan in
Ref.~\cite{Workman:2022ynf}. Note that, 
in the limit where the data sample is very large,
the distribution of $-2\ln\lambda(1)=\chi^2_{\rm SM}-\chi^2_{\rm min}$
approaches a $\chi^2$ distribution with the number of degrees of freedom 
being equal to the number of fitting parameters.}
\[
\mbox{goodness of fit (gof) } =
\int^\infty_{\chi^2_{\rm min} } f [x, n] \, d x \,; \ \
\mbox{$p$-value against the SM} =
\int^{\infty}_{\chi^2_{\rm SM}-\chi^2_{\rm min}} f[x, m] \, d x  \;,
\]
where $n$ is the degree of freedom (dof)
and $m$ the number of fitting parameters against the SM null hypothesis with $\mu=1$.
In our case, we have $n=76-m$.
The probability density function is given by
\[
f[x,l] = \frac{x^{l/2 - 1} e^{-x/2} }{ 2^{l/2} \Gamma(l/2) }\,,
\]
with $\Gamma(l/2)$ being the gamma function.
The goodness of fit approaches
to 1 when the value of $\chi^2$ per degree of freedom 
becomes smaller.  
On the other hand,
the $p$-value for compatibility with the SM hypothesis
approaches to 1 when the test hypothesis becomes more and more
SM-like.

\section{Global fits}
\label{sec:global_fits}

Now we are ready to perform global fits of the
Higgs boson couplings to the full Higgs datasets collected at the LHC.
Throughout this section, we use  the following short notations for 
the 125 GeV Higgs $H$ couplings to the SM particles:
\begin{eqnarray}
&&
C_W=g_{_{HWW}} \,, \ \ \
C_Z=g_{_{HZZ}} \,; \nonumber \\[2mm]
&&
C_t^{S,P}=g^{S,P}_{H\bar t t} \,, \ \ \
C_c^{S,P}=g^{S,P}_{H\bar c c} \,, \ \ \
C_b^{S,P}=g^{S,P}_{H\bar b b} \,, \ \ \
C_\tau^{S,P}=g^{S,P}_{H\bar \tau \tau} \,, \ \ \
C_\mu^{S,P}=g^{S,P}_{H\bar \mu \mu} \,.
\end{eqnarray}
Depending on specific models, all the Higgs couplings are not independent.
For example, the Higgs couplings to the massive vector bosons could be the same
as in the SM and the Yukawa couplings could be the same separately
in the up- and down-quark and charged-lepton sectors.
In this case, we denote the couplings as:
\begin{equation}
C_V=C_W=C_Z\,; \ \ \
C_u^{S,P}=C_t^{S,P}=C_c^{S,P}\,, \ \ \
C_d^{S,P}=C_b^{S,P}\,, \ \ \
C_\ell^{S,P}=C_\tau^{S,P}=C_\mu^{S,P}\,. 
\end{equation}
Further, some of the Yukawa couplings  could be the same like as in
the four types of two Higgs doublet models (2HDMs)
which are classified according to the Glashow-Weinberg condition
\cite{Glashow:1976nt} to avoid unwanted
tree-level Higgs-mediated flavor-changing neutral currents (FCNCs).
In this case, we denote the Higgs couplings as:
\footnote{In this work, we adopt the conventions
and notations of 2HDMs as in Ref.~\cite{Lee:2021oaj}.}
\begin{eqnarray}
&&
C_f^{S,P}=C_u^{S,P}=C_d^{S,P}=C_\ell^{S,P}\, 
({\rm as ~in~ type \,{\text -}\, I~ 2HDM})\,, \nonumber \\[2mm]
&&
C_u^{S,P}\,,\ C_{d\ell}^{S,P}=C_d^{S,P}=C_\ell^{S,P}\, 
({\rm as ~in~ type \,{\text -}\, II~ 2HDM})\,, \nonumber \\[2mm]
&&
C_{ud}^{S,P}=C_u^{S,P}=C_d^{S,P}\,, \ C_\ell^{S,P}\, 
({\rm as ~in~ type \,{\text -}\, III~2HDM})\,, \nonumber \\[2mm]
&&
C_{u\ell}^{S,P}=C_u^{S,P}=C_\ell^{S,P}\,,\ C_d^{S,P}\, 
({\rm as ~in~ type \,{\text -}\, IV~2HDM})\,.
\end{eqnarray}
Last but not least, when
all the Higgs couplings of $C_V$ and $C_f^S$
to the SM particles scale with a
single parameter as in, for example, 
Higgs-portal~\cite{Cheung:2015dta} and/or inert Higgs models,
we denote the coupling as:
\begin{eqnarray}
C_{Vf}=C_W=C_Z=C_t^{S}=C_c^{S}=C_b^{S}=C_\tau^S=C_\mu^{S}\,.
\end{eqnarray}

\medskip

Our fits are categorized into the CP-conserving (CPC) 
and CP-violating (CPV) fits as in the 
previous studies~\cite{Cheung:2013kla,Cheung:2014noa,Cheung:2018ave}.
The CPC fits have been performed assuming that
$C_f^P=0$ and $\Delta P^\gamma=\Delta P^g =0$ and, in this case,
we have 10 varying fitting parameters which might be
grouped into the non-SM and SM ones
as follow:
\begin{eqnarray}
{\rm CPC}\ {\rm parameters} &=&
\Bigg\{\Delta\Gamma_{\rm tot}\,, \Delta S^\gamma\,, \Delta S^g\,
\Bigg\}_{\rm non-SM} \nonumber \\[2mm]
&\oplus&
\Bigg\{
C_V=\left\{C_W\,, C_Z\right\}\,,
C^S_f=\Big\{
C^S_u=\left\{C^S_t\,, C^S_c\right\}\,,
C^S_d=\left\{C^S_b\right\}\,,
C^S_\ell=\left\{C^S_\tau\,, C^S_\mu\right\} \Big\}
\Bigg\}_{\rm SM}\,.
\end{eqnarray}
The non-SM parameters describe the variation of the signal strengths
due to the $H$ couplings to non-SM particles such as
light invisible and heavy charged/colored ones.
One the other hand, the SM parameters address the changes of
the signal strengths due to
the $H$ couplings to the SM particles of the massive vector bosons ($C_V$)
and the up- and down-type quarks ($C_{u,d}^S$), and the charged leptons ($C_\ell^S$).
When the normalized Yukawa couplings of $H$ to the SM fermions are generation and flavor
independent, we have only one Yukawa parameter
$C_f^S=C_u^S=C_d^S=C_\ell^S$.
To be more general, one may separately vary the $H$ couplings to
the $W$ and $Z$ bosons ($C_W$ and $C_Z$)
without keeping the custodial symmetry between them
and those to the top and charm quarks ($C^S_t$ and $C^S_c$).
The couplings to a pair of
tau leptons and muons
($C^S_\tau$ and $C^S_\mu$)
also can be separately varied
without assuming the lepton universality
of the normalized charged-lepton Yukawa couplings.
In the SM limit, the non-SM parameters are vanishing and
all the SM ones take the SM value of 1.
On the other hand, in the CPV fits under the assumption that
$H$ is a CP-mixed state, we have the following extended set
of fitting parameters containing 17 parameters:
\begin{eqnarray}
\label{eq:cpv_parameters}
{\rm CPV}\ {\rm parameters} &=&
\Bigg\{\Delta\Gamma_{\rm tot}\,, 
\left\{\Delta S^\gamma\,,\Delta P^\gamma\right\}\,, 
\left\{\Delta S^g\,,\Delta P^g\right\}
\Bigg\}_{\rm non-SM} \nonumber \\[2mm]
&\oplus&
\Bigg\{
C_V=\left\{C_W\,, C_Z\right\}\,,
C^S_f=\Big\{
C^S_u=\left\{C^S_t\,, C^S_c\right\}\,,
C^S_d=\left\{C^S_b\right\}\,,
C^S_\ell=\left\{C^S_\tau\,, C^S_\mu\right\} \Big\}
\nonumber \\
&&\hspace{2.9cm}
\,,C^P_f=\Big\{
C^P_u=\left\{C^P_t\,, C^P_c\right\}\,,
C^P_d=\left\{C^P_b\right\}\,,
C^P_\ell=\left\{C^P_\tau\,, C^P_\mu\right\} \Big\}
\Bigg\}_{\rm SM}\,.
\end{eqnarray}

\subsection{CP-conserving fits}
%
%
%
\begin{table}[!t]
\centering
\caption{\label{tab:CPCN}
{\bf CPCn} fits and their subfits 
considered in this work. 
Varied parameters are denoted by $\surd$ in each 
subfit of {\bf CPCn} and
the SM value of either 0 or 1 is assumed otherwise.
}  \vspace{1mm}
\begin{adjustbox}{width= 15cm}
\begin{tabular}{c|c|c|c|c|c|c|c|c|c|c|c|c|c}
\hline
\multicolumn{2}{c|}{\multirow{2}{*}{Parameters}} &
\multicolumn{4}{c|}{\bf CPC1} &
\multicolumn{8}{c}{\bf CPC2}   \\ \cline{3-14} 
\multicolumn{2}{c|}{}  &  IU & HC & IH & I &
IUHC & HCC & CSB & I & II & III & IV & HP \\ \hline
\multirow{3}{*}{non-SM} &
$\Delta\Gamma_{\rm tot}$ & 
$\surd$ & 0 & 0 & 0 &
$\surd$ & 0 & 0 & 0 & 0 & 0 & 0 & $\surd$ \\ 
&  $\Delta S^\gamma$ & 
0 & $\surd$ & 0 & 0 &
$\surd$ & $\surd$ & 0 & 0 & 0 & 0 & 0 & 0 \\ 
& 
$\Delta S^g$ & 
0 & 0 & 0 & 0 &
0 & $\surd$ & 0 & 0 & 0 & 0 & 0 & 0 \\  \hline
\multirow{4}{*}{SM} & \multirow{2}{*}{$C_V$} &
\multirow{4}{*}{1} & 
\multirow{4}{*}{1} & 
\multirow{4}{*}{$\surd\,(C_{Vf})$} & 
\multirow{2}{*}{1} & 
\multirow{4}{*}{1} & 
\multirow{4}{*}{1} & 
$\surd\,(C_W)$ & 
\multirow{2}{*}{$\surd\,(C_V)$} & 
\multirow{2}{*}{1} & 
\multirow{2}{*}{1} & 
\multirow{2}{*}{1} & 
\multirow{4}{*}{$\surd\,(C_{Vf})$}  \\
& &   &  &  & & & &
$\surd\,(C_Z)$ &  & 
& & &  \\   \cline{2-2} \cline {6-6} \cline{9-13}
& \multirow{2}{*}{$C_f^S$} & 
 &  &  & \multirow{2}{*}{$\surd\,(C_f^S)$} & & &
\multirow{2}{*}{1} &  
\multirow{2}{*}{$\surd\,(C_f^S)$} & 
$\surd\,(C_u^S)$ & $\surd\,(C_{ud}^S)$ & $\surd\,(C_{u\ell}^S)$ &  \\
& &  &  &  &  &  &  &  & &
$\surd\,(C_{d\ell}^S)$ & 
$\surd\,(C_{\ell}^S)$ & $\surd\,(C_{d}^S)$ &  \\   [1mm]
\hline
\end{tabular}
\end{adjustbox}
\\[5mm]
\begin{adjustbox}{width= 15cm}
\begin{tabular}{c|c|c|c|c|c|c|c|c|c|c|c}
\hline
\multicolumn{2}{c|}{\multirow{2}{*}{Parameters}} &
\multicolumn{5}{c|}{\bf CPC3} &
\multicolumn{2}{c|}{\bf CPC4} &
\multicolumn{2}{c|}{\bf CPC5} &
{\bf CPC6}   \\ \cline{3-12} 
\multicolumn{2}{c|}{}  &  IUHCC & II & III &
IV & HP & A & HP & AHC &  LUB & CSBLUB \\ \hline
\multirow{3}{*}{non-SM} &
$\Delta\Gamma_{\rm tot}$ &
$\surd$ & 0 & 0 &
0 & $\surd$ & 0 & $\surd$ & 0 & 0 & 0  \\
&  $\Delta S^\gamma$ &
$\surd$ & 0 & 0 &
0 & $\surd$ & 0 & $\surd$ & $\surd$  & 0 & 0  \\
&
$\Delta S^g$ &
$\surd$ & 0 & 0 &
0 & 0 & 0 & $\surd$ & 0 & 0 & 0   \\  \hline
\multirow{6}{*}{SM} & 
\multirow{2}{*}{$C_V$} &
\multirow{6}{*}{1} &
\multirow{2}{*}{$\surd\,(C_V)$} &
\multirow{2}{*}{$\surd\,(C_V)$} &
\multirow{2}{*}{$\surd\,(C_V)$} &
\multirow{6}{*}{$\surd\,(C_{Vf})$} &
\multirow{2}{*}{$\surd\,(C_V)$} &
\multirow{6}{*}{$\surd\,(C_{Vf})$} &
\multirow{2}{*}{$\surd\,(C_V)$} &
\multirow{2}{*}{$\surd\,(C_V)$} &
$\surd\,(C_W)$ \\
& &  &  &  & & & &  & & &
$\surd\,(C_Z)$  \\   \cline{2-2} \cline{4-6} \cline{8-8} \cline{10-12}
& \multirow{4}{*}{$C_f^S$} & &  
$\surd\,(C_u^S)$ & $\surd\,(C_{ud}^S)$ & $\surd\,(C_{u\ell}^S)$ & & 
$\surd\,(C_u^S)$ & &  $\surd\,(C_u^S)$ &
$\surd\,(C_{u}^S)$ & $\surd\,(C_{u}^S)$   \\
 & &  & $\surd\,(C_{d\ell}^S)$ &  &  $\surd\,(C_{d}^S)$ &  &  $\surd\,(C_d^S)$ &  & 
$\surd\,(C_d^S)$ & $\surd\,(C_d^S)$ & $\surd\,(C_d^S)$    \\   
 & &  &  & $\surd\,(C_{\ell}^S)$ &   &  &  $\surd\,(C_\ell^S)$ &  & $\surd\,(C_\ell^S)$ &
$\surd\,(C_\tau^S)$ & $\surd\,(C_\tau^S)$    \\   
 & &  &  & &   &  &   &  &   &
$\surd\,(C_\mu^S)$ & $\surd\,(C_\mu^S)$    \\   [1mm]
\hline
\end{tabular}
\end{adjustbox}
\end{table}

We generically label the CPC fits as {\bf CPCn} with {\bf n} standing for the
number of fitting parameters. Since there are 10 parameters to fit most generally,
each {\bf CPCn} contain several subfits.
One can not exhaust all the possibilities and the CPC fits considered in
this work are listed here:
\footnote{In each subfit of {\bf CPCn} fits, 
note that the parameters not mentioned are assumed to take
the SM value of either 0 or 1.
}
\begin{itemize}
\item{\bf CPC1}: in this fit, we consider the four subfits as follows:
\begin{itemize}
\item{IU}: vary $\Delta\Gamma_{\rm tot}$ to accommodate invisible Higgs decays into 
light non-SM and/or undetected particles 
\item{HC}: vary $\Delta S^\gamma$ to parametrize the contributions to
$H\to\gamma\gamma$ from the triangle loops in which
heavy electrically charged non-SM particles are running 
\item{IH}: vary $C_{Vf}$ to address the case in which all the normalized
Higgs couplings
to the SM particles are the same like as in inert Higgs models
\item{I}: vary $C_f^S$ to address the case in which all the normalized
Yukawa couplings are the same like as type-I 2HDM
\end{itemize}
\item{\bf CPC2}: in this fit, we consider the eight subfits as follows:
\begin{itemize}
\item{IUHC}: vary $\{\Delta\Gamma_{\rm tot}\,,\Delta S^\gamma\}$ for the case in which
the light and heavy (electrically charged) non-SM particles coexist
\item{HCC}: vary $\{\Delta S^\gamma\,,\Delta S^g\}$ for the contributions to
ggF, $H\to gg$, and $H\to\gamma\gamma$ from
heavy non-SM particles which are electrically charged {\it and} colored
\item{CSB}: vary $\{C_W\,,C_Z\}$ separately for the case in which
the custodial symmetry between the $W$ and $Z$ bosons is broken
\item{I}: vary $\{C_V\,,C_f^S\}$ for the case in which all the Yukawa couplings are 
as in type-I 2HDM
\item{II}: vary $\{C_{u}^S\,,C_{d\ell}^S\}$ for the case in which the Yukawa
couplings are as in type-II 2HDM
\item{III}: vary $\{C_{ud}^S\,,C_{\ell}^S\}$ for the case in which the Yukawa
couplings are as in type-III 2HDM
\item{IV}: vary $\{C_{u\ell}^S\,,C_{d}^S\}$ for the case in which the Yukawa
couplings are as in type-IV 2HDM
\item{HP}:  vary $\{\Delta\Gamma_{\rm tot}\,,C_{Vf}\}$ to address the Higgs-portal
case in which
Higgs decays invisibly and all the Higgs couplings to the SM particles scale with
a single parameter 
\end{itemize}
\item{\bf CPC3}: in this fit, we consider the five subfits as follows:
\begin{itemize}
\item{IUHCC}: vary $\{\Delta\Gamma_{\rm tot}\,,\Delta S^\gamma\,,\Delta S^g\}$ 
for the case in which
the light and heavy (charged {\it and} colored) non-SM particles coexist
\item{II}: vary $\{C_V\,,C_{u}^S\,,C_{d\ell}^S\}$ for the case in which the Yukawa
couplings are as in type-II 2HDM
\item{III}: vary $\{C_V\,,C_{ud}^S\,,C_{\ell}^S\}$ for the case in which the Yukawa
couplings are as in type-III 2HDM
\item{IV}: vary $\{C_V\,,C_{u\ell}^S\,,C_{d}^S\}$ for the case in which the Yukawa
couplings are as in type-IV 2HDM
\item{HP}:  vary $\{\Delta\Gamma_{\rm tot}\,,\Delta S^\gamma\,,C_{Vf}\}$ 
to address the Higgs-portal case when there exist
heavy electrically charged non-SM particles 
in addition to light particles into which $H$ could decay
\end{itemize}
\item{\bf CPC4}: in this fit, we consider the two subfits as follows:
\begin{itemize}
\item{A}: vary $\{C_V\,,C_{u}^S\,,C_{d}^S\,,C_{\ell}^S\}$ for the case in which the Yukawa
couplings are as in aligned 2HDM (A2HDM)~\cite{Pich:2009sp}
\item{HP}:  vary $\{\Delta\Gamma_{\rm tot}\,,\Delta S^\gamma\,,\Delta S^g\,,C_{Vf}\}$ 
to address the Higgs-portal case when 
heavy charged {\it and} colored non-SM particles exist
in addition to light particles into which $H$ could decay
\end{itemize}
\item{\bf CPC5}: in this fit, we consider the following two subfits:
\begin{itemize}
\item{AHC}: vary $\{\Delta S^\gamma\,,C_V\,,C_{u}^S\,,C_{d}^S\,,C_{\ell}^S\}$ 
like as in {\bf CPC4}-A in the presence of heavy electrically charged particles such as
charged Higgs bosons contributing to $H\to\gamma\gamma$
\item{LUB}: vary 
$\{C_V\,,C_{u}^S\,,C_d^S\,,C_\tau^S\,,C_\mu^S\}$ to address the case in which
the lepton universality of the normalized Yukawa couplings to charged leptons is broken 
\end{itemize}
\item{\bf CPC6}: in this fit, we consider the following scenario:
\begin{itemize}
\item{CSBLUB}: vary 
$\{C_W\,,C_Z\,,C_{u}^S\,,C_d^S\,,C_\tau^S\,,C_\mu^S\}$
to address the most general case of the Higgs couplings to the 
SM particles involved under the constraint of $C_t^S=C_c^S=C_u^S$
\end{itemize}
\end{itemize}
We provide Table~\ref{tab:CPCN} to summarize all the {\bf CPCn} fits together 
with their subfits.
Note that we do not address the case in which the charm- and top-quark 
Yukawa couplings are different from each other in this work.
If only $C_c^S$ is fitted while all the other non-SM and SM parameters are 
fixed at their SM values, we have  $|C_c^S|\lsim 2$ at 95\% CL. 
\footnote{See Appendix~\ref{sec:appendix_B}.}
But, if other gauge-Higgs and Yukawa couplings are simultaneous varied
taking $C_c^S\neq C_t^S$, fitting 
to the 76 signal strengths considered in this work does 
not lead to the bounded results for the couplings.
From
a search for the Higgs boson decaying into a pair of charm quarks,
the ATLAS collaboration gives the observed (expected) constraints
of $|C_c^S|<8.5\,(12.4)$ at 95\% CL and
$|C_c^S/C_b^S|<4.5$ at 95\% CL (5.1 expected)~\cite{ATLAS:2022ers}.
The CMS collaboration gives the
observed (expected) 95\% CL value of
$1.1<|C_c^S|<5.5\,(|C_c^S|<3.40)$~\cite{CMS:2022psv}.
\begin{figure}[t!]
\vspace{-1.0cm}
\begin{center}
\includegraphics[width=13.5cm]{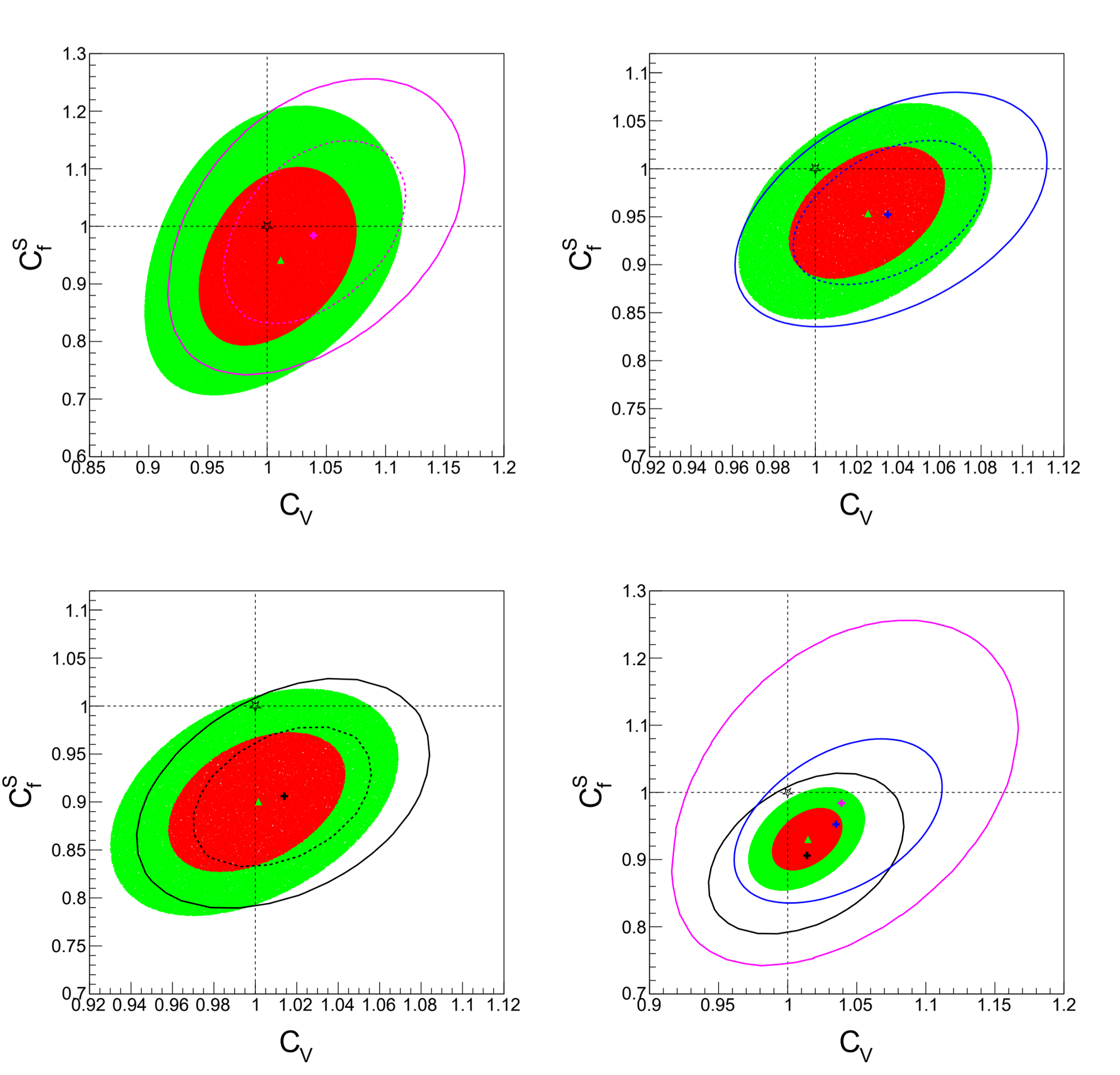}
\end{center}
\vspace{-0.5cm}
\caption{\it
The 68\% (red) and 95\% (green) CL regions 
in the $(C_V,C_f^S)$ planes
obtained using the ATLAS$\oplus$CMS Run 1 (upper-left),
ATLAS Run 2 (upper-right), and CMS Run 2 (lower-left) 
experimental signal strengths in
Tables~\ref{tab:78all}, \ref{tab:ATLAS13}, and \ref{tab:CMS13},
respectively, and the theoretical ones in
subsection~\ref{subsec:throretical_signal_strength}.
The couplings of $C_V$ and $C_f^S$ are varied as in {\bf CPC2}-I.
Also shown are the dashed and solid ellipses 
enclosing the 68\% and 95\% CL regions, respectively,
presented in
Refs.~\cite{ATLAS:2016neq} (upper-left), 
\cite{ATLAS:2022vkf} (upper-right), and \cite{CMS:2022dwd} (lower-left).
In the lower-right frame, the CL regions obtained 
using the full LHC Run 1 and Run 2 data 
are shown together with the solid magenta, blue, and
black ellipses for the 95\% CL regions presented in
Refs.~\cite{ATLAS:2016neq}, \cite{ATLAS:2022vkf}, 
and \cite{CMS:2022dwd}.
The colors and lines are the same in all the frames
and
the vertical and horizontal lines denote the SM values
of $C_V=1$ and $C_f^S=1$ with
the best-fit points denoted by triangles (colored regions) 
and pluses (ellipses).
The SM points where $C_V=C_f^S=1$ are denoted by stars.
} 
\label{fig:CPC2-I.comparisons}
\end{figure}

\medskip

Before presenting the results of the {\bf CPCn} fits and their subfits and discussing 
details of them separately, 
we make comparisons of the 68\% and 95\% confidence-level (CL) regions
presented in 
Refs.~\cite{ATLAS:2016neq}, \cite{ATLAS:2022vkf}, and \cite{CMS:2022dwd}
with those obtained by using the LHC Run 1 and Run 2 experimental
signal strengths taken in this work
\footnote{Precisely, we mean the 76 signal strengths in
Tables~\ref{tab:tev}, \ref{tab:78all}, \ref{tab:ATLAS13}, and \ref{tab:CMS13}.}
and the theoretical ones elaborated in
subsection~\ref{subsec:throretical_signal_strength}.
To be specific, we have taken {\bf CPC2}-I subfit in which
the couplings $C_V$ and $C_f^S$ are varied and 
the ATLAS$\oplus$CMS Run 1, ATLAS Run 2, and CMS Run 2 CL regions
are taken from 
Fig.~26 in Ref.~\cite{ATLAS:2016neq}, 
Fig.~4 in Ref.~\cite{ATLAS:2022vkf}, and 
Fig.~3 in Ref.~\cite{CMS:2022dwd}, respectively:
see the regions inside the dashed (68\%) and solid (95\%) ellipses in
the upper and lower-left frames of Fig.~\ref{fig:CPC2-I.comparisons}.
The CL regions obtained in this work
are colored in red (68\%) and green (95\%).
We observe that the best-fit values for $C_f^S$ agrees excellently for
ATLAS Run 2 and CMS Run 2 while, for ATLAS$\oplus$CMS Run 1, our value is 
smaller by the amount of about $0.04$ which corresponds to 
about 0.5-$\sigma$ level. 
On the other hand, our best-fit values for $C_V$ are nearer to the SM point
and the differences are at the level 
below $0.7\sigma$ (Run 1) and $0.5\sigma$ (Run 2).
The $1\sigma$ errors agree well except $C_V$ of ATLAS Run 2 for which
we have obtained 
$C_V^{\rm this\,work}=1.025\pm 0.025$ while
$C_V^{\rm ATLAS\text{\cite{ATLAS:2022vkf}}}=1.035\pm 0.031$:
the lower edges of the two $1\sigma$ regions are around 1 while
our upper edge reduces to the SM direction by the amount
$0.016$  which corresponds to about 0.5-$\sigma$ level.
From these critical comparisons, we conclude that
our global fits to the Higgs signal strengths in
Tables~\ref{tab:78all}, \ref{tab:ATLAS13}, and \ref{tab:CMS13}
using the theoretical signal strengths given in 
subsection~\ref{subsec:throretical_signal_strength} 
remarkably reproduces the fitting results in 
Ref.~\cite{ATLAS:2016neq} (Run 1) and Refs.~\cite{ATLAS:2022vkf,CMS:2022dwd} (Run 2) 
within the 0.5-$\sigma$ level.
Further we have arrived at the conclusion that
the combined results of our precision analysis of the full LHC Run 1 
and Run 2 data should be reliable
better than the 0.5-$\sigma$ level
since Run 2 data are now statistically dominant and
our Run 2 results are more consistent with those in 
Refs.~\cite{ATLAS:2022vkf,CMS:2022dwd}.
Lastly, we present the fully combined results 
in the lower-right frame of Fig.~\ref{fig:CPC2-I.comparisons}:
the 68\% (red) and 95\% (green) CL regions are
obtained from the full LHC Run 1 and Run 2 data.
For comparisons, we also show
the 95\% CL magenta, blue, and black solid ellipses from
Refs.~\cite{ATLAS:2016neq}, \cite{ATLAS:2022vkf}, and \cite{CMS:2022dwd},
respectively, which are the same as in the
upper-left, upper-right, and lower-left frames, respectively.
From Run 1 to Run 2, we observe that $C_V$ approaches to the SM value of 1
while $C_f^S$ deviates from it, see the points marked by pluses
in the lower-right frame. The combined results gives
$C_V=1.015\pm 0.017$ and $C_f^S=0.930\pm 0.031$ and the SM point
denoted by a star locates just outside of the 95\% CL region.
The deviation  of $C_f^S$ from its SM value of 1 has been noticed 
not only in Refs.~\cite{ATLAS:2022vkf,CMS:2022dwd} but also
in Ref.~\cite{Biekotter:2022ckj}
and our combined analysis strengthens the observation
by showing that its best-fitted value is more than
2 standard deviations below the SM prediction.

\subsubsection{{\bf CPC1}}
%
\begin{table}[!b]
\centering
\caption{\label{tab:CPC1}
{\bf CPC1}: The best-fitted values in the four {\bf CPC1} subfits. 
Also shown are the corresponding minimal
chi-square per degree of freedom ($\chi^2_{\rm min}$/dof), 
goodness of fit (gof), and 
$p$-value against the SM for compatibility
with the SM hypothesis.
For the SM, we obtain $\chi^2_{\rm SM}/{\rm dof}=82.3480/76$
and gof $=0.2895$.
}  \vspace{1mm}
\begin{adjustbox}{width=12cm}
\begin{tabular}{c|c|c|c|c|c}
\hline
\multicolumn{2}{c|}{\multirow{2}{*}{Parameters}} &
\multicolumn{4}{c }{\bf CPC1} \\ \cline{3-6}
\multicolumn{2}{c|}{}  &  IU & HC & IH & I \\ \hline
\multirow{3}{*}{non-SM} &
$\Delta\Gamma_{\rm tot}/$MeV& 
$-0.042^{+0.142}_{-0.132}$ & 0 & 0 & 0 \\
&  $\Delta S^\gamma$ & 
0 & $-0.313^{+0.176}_{-0.176}$ & 0 & 0 \\
& $\Delta S^g$ & 0 & 0 & 0 & 0 \\ \hline
\multirow{4}{*}{SM} & \multirow{2}{*}{$C_V$} &
\multirow{4}{*}{1} & 
\multirow{4}{*}{1} & 
\multirow{4}{*}{$C_{Vf}=1.005^{+0.017}_{-0.017}$} &
\multirow{2}{*}{1}  \\
& &   &  &  & \\ \cline{2-2} \cline{6-6}
& \multirow{2}{*}{$C_f^S$} & &  &  &
\multirow{2}{*}{$C_f^S=0.920^{+0.029}_{-0.029}$}  \\
& &  &  &  & \\ [1mm]
\hline
\multicolumn{2}{c|}{$\chi^2_{\rm min}$/dof} & $82.2540/75$ & $79.2183/75$ & $82.2568/75$ &
$75.0931/75$ \\
\multicolumn{2}{c|}{goodness of fit (gof)} & $0.2649$ & $0.3474$ & $0.2649$ &
$0.4753$ \\
\multicolumn{2}{c|}{$p$-value against the SM} & $0.7590$ & $0.0769$ & $0.7626$ &
$0.0071$ \\
\hline
\end{tabular}
\end{adjustbox}
\end{table}
\begin{figure}[t!]
\begin{center}
\includegraphics[width=\textwidth]{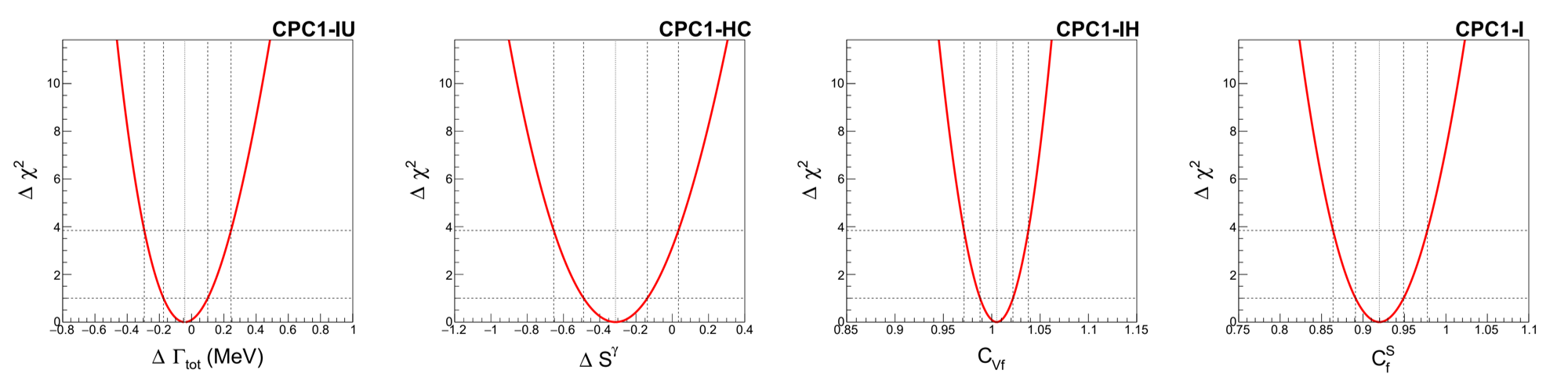}
\end{center}
\vspace{-0.5cm}
\caption{\it
{\bf CPC1}: $\Delta\chi^2$ above the minimum versus
$\Delta\Gamma_{\rm tot}$ in MeV (left),
$\Delta S^\gamma$ (middle-left), 
$C_{Vf}$ (middle-right), and $C_f^S$ (right) in the IU, HC, IH, and I subfits, respectively.
The vertical and horizontal lines find the $1\sigma$ errors and 
95\% confidence intervals.
} 
\label{fig:CPC1}
\end{figure}

We show the fitting results for the four subfits of {\bf CPC1} 
and $\Delta\chi^2$ above each minimum 
in Table~\ref{tab:CPC1} and Fig.~\ref{fig:CPC1}, respectively.
The $p$-values against the SM 
for compatibility with the SM hypothesis
are high for IU and IH but they are only 8\% (HC) and 1\% (I).

\medskip

In IU where we consider the case in which there exist 
light non-SM 
invisible and/or undetected particles and the 125 GeV Higgs boson $H$ 
can decay into them, we obtain
\footnote{
Note that the asymmetric upper and lower $1\sigma$
errors indicate non-Gaussianity in the fitting results,
as shown in the left frame of Fig.~\ref{fig:CPC1}.
For each one-parameter fit, we derive the 
$1\sigma$ errors and the 95\% CL range from
the $\Delta\chi^2=1$ and $3.84$ lines, respectively, in the plot of
$\Delta\chi^2$ distribution above the corresponding minimum, see Fig.~\ref{fig:CPC1}.
In fits with more than two parameters, even with non-Gaussianities, we derive 
the 68.27\%, 95\%, and 99.73\% CL regions in two-parameter planes 
from the contours of $\Delta\chi^2=2.3$, $5.99$, and $11.83$,  assuming
Gaussian distributions, see Fig.~\ref{fig:CPC2_CLregions} for example. 
Therefore, be cautious when interpreting our fitting results if
non-Gaussianities are significant.}
\begin{equation}
\Delta\Gamma_{\rm tot} \ = \ -0.042^{+0.142}_{-0.132} \ {\rm MeV}\,.
\end{equation}
The central value 
is below zero by the amount of about 1\% of
the total SM Higgs decay width $\Gamma_{\rm tot}(H_{\rm SM})=4.059-4.128$ MeV
in the range between $M_H=125.0-125.5$ GeV.
Considering the current theoretical and parametric uncertainties of
$0.6 \sim 1$\%~\cite{LHCHiggsCrossSectionWorkingGroup:2016ypw} involved in
the calculation of the total SM Higgs decay width around $M_H=125$ GeV, 
we observe that it is consistent with zero. 
The gof value is $0.2649$ which is a little bit worse than the SM (gof $=0.2895$)
as consistently indicated by chi-square per degree of freedom:
$\left(\chi^2_{\rm min}/{\rm dof}\right)^{\rm{\bf CPC1}-IU}=82.2540/75=1.09672$ and
$\left(\chi^2_{\rm SM}/{\rm dof}\right)^{\rm SM}=82.3480/76=1.08353$.
The 95\% CL range is given by $\Delta\Gamma_{\rm tot}=-0.042^{+0.287}_{-0.251}$ MeV, 
see the left frame of Fig.~\ref{fig:CPC1}.
Since the negative central value of $\Delta\Gamma_{\rm tot}$ is unphysical,
we take the upper error of $0.287$ MeV as the conservative upper limit to obtain
the following limit on the non-SM branching ratio at 95\% CL:
\begin{equation}
B(H \to {\rm non\!\!-\!\!SM}) < 7.1\%\,,
\end{equation}
which is better than the combined 95\% CL limit
of either 10.7\% (ATLAS: 7.7\% expected)~\cite{ATLAS:2023tkt} or 
15\% (CMS: 8\% expected)~\cite{CMS:2023sdw} observed
in searches for decays of the Higgs boson to
invisible particles.

\medskip

In HC where the gauge-Higgs and Yukawa couplings
of the Higgs boson are the same as in the SM while there exist heavy electrically charged
non-SM particles which could modify the loop-induced Higgs couplings to two photons,
we obtain
\begin{equation}
\Delta S^\gamma = -0.313\pm 0.176\,,
\end{equation}
which shows a $1.8\sigma$ deviation from the SM. This could be understood by observing
that the combined decay signal strength of the $H\to\gamma\gamma$ mode is
$1.10\pm 0.07$, see Table~\ref{tab:pdss76}.
Note that $S^\gamma_{\rm SM}=-6.542 + 0.046\,i$
and $\Delta S^\gamma/S^\gamma_{\rm SM} \simeq 0.048 \pm 0.027$.
The gof value is $0.3474$ which is definitely better than the SM.
Incidentally, we obtain the following 95\% CL region and limit:
\begin{equation}
-0.654 < \Delta S^\gamma < 0.034\,; \ \ \
\left|\frac{\Delta S^\gamma}{S^\gamma_{\rm SM}}\right| < 0.1\,,
\end{equation}
see the middle-left frame of Fig.~\ref{fig:CPC1}.

\medskip

In IH where all the Higgs couplings to the SM particles scale
with a single coupling 
$C_{Vf}=C_W=C_Z=C_t^{S}=C_c^{S}=C_b^{S}=C_\tau^S=C_\mu^{S}$,
we obtain
\begin{equation}
C_{Vf} = 1.005 \pm 0.017.
\end{equation}
The gof value is the same as in  {\bf CPC1}-IU.
Actually, in {\bf CPC1}-IH and {\bf CPC1}-IU, 
all the production and decay processes scale with the overall 
single theoretical signal strength as follows:
\begin{equation}
\mu({\cal P},{\cal D})^{{\bf CPC1}-{\rm IH}} =  C_{Vf}^2\,; \ \ \
\mu({\cal P},{\cal D})^{{\bf CPC1}-{\rm IU}} = 
\frac{1}{1+\Delta\Gamma_{\rm tot}/\Gamma_{\rm tot}(H_{\rm SM})} \,.
\end{equation}
Accordingly,
the best-fitted values are consistent with the 
global signal strength of
$\mu^{\,\rm Global}_{\,\rm 76\,signal\,strengths} = 1.012 \pm 0.034$,
see Eq.~(\ref{eq:globalss_76}), which leads to the best-fit point 
deviated from the SM one
by the amount of about $+1$\% with about $\pm 3$\% error in terms of
signal strength.

\medskip

In I where all the Yukawa couplings to the SM fermions scale with the same
coupling parameter $C_f^S=C_u^S=C_d^S=C_\ell^S$
like as in type-I 2HDM but $C_V$ is fixed at its SM value of 1, we obtain
\begin{equation}
C_f^S = 0.920 \pm 0.029\,.
\end{equation}
Note that we have the highest gof value of $0.4753$ which is
larger than the {\bf CPC2}-IV gof value of $0.4699$ though slightly,
see Table~\ref{tab:CPC2}.
We find that this simple one-parameter fit gives the best 
gof value among the {\bf CPC} and {\bf CPV} fits
considered in this work, see Fig.~\ref{fig:GOF}.

\subsubsection{{\bf CPC2}}
%
\begin{table}[!b]
\caption{\it
\label{tab:CPC2}
{\bf CPC2}: The best-fitted values in the eight {\bf CPC2} subfits. 
Also shown are the corresponding minimal
chi-square per degree of freedom ($\chi^2_{\rm min}$/dof), 
goodness of fit (gof), and 
$p$-value against the SM for compatibility
with the SM hypothesis.
For the SM, we obtain $\chi^2_{\rm SM}/{\rm dof}=82.3480/76$
and gof $=0.2895$.
}  \vspace{1mm}
\renewcommand{\arraystretch}{1.1}
\begin{adjustbox}{width= \textwidth}
\begin{tabular}{c|c|c|c|c|c|c|c|c|c}
\hline
\multicolumn{2}{c|}{\multirow{2}{*}{Parameters}} &
\multicolumn{8}{c}{\bf CPC2} \\ \cline{3-10}
\multicolumn{2}{c|}{} & IUHC & HCC & CSB & I & II & III & IV & HP \\ \hline
\multirow{3}{*}{non-SM} &
$\Delta\Gamma_{\rm tot}$/MeV &
$~~0.090^{+0.168}_{-0.157}$ & 0 & 0 & 0 & 0 & 0 & 0 & $0.0^{+0.105}$ \\
&  $\Delta S^\gamma$ &
$-0.369^{+0.202}_{-0.207}$ & $-0.400^{+0.196}_{-0.196}$ & 0 & 0 & 0 & 0 & 0 & 0 \\
&
$\Delta S^g$ &
0 & $-0.032^{+0.031}_{-0.031}$ & 0 & 0 & 0 & 0 & 0 & 0 \\  \hline
\multirow{4}{*}{SM} & \multirow{2}{*}{$C_V$} &
\multirow{4}{*}{1} &
\multirow{4}{*}{1} &
$C_W = 1.038^{+0.018}_{-0.018}$ &
\multirow{2}{*}{$1.015^{+0.017}_{-0.017}$} &
\multirow{2}{*}{1} &
\multirow{2}{*}{1} &
\multirow{2}{*}{1} &
\multirow{4}{*}{$C_{Vf}=1.0_{-0.013}$} \\
& &   &  &$C_Z = 0.999^{+0.030}_{-0.030}$  &  &  &  &  &  \\   \cline{2-2} \cline{5-9}
& \multirow{2}{*}{$C_f^S$} &
&  & \multirow{2}{*}{ 1 } &
\multirow{2}{*}{$0.930^{+0.031}_{-0.031}$} &
$C^S_u = 0.931^{+0.032}_{-0.032}$ &
$C^S_{ud} = 0.919^{+0.037}_{-0.035}$ &
$C^S_{u\ell} = 0.920^{+0.029}_{-0.028}$ &  \\
& &  &  &  &  &
$C^S_{d\ell} = 0.907^{+0.032}_{-0.032}$ &
$C^S_\ell = 0.921^{+0.038}_{-0.038}$ &
$C^S_{d} = 0.894^{+0.040}_{-0.039}$  &  \\   [1mm]
\hline
\multicolumn{2}{c|}{$\chi^2_{\rm min}$/dof}& 78.8971/74 & 78.1906/74 & 77.8970/74 &
74.3664/74 & 74.3154/74 & 75.0916/74 & 74.2510/74 & 82.3480/74  \\
\multicolumn{2}{c|}{goodness of fit (gof)}   &  0.3267 &  0.3473 &  0.3559 &  0.4662 &  0.4678
&  0.4427 &  0.4699 & 0.2369  \\
\multicolumn{2}{c|}{$p$-value against the SM}         &  0.1781 &  0.1252 &  0.1080 &  0.0185 &  0.0180
&  0.0266 &  0.0174 & 1.0  \\ \hline
\end{tabular}
\end{adjustbox}
\end{table}
%
%
\begin{figure}[b!]
\begin{center}
\includegraphics[width=9.1cm]{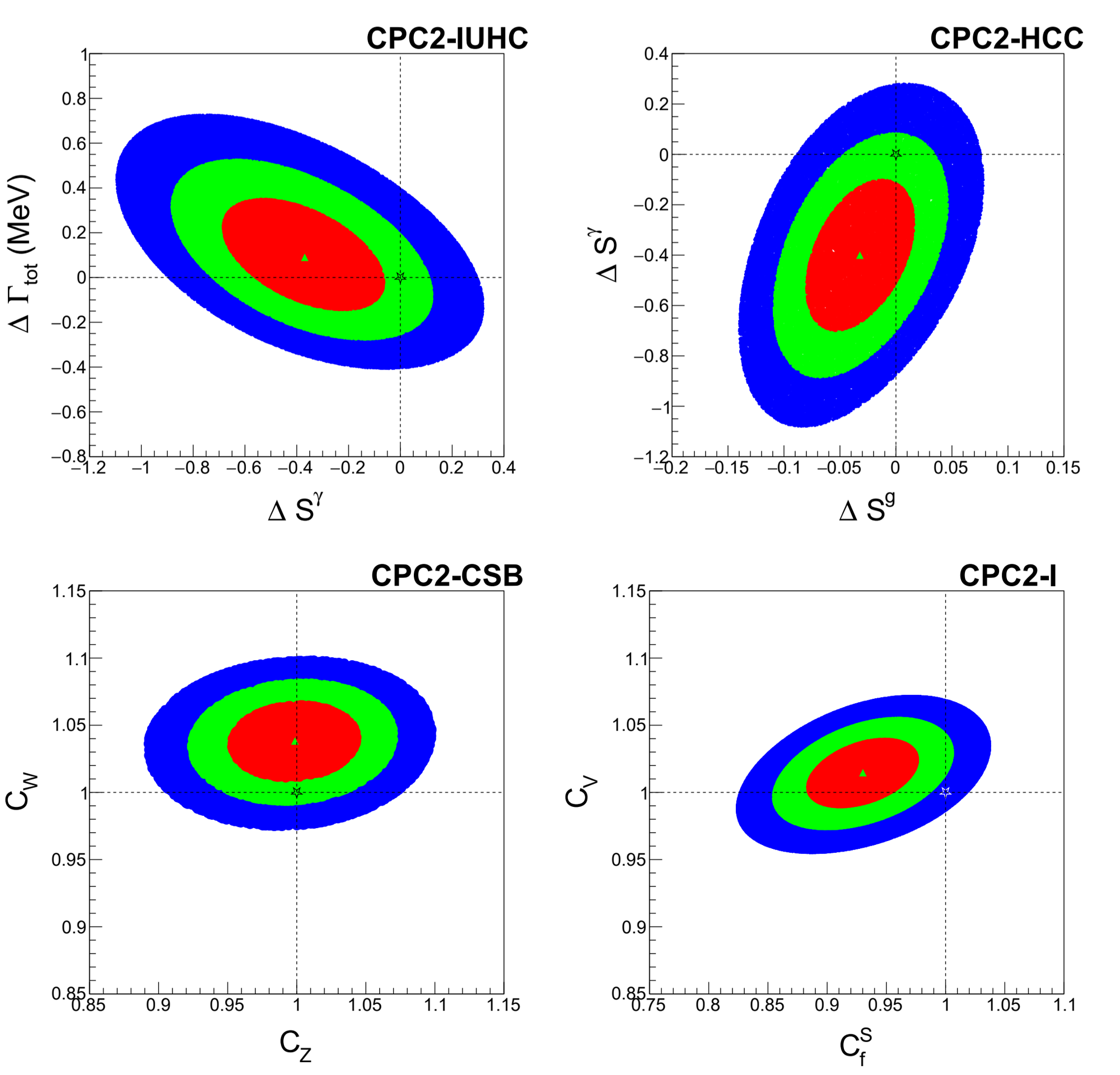}
\includegraphics[width=9.1cm]{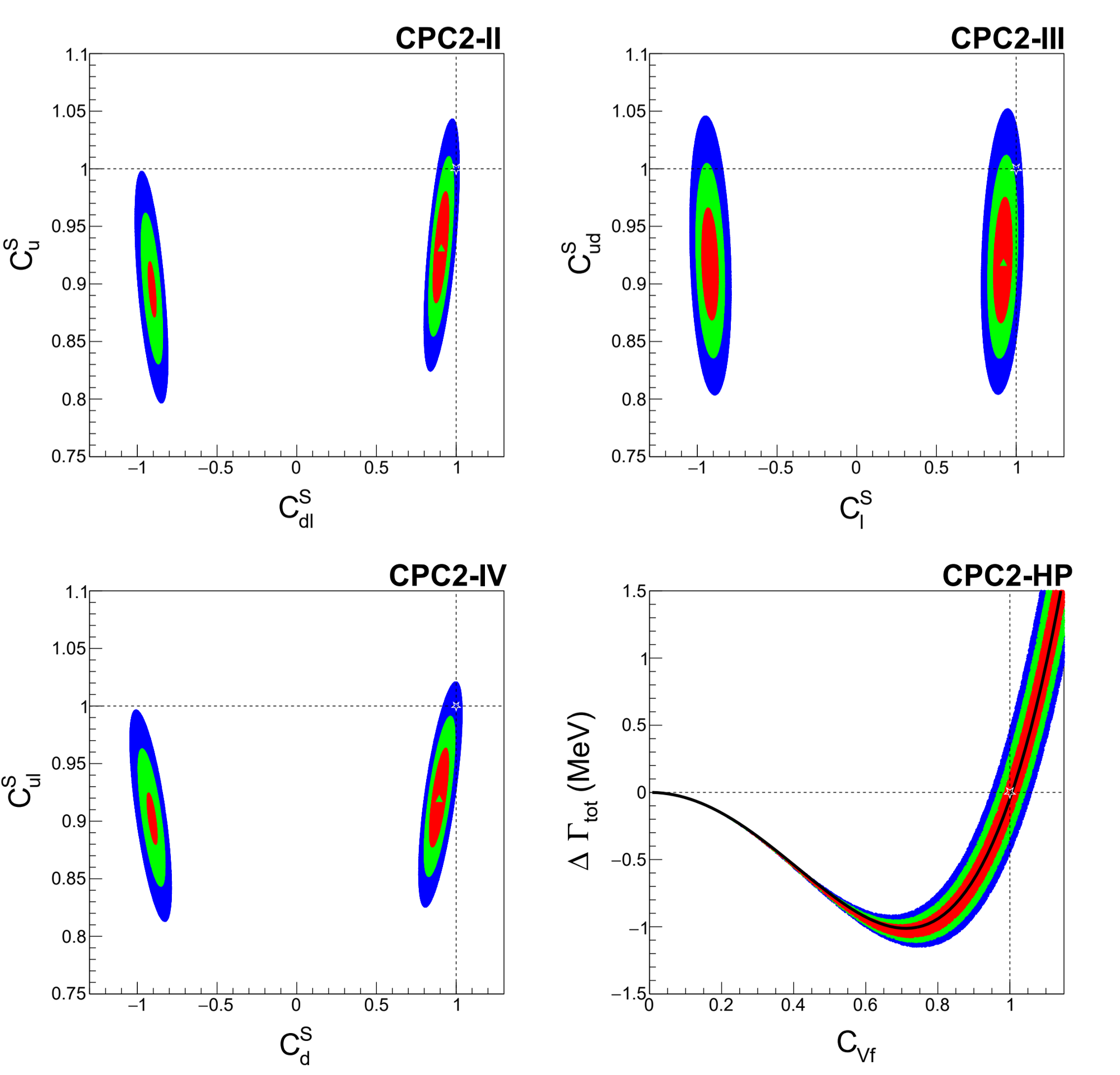}
\end{center}
\vspace{-0.5cm}
\caption{\it {\bf CPC2}: 
The CL regions of the eight {\bf CPC2} subfits in two-parameter planes.
The contour regions shown are for 
$\Delta\chi^2\leq 2.3$ (red), 
$\Delta\chi^2\leq 5.99$ (green), 
$\Delta\chi^2\leq 11.83$ (blue)
above the minimum, which correspond to
confidence levels of 68.27\%, 95\%, and 99.73\%, respectively. 
In each frame, the vertical and horizontal lines locate the SM point denoted
by a star and the best-fit point is denoted by a triangle.
}
\label{fig:CPC2_CLregions}
\end{figure}

We show the fitting results for the eight subfits of {\bf CPC2} 
in Table~\ref{tab:CPC2} and depict their CL regions
in Fig.~\ref{fig:CPC2_CLregions}.
In HP, we perform the fit under the constraints of
$\Delta\Gamma_{\rm tot}\geq 0$ and
$C_{Vf}=C_W=C_Z\leq 1$ and, only for this, we have the gof value worse than the SM.
Otherwise, the gof values range between $0.3267$ (IUHC) and $0.4699$ (IV) which
are indeed better than the SM.
The $p$-values against the SM 
for compatibility with the SM hypothesis
are only a few \% for I, II, III, and IV.

\medskip

In IUHC, we assume the simultaneous existence of the light non-SM particles into which
the Higgs boson $H$ decays and the heavy electrically charged non-SM particles
contributing to $H\to\gamma\gamma$ through the triangle loops. The gof value is better
than {\bf CPC1}-IU but a little bit worse than {\bf CPC1}-HC. Otherwise, the best-fitted
values are similar to those in {\bf CPC1} with a bit larger $1\sigma$ errors.
The SM point lies outside the 68\% CL region, see the upper-left frame of the left panel
of Fig.~\ref{fig:CPC2_CLregions}.

\medskip

In HCC, we assume the existence of the heavy electrically charged {\it and} colored
non-SM particles contributing to ggF and  $H\to\gamma\gamma$
through the triangle loops. 
The scalar form factor $\Delta S^\gamma$ deviates from the SM by
$2\sigma$ similarly as in {\bf CPC1}-HC and {\bf CPC2}-IUHC
while $\Delta S^g$ is consistent with the SM value of 0 within $1\sigma$.
The SM point
lies outside of the 68\% CL region, see the upper-right frame of the left panel
of Fig.~\ref{fig:CPC2_CLregions}. The negative central value of $-0.032$ of
$\Delta S^g$ decreases $|\Delta S^g/S^g_{\rm SM}|$ by the amount of 
about 5\% with $S^g_{\rm SM}=0.636+0.071\,i$
which contributes to the 1\% increment of 
the global signal strength
$\mu^{\,\rm Global}_{\,\rm 76\,signal\,strengths}$ since
\begin{equation}
\widehat\mu({\cal D}\neq\gamma\gamma\,,gg)^{\rm{\bf CPC2}-HCC}
= \frac{\Gamma_{\rm tot}(H_{\rm SM})}{\Gamma_{\rm tot}(H)}\simeq
\frac{1}{0.92+0.08\left(1+\Delta S^g/|S^g_{\rm SM}|\right)^2}\,.
\end{equation}

\medskip

In CSB, $C_W$ is $2\sigma$ above the SM with the $1\sigma$ error of 2\%,
while $C_Z$ is consistent with the SM with
the $1\sigma$ error of 3\%. This is understood by 
comparing the WH production and $H\to WW^*$ decay signal strengths of 
$\mu({\rm WH},\sum{\cal D})=1.20\pm 0.15$ and $\mu(\sum{\cal P},WW^*)=1.04\pm 0.07$
to ZH production and $H\to ZZ^*$ decay signal strengths of 
$\mu({\rm ZH},\sum{\cal D})=1.03\pm 0.14$ and $\mu(\sum{\cal P},ZZ^*)=0.97\pm 0.08$,
see Table~\ref{tab:pdss76}.
The SM point lies outside of the 68\% CL region, see the lower-left frame of the left panel
of Fig.~\ref{fig:CPC2_CLregions}.

\medskip
In I, we assume all the Yukawa couplings to the SM particles are the
same like as in type-I 2HDM but, compared to {\bf CPC1}-I, $C_V$ is also varied.
We obtain that
\begin{equation}
\label{eq:CPC2I_best}
C_V = 1.015 \pm 0.017\,; \ \ \
C_f^S = 0.930 \pm 0.031\,. 
\end{equation}
While $C_V$ is consistent with the SM with the $1\sigma$ error of about 2\%,
$C_f^S$  deviates from the SM by the amount of more than $2\sigma$ resulting
in that the SM point lies just outside of the 95\% CL region, see the lower-right 
frame of the left panel of Fig.~\ref{fig:CPC2_CLregions}. 
We observe that this is a combined result of
$\mu(\sum{\cal P},\gamma\gamma)=1.10\pm 0.07$,
$\mu(\sum{\cal P},ZZ^*)=0.97\pm 0.08$,
$\mu(\sum{\cal P},WW^*)=1.04\pm 0.07$,
$\mu(\sum{\cal P},bb)=0.90\pm 0.12$, and
$\mu(\sum{\cal P},\tau\tau)=0.87\pm 0.08$, see Table~\ref{tab:pdss76}.
More precisely, we find that
the central value $1.10$ of $\mu(\sum{\cal P},\gamma\gamma)$ correlates
$C_V$ and $C_f^S$ as $C_f^S \sim 3\,C_V - 2.1$ 
\footnote{See Appendix~\ref{sec:appendix_C}.}
under which the Yukawa coupling
$C_f^S$ is driven to give the branching ratios 10\% below the SM 
by $\mu(\sum{\cal P},bb)$ and $\mu(\sum{\cal P},\tau\tau)$
while the gauge coupling $C_V$ near to the SM value of 1 by 
$\mu(\sum{\cal P},ZZ^*)$ and $\mu(\sum{\cal P},WW^*)$.
Indeed, we find that
\begin{equation}
(C_V)^{H\to\gamma\gamma} = 1.038^{+0.041}_{-0.039}\,; \ \ \
(C_f^S)^{H\to\gamma\gamma}= 0.999^{+0.114}_{-0.098} \,, 
\end{equation}
by fitting to the $\gamma\gamma$ signal strengths only and
\begin{equation}
(C_V)^{H\to ff} = 1.022^{+0.078}_{-0.072}\,; \ \ \
(C_f^S)^{H\to ff} = 0.910^{+0.042}_{-0.044} \,, 
\end{equation}
by fitting to the fermionic signal strengths only.
Incidentally, we obtain
\begin{equation}
(C_V)^{H\to\gamma\gamma,WW^*,ZZ^*} = 1.020\pm 0.020\,; \ \ \
(C_f^S)^{H\to\gamma\gamma,WW^*,ZZ^*}= 0.956^{+0.052}_{-0.049} \,, 
\end{equation}
by fitting to the bosonic signal strengths only.

\begin{figure}[t!]
\begin{center}
\includegraphics[width=14.5cm]{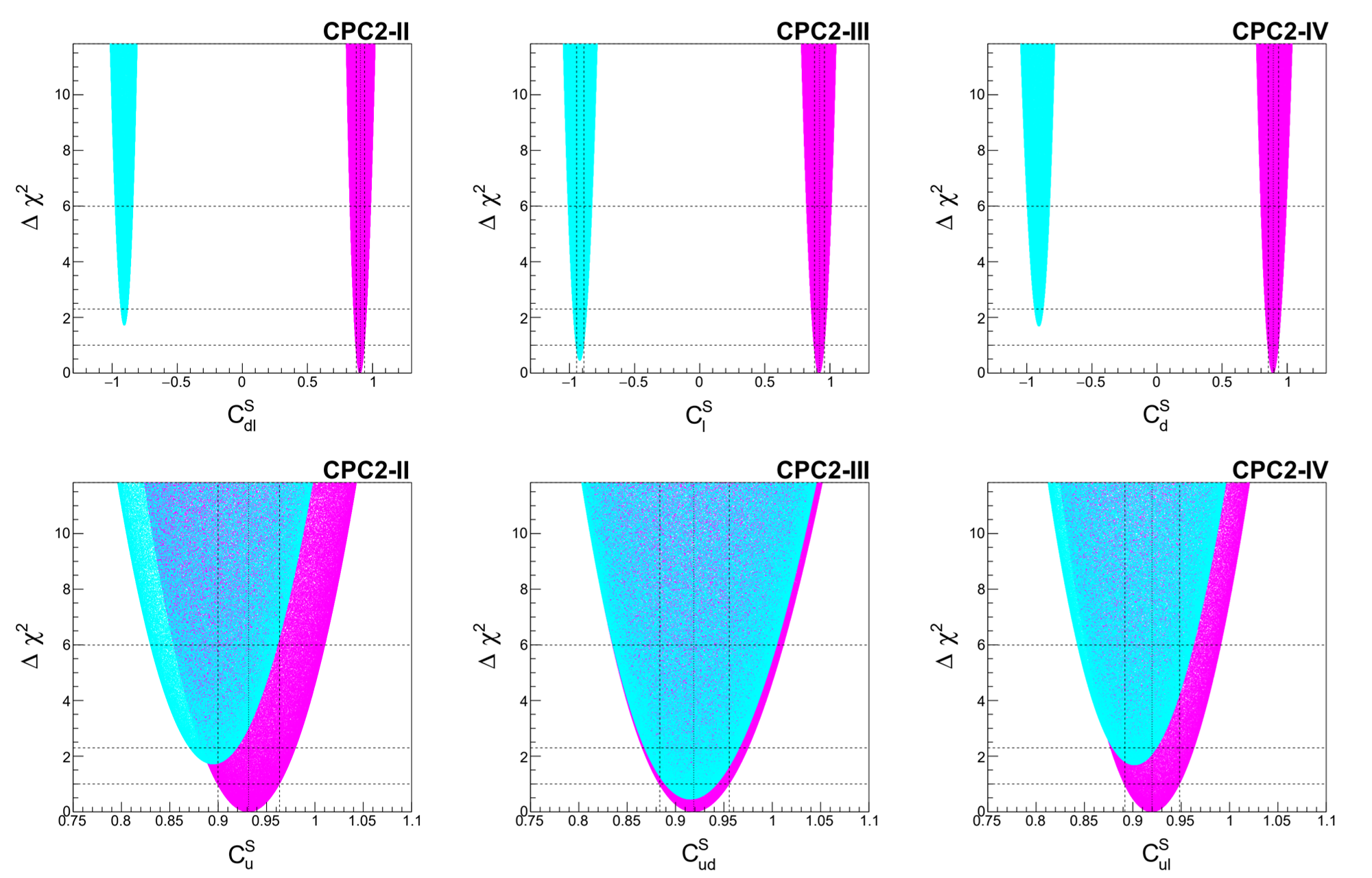}
\end{center}
\vspace{-0.5cm}
\caption{\it {\bf CPC2}:
$\Delta\chi^2$ above the minimum versus Yukawa couplings
in the II (left), III, (middle), and IV (right) subfits.
The magenta (cyan) points are for the positive (negative) values of
the  Yukawa couplings of the down-type quarks and/or charged leptons.
In each frame, the horizontal lines finds the $1\sigma$ errors and
the 68.27\% and 95\% CL regions in two-parameter planes.
}
\label{fig:CPC2_dxsq}
\end{figure}
\medskip
In II, III, and IV, taking $C_V=1$,
we assume the Yukawa couplings to the SM particles
behave like as in type-II, type-III, and type-IV 2HDMs, respectively.
We obtain
\begin{eqnarray}
(C_{u}^S)^{\rm II} &=&    0.931 \pm 0.032\,, \ \ \ 
(C_{d\ell}^S)^{\rm II} ~= 0.907 \pm 0.032\,; \nonumber \\[2mm]
(C_{ud}^S)^{\rm III} &=&    0.919^{+0.037}_{-0.035} \,, \ \ \ 
\hspace{0.3cm}
(C_{\ell}^S)^{\rm III} ~= 0.921 \pm 0.038 \,; \nonumber \\[2mm]
(C_{u\ell}^S)^{\rm IV} &=&  0.920^{+0.029}_{-0.028} \,, \ \ \ 
\hspace{0.3cm}
(C_{d}^S)^{\rm IV} \,~=   0.894^{+0.040}_{-0.039} \,.
\end{eqnarray}
We note that $C_{\ell}^S$ is basically determined by
$\mu(\sum{\cal P},\tau\tau)=0.87\pm 0.08$, 
see Table~\ref{tab:pdss76}.
Otherwise, all the Yukawa couplings
deviate from the SM by the amount of more than $2\sigma$ like as in {\bf CPC2}-I
with the $1\sigma$ errors of 3\%-4\%.
The SM point lies around the boundary between the 95\% and
99.73\% CL regions, 
see the upper-left (II), upper-right (III), and lower-left (IV)
frames of the right panel of Fig.~\ref{fig:CPC2_CLregions}. 
We further note that all the best-fitted values are positive and
the negative values of the Yukawa couplings 
$C_{d\ell}^S$, $C_\ell^S$, and $C_d^S$
around $-1$ are a bit less favored.
In Fig.~\ref{fig:CPC2_dxsq}, we show
$\Delta\chi^2$ above the positive minimum versus the down-type Yukawa couplings
in the II (left), III, (middle), and IV (right) subfits.
For $C_{d\ell}^S$, $C_\ell^S$, and $C_d^S$,
we observe that the data prefer the positive minima to the negative
ones by $\Delta\chi^2\sim 1.5$ ($C_{d\ell}^S$ and $C_d^S$)
and $\Delta\chi^2\sim 0.5$ ($C_\ell^S$), see the upper frames of Fig.~\ref{fig:CPC2_dxsq}.
This could be understood by observing that
$\widehat\mu({\rm ggF}+{\rm bbH})$ increases by the amount of about 10\%
by changing $C_b^S$ from $+1$ to $-1$, see
Eq.~(\ref{eq:ggF_coeff}) and Table~\ref{tab:ggF_coeff}.
\footnote{It is worthwhile to note that the increment 
of $\widehat\mu({\rm ggF}+{\rm bbH})$ due to the sign flip amounts to
about 30\% if we consider ggF in LO. Considering ggF beyond LO, we lose
a power to reject the negative bottom-quark Yukawa coupling.}
Similarly, $\widehat\mu(\gamma\gamma)$
is also sensitive to the sign of $C_\tau^S$ but the sign dependence
is weaker due to the dominance of the $W$-boson loop contribution to $S^\gamma$,
see Eq.~(\ref{eq:spaa_numeric}).
In fact, $\widehat\mu(\gamma\gamma)$ is powerful to reject the wrong sign of the 
top-quark Yukawa coupling and we see that negative $C_t^S$ is completely 
ruled out, see the lower frames of Fig.~\ref{fig:CPC2_dxsq}.

\medskip
In II, III, and IV,
we scrutinize that the fitting results are consistent with the pattern of the Yukawa
couplings predicted in each model.
In type-II, type-III, and type-IV 2HDMs, the Yukawa couplings 
are correlated as follows:
\footnote{Note again that we adopt the conventions
and notations of 2HDMs as in Ref.~\cite{Lee:2021oaj}.}
\begin{eqnarray}
{\rm II}&:& \ \ \ C_u^S  =  \cos\gamma - 1/\tan\beta\,\sin\gamma\,, \ \ \ 
C_{d\ell}^S = \cos\gamma + \tan\beta\,\sin\gamma\,, \nonumber \\[2mm]
{\rm III}&:& \ \ \ C_{ud}^S  =  \cos\gamma - 1/\tan\beta\,\sin\gamma\,, \ \ \ 
C_{\ell}^S = \cos\gamma + \tan\beta\,\sin\gamma\,, \nonumber \\[2mm]
{\rm IV}&:& \ \ \ C_{u\ell}^S  =  \cos\gamma - 1/\tan\beta\,\sin\gamma\,, \ \ \ 
C_{d}^S = \cos\gamma + \tan\beta\,\sin\gamma\,. 
\end{eqnarray}
Note that, in each model, only one of the two couplings
could be larger or smaller than 1 
depending on the sign of $\sin\gamma$ when $\cos\gamma=C_V\sim 1$.
It is impossible to have both the
couplings larger or smaller than 1 likes as in type-I 2HDM.
In the upper-left (II), upper-right (III), and lower-left (IV)
frames of the right panel of Fig.~\ref{fig:CPC2_CLregions},
we note that most of the 95\% CL regions locate  where 
both of the couplings are smaller than 1.
The situation will be clearer in {\bf CPC3} by varying $C_V$ also and
in {\bf CPC4}-A by varying 
the Yukawa couplings of the up- and down-type quarks and
the charged leptons separately.

\medskip

In HP, all the production and decay processes scale with the overall 
single theoretical signal strength of:
\begin{equation}
\mu({\cal P},{\cal D})^{{\bf CPC2}-{\rm HP}} = 
\frac{C_{Vf}^4}{C_{Vf}^2+\Delta\Gamma_{\rm tot}/\Gamma_{\rm tot}(H_{\rm SM})} \,,
\end{equation}
which leads to the relation
\begin{equation}
\frac{\Delta\Gamma_{\rm tot}}{\Gamma_{\rm tot}(H_{\rm SM})} \ = \
C_{Vf}^2\,\left(\frac{C_{Vf}^2}{\mu^{\rm Global}}-1\right).
\end{equation}
In the lower-right frame of the right panel of Fig.~\ref{fig:CPC2_CLregions},
the black line passing the origin $(\Delta\Gamma_{\rm tot},C_{Vf})=(0,0)$
and the SM point $(\Delta\Gamma_{\rm tot},C_{Vf})=(0,1)$ represents the above relation
when $\mu^{\rm Global}=1$.
In Higgs-portal models, the varying parameters are physically constrained by
$\Delta\Gamma_{\rm tot}\geq 0$ and $C_{Vf}\leq 1$. Imposing 
these conditions, we find the following best-fitted values
\begin{equation}
\Delta\Gamma_{\rm tot}/{\rm MeV} = 0.0^{+0.105}\,, \ \ \
C_{Vf} = 1.0_{-0.013}\,.
\end{equation}
We consider some extended HP scenarios in {\bf CPC3} and {\bf CPC4}.

\subsubsection{{\bf CPC3} and {\bf CPC4}}
%
\begin{table}[!t]
\caption{\it
\label{tab:CPC34}
{\bf CPC3} and {\bf CPC4}: The best-fitted values in the five {\bf CPC3} 
and two {\bf CPC4} subfits. 
Also shown are the corresponding minimal
chi-square per degree of freedom ($\chi^2_{\rm min}$/dof), 
goodness of fit (gof), and 
$p$-value against the SM for compatibility
with the SM hypothesis.
For the SM, we obtain $\chi^2_{\rm SM}/{\rm dof}=82.3480/76$
and gof $=0.2895$.
}  \vspace{1mm}
\renewcommand{\arraystretch}{1.1}
\begin{adjustbox}{width= \textwidth}
\begin{tabular}{c|c|c|c|c|c|c|c|c}
\hline
\multicolumn{2}{c|}{\multirow{2}{*}{Parameters}} &
\multicolumn{5}{c|}{\bf CPC3} &
\multicolumn{2}{c}{\bf CPC4} \\ \cline{3-9}
\multicolumn{2}{c|}{} & IUHCC & II & III & IV & HP & A & HP \\ \hline
\multirow{3}{*}{non-SM} &
$\Delta\Gamma_{\rm tot}$/MeV &
$-0.029^{+0.215}_{-0.191}$ & 0 & 0 & 0 & $0.0^{+0.255}$ & 0 & $0.0^{+0.186}$ \\
&  $\Delta S^\gamma$ &
$-0.392^{+0.204}_{-0.206}$ & 0 & 0 & 0 & $-0.366^{+0.197}_{-0.209}$ & 0 &
$-0.400^{+0.181}_{-0.200}$ \\
&
$\Delta S^g$ &
$-0.036^{+0.042}_{-0.039}$ & 0 & 0 & 0 & 0 & 0 & $-0.032^{+0.039}_{-0.030}$ \\ \hline
\multirow{7}{*}{SM} &
\multirow{2}{*}{$C_V$} &
\multirow{5}{*}{1} &
\multirow{2}{*}{$1.007^{+0.026}_{-0.026}$} &
\multirow{2}{*}{$1.015^{+0.017}_{-0.017}$} &
\multirow{2}{*}{$1.004^{+0.034}_{-0.035}$} &
\multirow{5}{*}{$C_{Vf} = 0.989^{+0.011}_{-0.019}$} &
\multirow{2}{*}{$1.002^{+0.034}_{-0.035}$} &
\multirow{5}{*}{$C_{Vf} = 1.0_{-0.021}$} \\ & & & & & & & \\ \cline{2-2} \cline{4-6}
\cline{8-8}
&
\multirow{3}{*}{$C_f^S$} & &
$C^S_u = 0.932^{+0.032}_{-0.032}$ & \multirow{2}{*}{ $C^S_{ud} = 0.933^{+0.039}_{-0.039}$
} & \multirow{2}{*}{ $C^S_{u\ell} = 0.923^{+0.038}_{-0.037}$ } & &
$C^S_u = 0.927^{+0.040}_{-0.040}$ & \\
& & & \multirow{2}{*}{$C^S_{dl} = 0.917^{+0.048}_{-0.047}$}  &  &  &  &  $C^S_d =
0.902^{+0.082}_{-0.081}$ & \\
& & &  & $C^S_\ell = 0.928^{+0.038}_{-0.039}$ & $C^S_d = 0.902^{+0.084}_{-0.084}$ &  &
$C^S_\ell = 0.916^{+0.047}_{-0.046}$ & \\
\hline
\multicolumn{2}{c|}{$\chi^2_{\rm min}$/dof} & 78.1707/73 & 74.2343/73 & 74.3554/73 &
74.2386/73 & 78.8928/73 & 74.1893/72 & 78.1906/72  \\
\multicolumn{2}{c|}{goodness of fit (gof)}    &  0.3181 &  0.4377 &  0.4338 &  0.4376 &  0.2981
&  0.4067 & 0.2883  \\
\multicolumn{2}{c|}{$p$-value against the SM}          &  0.2429 &  0.0437 &  0.0462 &  0.0438 &  0.3265
&  0.0859 & 0.3874  \\ \hline
\end{tabular}
\end{adjustbox}
\end{table}
\begin{figure}[t!]
\begin{center}
\includegraphics[width=9.1cm]{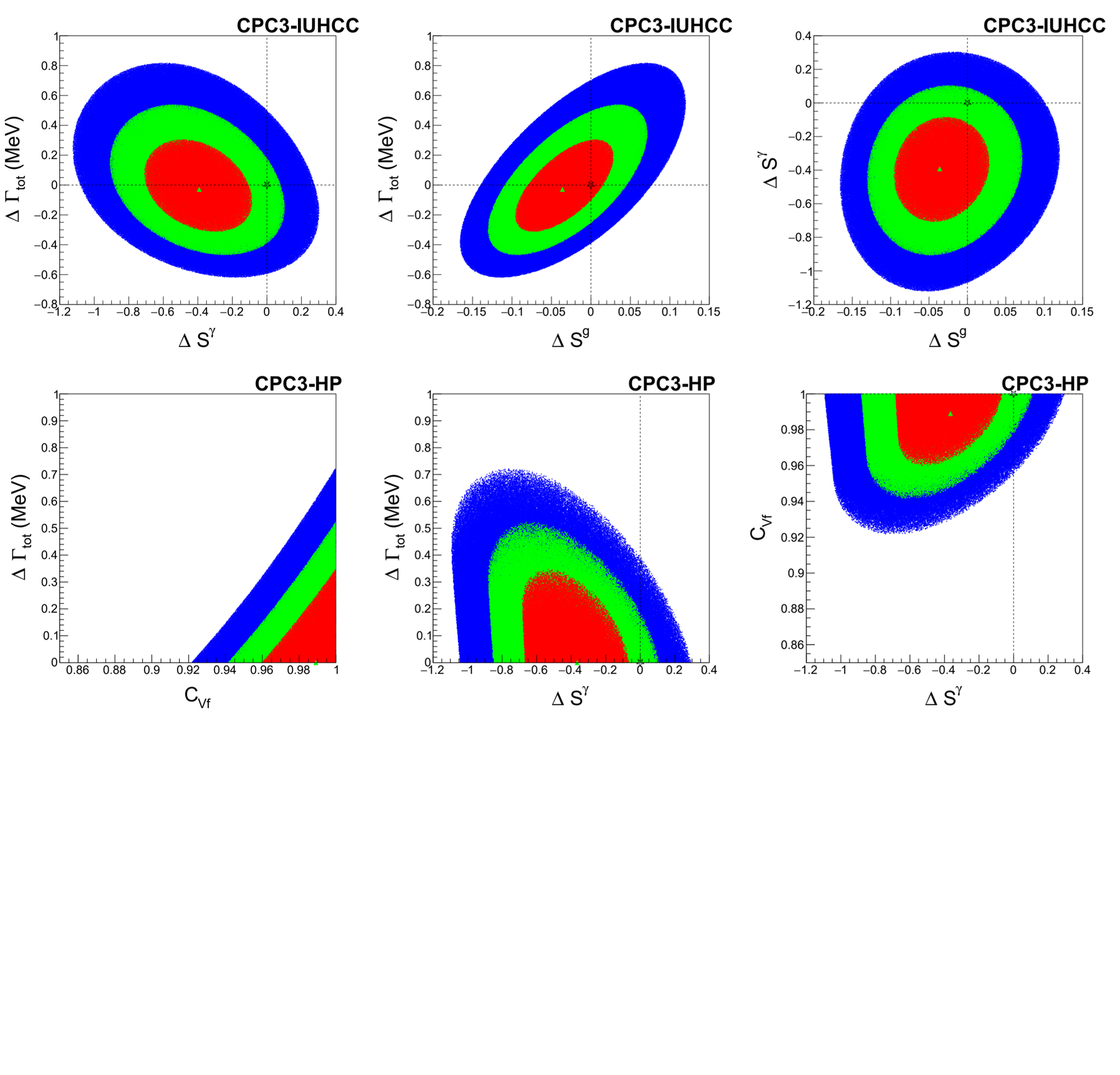}
\includegraphics[width=9.1cm]{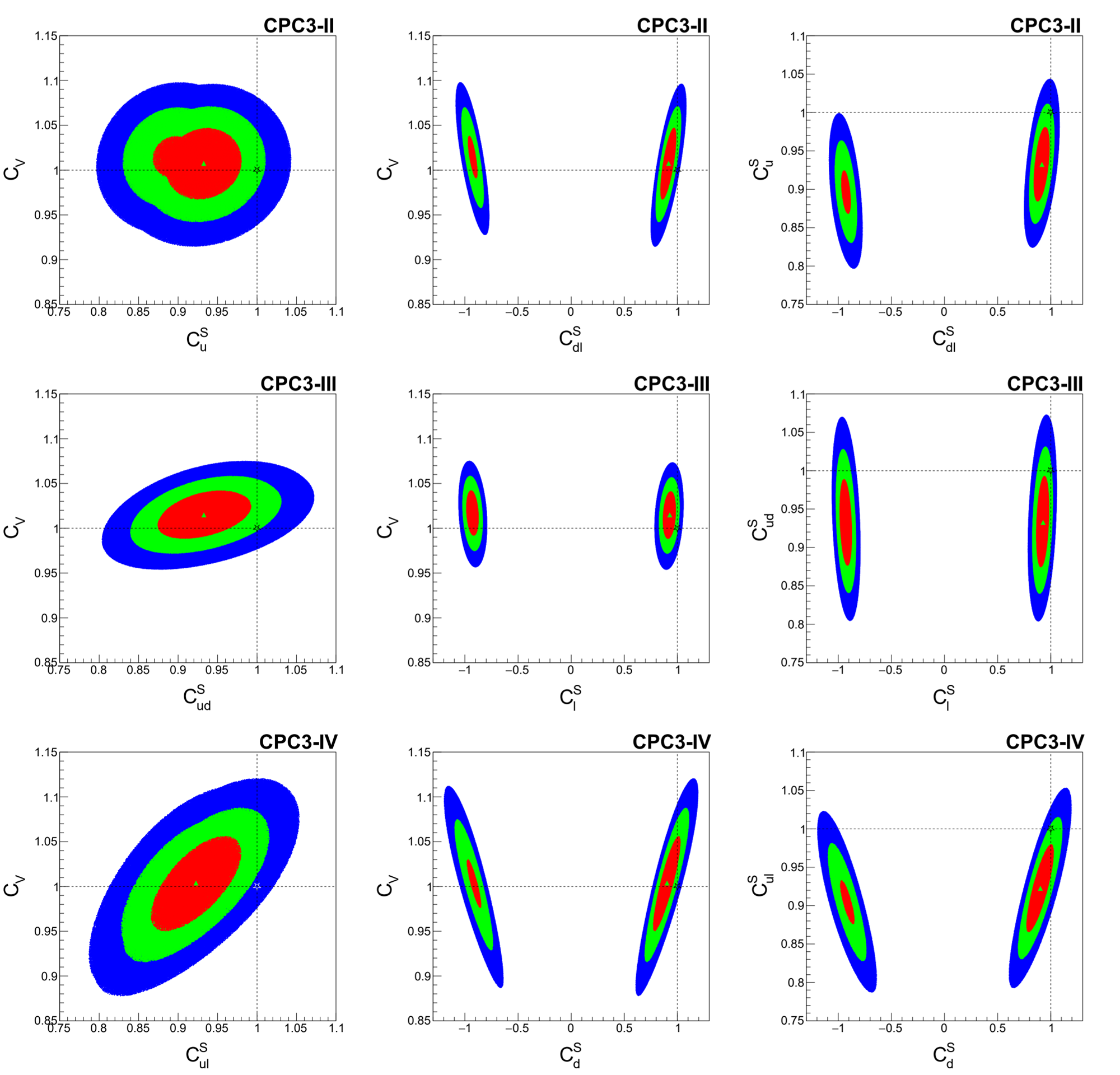}
\end{center}
\caption{\it {\bf CPC3}:
The CL regions of the five {\bf CPC3} subfits in two-parameter planes:
[Left] IUHCC (upper) and HP (middle)
[Right] II (upper), III (middle), and IV (lower).
The contour regions shown are for
$\Delta\chi^2\leq 2.3$ (red),
$\Delta\chi^2\leq 5.99$ (green),
$\Delta\chi^2\leq 11.83$ (blue)
above the minimum, which correspond to
confidence levels of 68.27\%, 95\%, and 99.73\%, respectively.
In each frame, the vertical and horizontal lines locate the SM point denoted
by a star and the best-fit point is denoted by a triangle.
}
\label{fig:CPC3_CLregions}
\end{figure}
%
\begin{figure}[b!]
\begin{center}
\includegraphics[width=9.1cm]{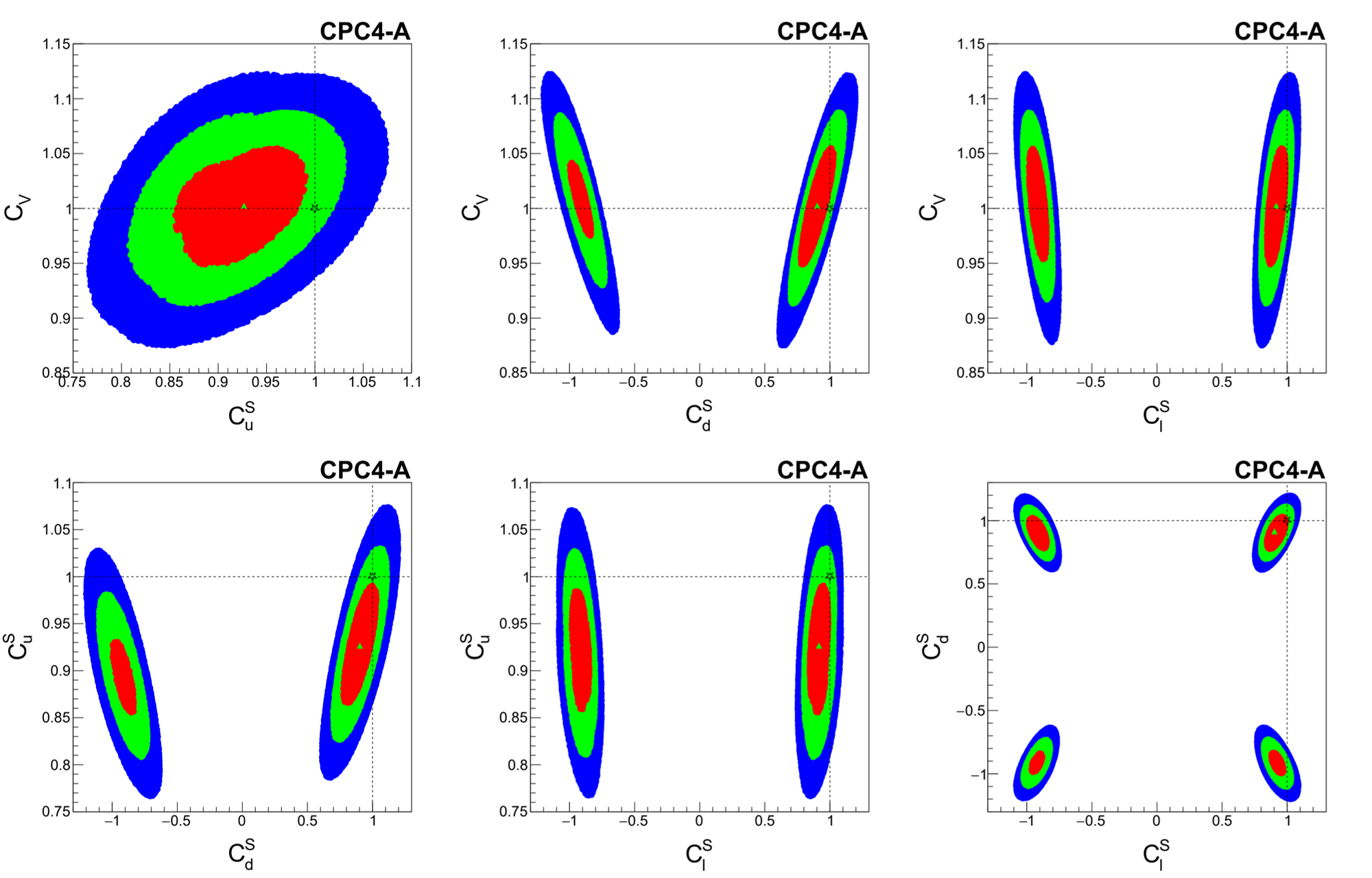}
\includegraphics[width=9.1cm]{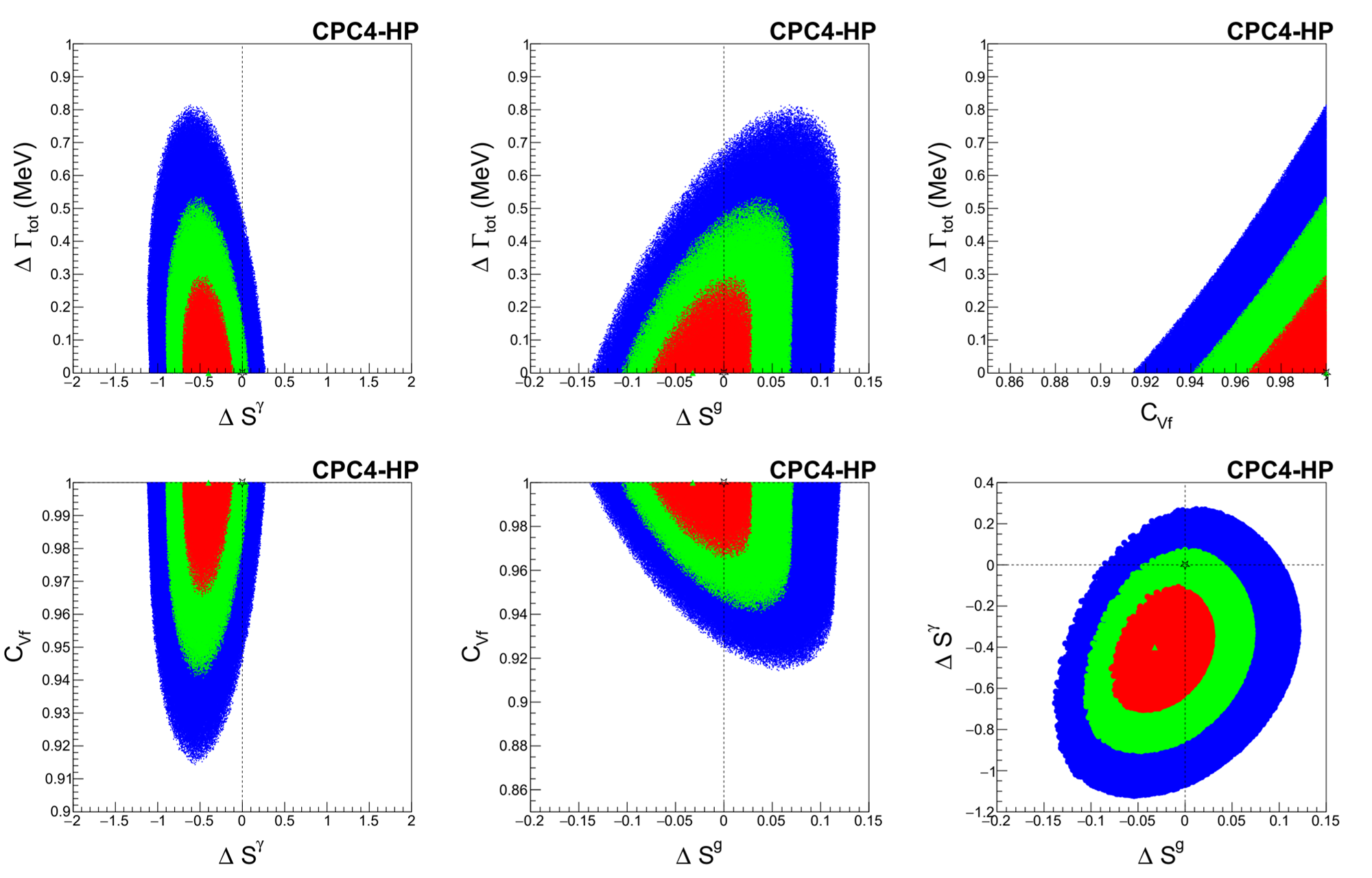}
\end{center}
\vspace{-0.5cm}
\caption{\it {\bf CPC4}:
The CL regions of {\bf CPC4}-A [Left] and {\bf CPC4}-HP [Right] subfits 
in two-parameter planes:
The contour regions shown are for
$\Delta\chi^2\leq 2.3$ (red),
$\Delta\chi^2\leq 5.99$ (green),
$\Delta\chi^2\leq 11.83$ (blue)
above the minimum, which correspond to
confidence levels of 68.27\%, 95\%, and 99.73\%, respectively.
In each frame, the vertical and horizontal lines locate the SM point denoted
by a star and the best-fit point is denoted by a triangle.
}
\label{fig:CPC4_CLregions}
\end{figure}

We show the fitting results for the five {\bf CPC3} and two {\bf CPC4} subfits
in Table~\ref{tab:CPC34}.
In the HP scenarios, we perform the fit under the constraints of
$\Delta\Gamma_{\rm tot}\geq 0$ and
$C_{Vf}=C_W=C_Z\leq 1$ and we
have the gof values similar to the SM.
Otherwise, the gof values range between 
$0.3181$ ({\bf CPC3}-IUHCC) and $0.4377$ ({\bf CPC3}-II) which
are better than the SM but slightly worse than the corresponding {\bf CPC2} fits.
The gof value of {\bf CPC4}-A is also larger than $0.4$.
The CL regions in two-parameter planes are depicted in 
Fig.~\ref{fig:CPC3_CLregions} and Fig.~\ref{fig:CPC4_CLregions} 
for {\bf CPC3} and {\bf CPC4}, respectively.
Note that, in the HP scenarios, we show the
parameter spaces in which the fitting constraints of $\Delta\Gamma_{\rm tot}\geq 0$
and $C_{Vf}\leq 1$ are fulfilled.
The $p$-values against the SM 
for compatibility with the SM hypothesis
are smaller than  10\% except 
{\bf CPC3}-IUHCC, {\bf CPC3}-HP, and {\bf CPC4}-HP.

\medskip

In {\bf CPC3}-IUHCC, we vary all the three non-SM parameters.
The best-fitted values for
$\Delta\Gamma_{\rm tot}$ and $\Delta S^g$ are consistent with the SM 
within $1\sigma$ while
$\Delta S^\gamma$ deviates from the SM by about $1.5\sigma$.
The $1\sigma$ errors are about 
5\% for $\Delta\Gamma_{\rm tot}/\Gamma_{\rm tot}(H_{\rm SM})$,
3\% for $|\Delta S^\gamma/S^\gamma_{\rm SM}|$, and
6\% for $|\Delta S^g/S^g_{\rm SM}|$ which are slightly larger than
those found in {\bf CPC1}-IU, {\bf CPC2}-IUHC, and {\bf CPC2}-HCC.
The SM points lies outside of the 68\% CL regions in
the $(\Delta S^\gamma,\Delta\Gamma_{\rm tot})$ and
$(\Delta S^g,\Delta S^\gamma)$ planes, see the upper frames
of the left panel of Fig.~\ref{fig:CPC3_CLregions}.

\medskip

In II, III, IV subfits of {\bf CPC3}, we additionally vary $C_V$ compared to
the corresponding {\bf CPC2} subfits.
We observe that $C_V$ is consistent with SM within $1\sigma$ 
errors of about 2\%-3\%. In contrast, the Yukawa couplings are about $2\sigma$
below the SM except for $C_d^S$ (IV).
The $1\sigma$ errors of 
$C_u^S$ (II), $C_{ud}^S$ (III), and $C_{u\ell}^S$ (IV)
grouped with the up-type quarks
are 3\%-4\% and those of
$C_{d\ell}^S$ (II), $C_{\ell}^S$ (III), and $C_{d}^S$ (IV)
for the down-type fermions 4-8\%. The larger error
of the down-type fermions
could be understood from the positive correlations between $C_V$ and 
the absolute values of them,
see the upper-, middle-, and lower-middle frames of the 
right panel of Fig.~\ref{fig:CPC3_CLregions}. The stronger correlation leads to the
the larger errors.
We find that the minima for the negative  values of
$C_{d\ell}^S$ (II), $C_{\ell}^S$ (III), and $C_{d}^S$ (IV) are 
above the corresponding positive ones by the amount of
$\Delta\chi^2\sim 1.5$, $\sim 0.3$, and $\sim 1.7$, respectively.
Incidentally, we observe that the distortion of the CL regions
in $(C_u^S,C_V)$ (II) plane appearing  the upper-left frame of the 
right panel of Fig.~\ref{fig:CPC3_CLregions} is due to the minimum
around $C_u^S=0.9$ for the negative values of $C_{d\ell}^S$,
see the upper-right frame of the same panel
in the $(C_{d\ell}^S,C_u^S)$ plane.
Finally, we observe that the 68\% CL regions locate  where 
both of the Yukawa couplings are smaller than 1 indicating deviation from
the conventional type-II, type-III, and type-IV 2HDMs, 
see the upper-, middle-, and lower-right frames of the 
right panel of Fig.~\ref{fig:CPC3_CLregions}.

\medskip

In {\bf CPC3}-HP where we add the non-SM contribution to $H\to\gamma\gamma$
compared to {\bf CPC2}-HP, $\Delta S^\gamma$ is fitted to accommodate 
$\widehat\mu(\gamma\gamma)=1.1\pm 0.07$ like as in
{\bf CPC1}-HC. 
The parameters $\Delta\Gamma_{\rm tot}$ and $C_{Vf}$ are fitted to
have the SM values like as in {\bf CPC2}-HP but with a bit larger $1\sigma$ errors
under the constraints of $\Delta\Gamma_{\rm tot}>0$ and $C_{Vf}<1$.
In {\bf CPC4}-HP, we further add the non-SM contribution also to $H\to gg$
assuming non-SM particles such as vector-like quarks. 
For $\Delta S^\gamma$ and $\Delta S^g$,
the fitting results are very similar to {\bf CPC2}-HCC and
the parameters $\Delta\Gamma_{\rm tot}$ and $C_{Vf}$ are 
again fitted to have the SM values under the constraints of 
$\Delta\Gamma_{\rm tot}>0$ and $C_{Vf}<1$
like as in other HP scenarios.
The SM points lie outside of the 68\% CL regions in the
$(\Delta\Gamma_{\rm tot},\Delta S^\gamma)$,
$(C_{Vf},\Delta S^\gamma)$, and
$(\Delta S^g,\Delta S^\gamma)$ planes, 
see the right panel of Fig.~\ref{fig:CPC4_CLregions}.

\medskip

In {\bf CPC4}-A, we vary the Yukawa couplings of the up- and down-type quarks and
the charged leptons separately together with $C_V$.
This scenario does not alter our previous observation made in {\bf CPC3}
for the Yukawa couplings: they are about $1\sigma$ ($C_d^S$) and
$2\sigma$ ($C_u^S$ and $C_\ell^S$) below the SM.
The $C_V$ is very consistent with the SM with the $1\sigma$ error of about 3\%
and the $1\sigma$ errors of $C_u^S$, $C_\ell^S$, and $C_d^S$ are
4\%, 5\%, and 8\%, respectively.
And, from the CL regions in the 
$(C_d^S,C_u^S)$, $(C_\ell^S,C_u^S)$, and $(C_\ell^S,C_{d}^S)$ planes
shown in the lower frames of the left panel of Fig.~\ref{fig:CPC4_CLregions},
we see that the data favor the type-I 2HDM over the other three models.
Incidentally, we find that the minima for the negative values of
$C_{d}^S$ and $C_{\ell}^S$ are 
above the positive ones by the amount of
$\Delta\chi^2\sim 1.5$ and $\sim 0.3$, respectively.

\subsubsection{{\bf CPC5} and {\bf CPC6} }
%
\begin{table}[!t]
\caption{\it
\label{tab:CPC56}
{\bf CPC5} and {\bf CPC6}: The best-fitted values in the two {\bf CPC5} 
and one {\bf CPC6} subfits. 
Also shown are the corresponding minimal
chi-square per degree of freedom ($\chi^2_{\rm min}$/dof), 
goodness of fit (gof), and 
$p$-value against the SM for compatibility
with the SM hypothesis.
For the SM, we obtain $\chi^2_{\rm SM}/{\rm dof}=82.3480/76$
and gof $=0.2895$.
Note that there are two degenerate minima for the positive and negative
values of either $C_\ell^S$ ({\bf CPC5}-AHC) or 
$C_\mu^S$ ({\bf CPC5}-LUB  and {\bf CPC6}-CSBLUB).
}  \vspace{1mm}
\renewcommand{\arraystretch}{1.1}
\begin{adjustbox}{width=12cm}
\begin{tabular}{c|c|c|c|c}
\hline
\multicolumn{2}{c|}{\multirow{2}{*}{Parameters}} &
\multicolumn{2}{c|}{\bf CPC5} & {\bf CPC6}   \\ \cline{3-5} 
\multicolumn{2}{c|}{}  &  AHC & LUB &  CSBLUB\\ \hline
\multirow{3}{*}{non-SM} &
$\Delta\Gamma_{\rm tot}$/MeV & 0 & 0 & 0 \\
&  $\Delta S^\gamma$ & $-0.102^{+0.217}_{-0.212}\,(C_\ell^S>0)$ ,
$-0.146^{+0.219}_{-0.208}\,(C_\ell^S<0)$ & 0 & 0 \\
&
$\Delta S^g$ & 0 & 0 & 0 \\ \hline
\multirow{7}{*}{SM} &
\multirow{2}{*}{$C_V$} &
\multirow{2}{*}{$0.998^{+0.034}_{-0.035}\,(C_\ell^S>0)$  ,
$0.998^{+0.035}_{-0.034}\,(C_\ell^S<0)$} &
\multirow{2}{*}{$1.003^{+0.033}_{-0.030}$} &
$C_W = 1.012^{+0.033}_{-0.033}$ \\
& & & & $C_Z = 0.987^{+0.037}_{-0.038}$ \\ \cline{2-5}
& \multirow{5}{*}{$C_f^S$} &
$C^S_u = 0.929^{+0.041}_{-0.040}\,(C_\ell^S>0)$ ,
$C^S_u = 0.929^{+0.040}_{-0.038}\,(C_\ell^S<0)$ & 
$C^S_u = 0.925^{+0.036}_{-0.035}$ & 
$C^S_u = 0.933^{+0.035}_{-0.035}$ \\
& & 
$C^S_d = 0.907^{+0.079}_{-0.078}\,(C_\ell^S>0)$ ,
$C^S_d = 0.907^{+0.081}_{-0.082}\,(C_\ell^S<0)$ 
& 
$C^S_d = 0.902^{+0.073}_{-0.076}$ & 
$C^S_d = 0.913^{+0.066}_{-0.067}$ \\
& & \multirow{2}{*}{$C^S_\ell = 0.920^{+0.046}_{-0.043}\,,
-0.920^{+0.045}_{-0.046}$} & 
$C^S_\tau = 0.910^{+0.042}_{-0.044}$ & 
$C^S_\tau = 0.915^{+0.039}_{-0.039}$ \\
& & & $C^S_\mu = \pm\, 1.057^{+0.134}_{-0.151}$ & $C^S_\mu = \pm\, 1.061^{+0.118}_{-0.131}$  \\   [1mm]
\hline
\multicolumn{2}{c|}{$\chi^2_{\rm min}$/dof} & 73.9870/71 & 73.4684/71 & 72.9828/70   \\
\multicolumn{2}{c|}{goodness of fit (gof)} & 0.3809& 0.3972 & 0.3803   \\
\multicolumn{2}{c|}{$p$-value against the SM} & 0.1374& 0.1140 & 0.1541  \\ \hline
\end{tabular}
\end{adjustbox}
\end{table}
\begin{figure}[t!]
\begin{center}
\includegraphics[width=9.1cm]{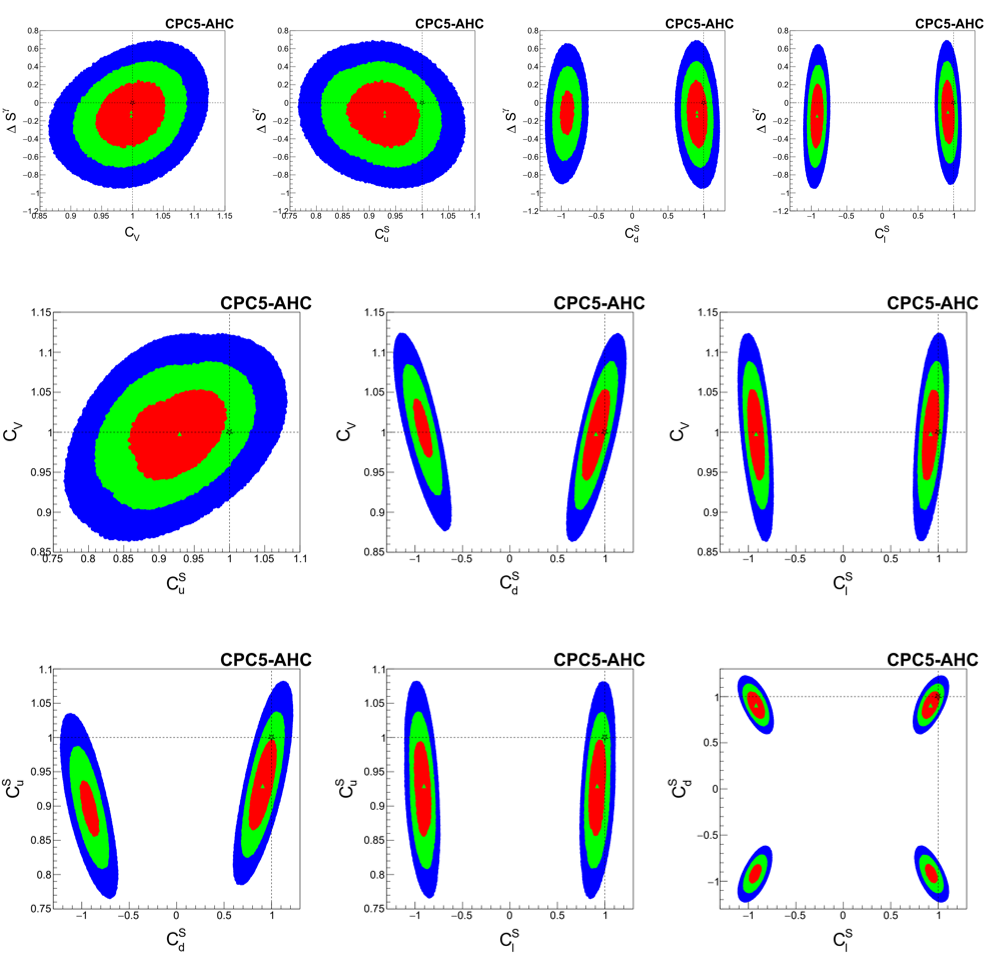}
\includegraphics[width=9.1cm]{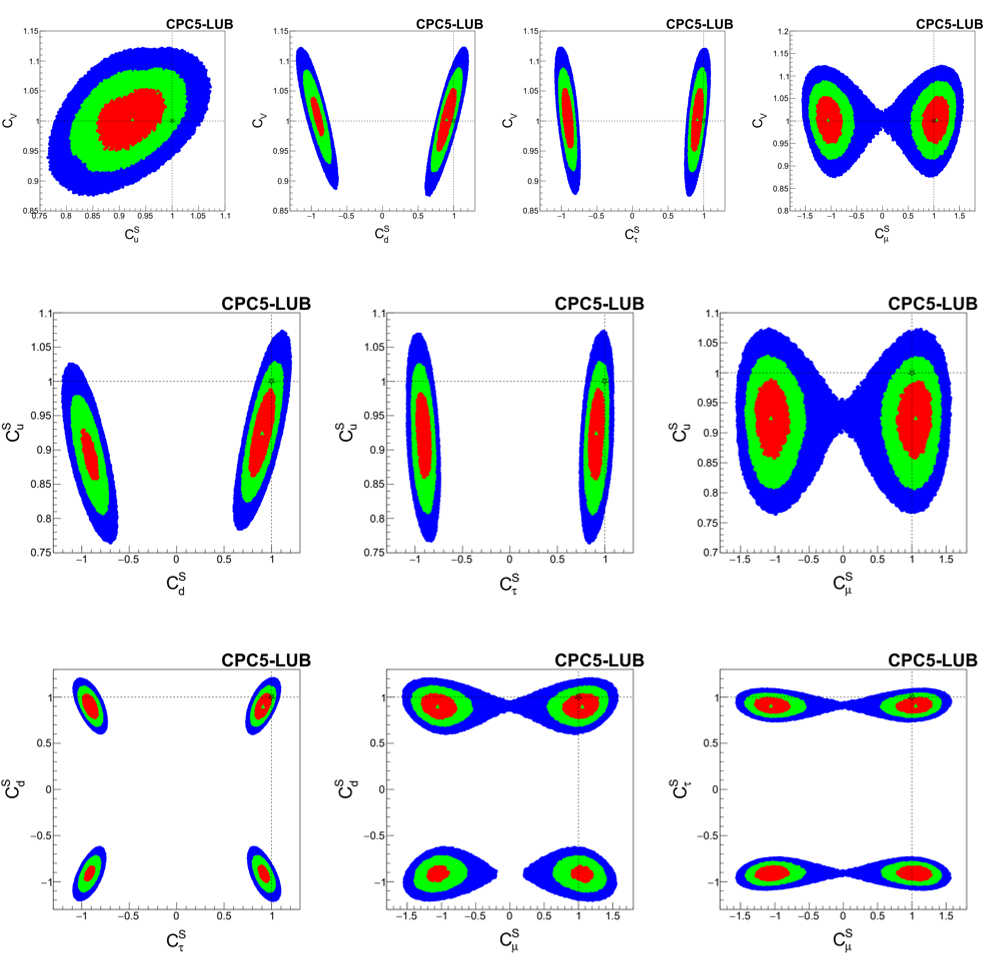}
\end{center}
\caption{\it {\bf CPC5}:
The CL regions of the two {\bf CPC5} subfits in two-parameter planes:
[Left] AHC [Right] LUB.
The contour regions shown are for
$\Delta\chi^2\leq 2.3$ (red),
$\Delta\chi^2\leq 5.99$ (green),
$\Delta\chi^2\leq 11.83$ (blue)
above the minimum, which correspond to
confidence levels of 68.27\%, 95\%, and 99.73\%, respectively.
In each frame, the vertical and horizontal lines locate the SM point denoted
by a star and the best-fit is denoted by a triangle.
}
\label{fig:CPC5_CLregions}
\end{figure}
\begin{figure}[t!]
\begin{center}
\includegraphics[width=9.1cm]{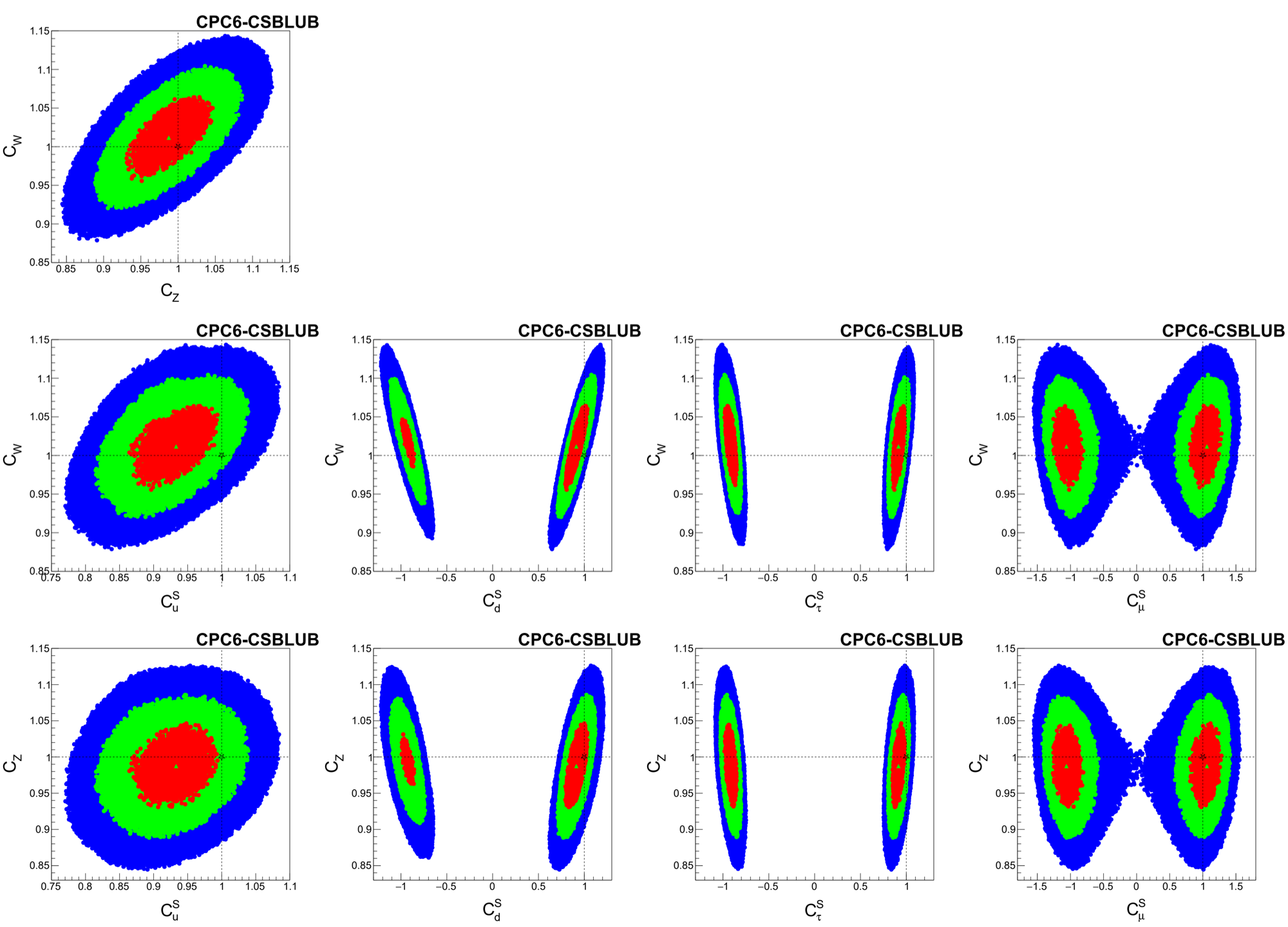}
\includegraphics[width=9.1cm]{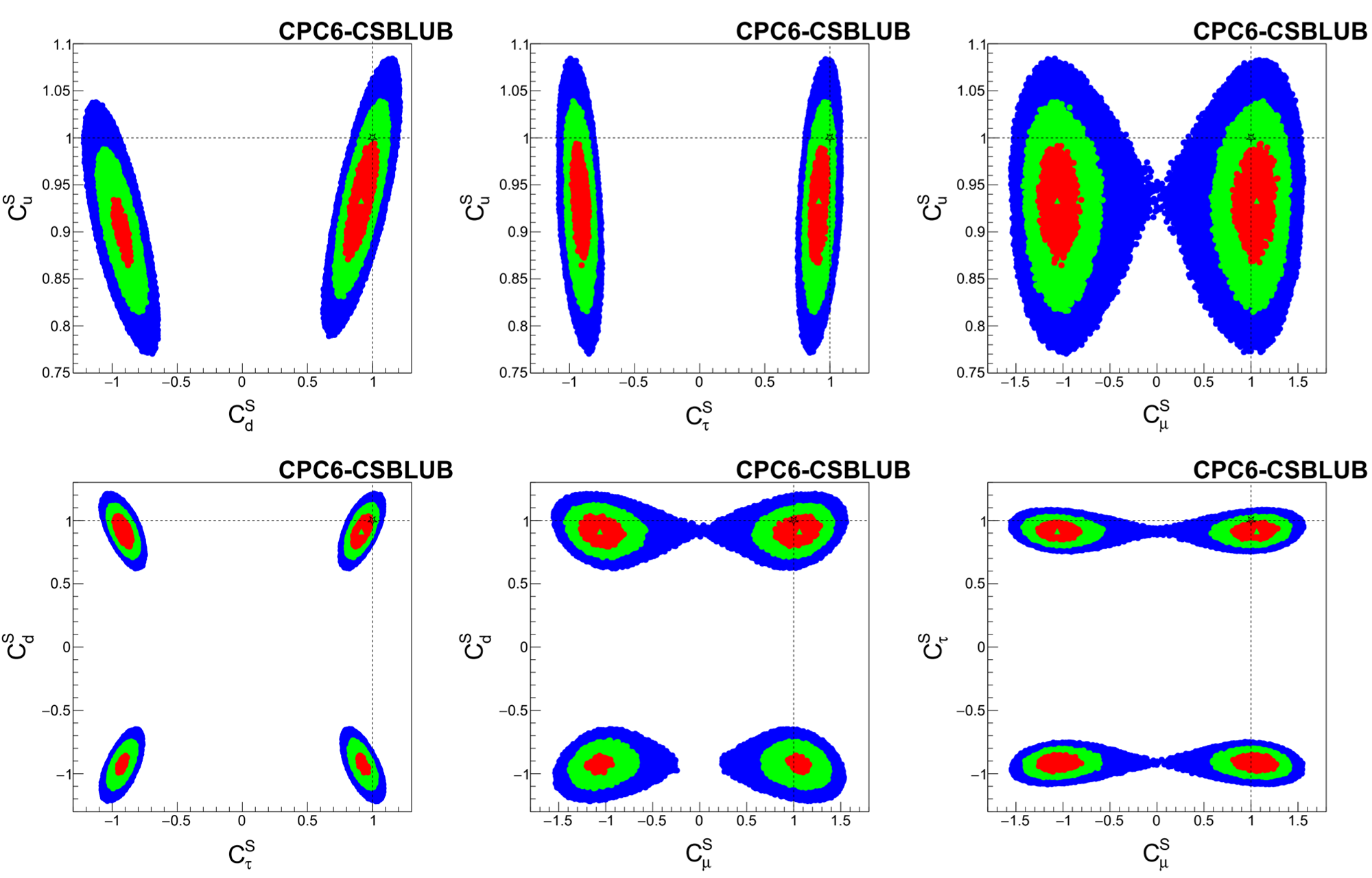}
\end{center}
\caption{\it {\bf CPC6}:
The CL regions of {\bf CPC6}-CSBLUB.
The contour regions shown are for
$\Delta\chi^2\leq 2.3$ (red),
$\Delta\chi^2\leq 5.99$ (green),
$\Delta\chi^2\leq 11.83$ (blue)
above the minimum, which correspond to
confidence levels of 68.27\%, 95\%, and 99.73\%, respectively.
In each frame, the vertical and horizontal lines locate the SM point denoted
by a star and the best-fit point is denoted by a triangle.
}
\label{fig:CPC6_CLregions}
\end{figure}

We show the fitting results for the two {\bf CPC5} and one {\bf CPC6} subfits
in Table~\ref{tab:CPC56}.
The gof values are
$0.3809$ ({\bf CPC5}-AHC), $0.3972$ ({\bf CPC5}-LUB) and $0.3803$ ({\bf CPC6}-CSBLUB) which
are better than the SM.
The $p$-values against the SM 
for compatibility with the SM hypothesis
are low and it is 15\% for {\bf CPC6}-CSBLUB.
The CL regions in two-parameter planes are depicted in 
Fig.~\ref{fig:CPC5_CLregions} and Fig.~\ref{fig:CPC6_CLregions} 
for {\bf CPC5} and {\bf CPC6}, respectively.

\medskip

In {\bf CPC5}-AHC, compared to {\bf CPC4}-A,
we add the contribution to $H\to\gamma\gamma$ from
heavy electrically charged particles.
First of all,
we find that the minima for the positive and negative values of 
$C_{\ell}^S$ are degenerate with the change of $\Delta S^\gamma$
by the amount of $0.044$ compensating the effects of the flipped sign of 
$C_\ell^S$, see Eq.~(\ref{eq:spaa_numeric}).
The parameter $\Delta S^\gamma$ is consistent with the SM:
$\Delta S^\gamma/S^\gamma_{\rm SM}\sim 0.016 \pm 0.03$ 
and $0.022 \pm 0.03$ for the positive and negative values
of $C_\ell^S$, respectively, and 
the two minima of $\Delta S^\gamma$ are very near to each other
separated by only $\sim 0.2\sigma$.
The gauge-Higgs coupling $C_V$
and the Yukawa couplings
of $C_u^S$, $C_d^S$, and $|C_\ell^S|$
are fitted similarly as in {\bf CPC4}-A.
\footnote{
Note that, in {\bf CPC5}-AHC, the best-fitted values 
of the fitting parameters at the two degenerate minima are 
almost the same except for $\Delta S^\gamma$:
the $1\sigma$ errors of 
$C_V$ and $C_{u,d}^S$ depend on the sign of $C_\ell^S$  very weakly with
their central values untouched, see Table~\ref{tab:CPC56}.}
The SM points are now near to or in the 68\% CL regions,
see the left panel of Fig.~\ref{fig:CPC5_CLregions}.
We find that 
the negative minimum of $C_d^S$ is 
above the positive one by the amount of $\Delta\chi^2\sim 1.5$.

\medskip
In {\bf CPC5}-LUB, compared to {\bf CPC4}-A,
we fit the tau-lepton- and muon-Yukawa couplings separately.
We obtain that the gauge-Higgs coupling is consistent with 
the SM with the $1\sigma$ error of about 3\% and
the up-quark and tau-lepton Yukawa couplings are about $2\sigma$ below the SM
with the $1\sigma$ errors of about 4\%.
The down-quark  Yukawa coupling is about $1\sigma$ below the SM
with the $1\sigma$ error of about 7\%-8\%.
We find that the minima for the positive and negative values of
$C_{\mu}^S$ are degenerate and, at the positive minimum,
the muon-Yukawa coupling is consistent with the SM with the $1\sigma$ error of 
13\%-15\%.
The best-fitted values of
the gauge-Higgs coupling $C_V$ and  the Yukawa couplings of $C^S_{u,d,\tau}$
are the same at the two generate minima.
Around the SM values of $C_\tau^S=C_\mu^S=1$,
the $1\sigma$ regions of $\left(C_\tau^S\right)_{1\sigma}=[0.866\,,0.952]$ 
and $\left(C_\mu^S\right)_{1\sigma}=[0.906\,,1.191]$ overlap
with no violation of lepton universality.
We find that the negative minima of $C_d^S$ and $C_\tau^S$ are 
above the positive ones by the amount of 
$\Delta\chi^2\sim 1.5$ and $\sim 0.2$, respectively.
The negative and positive regions of $C_\mu^S$ are connected at
99.73\% CL, see the right panel of Fig.~\ref{fig:CPC5_CLregions}.

\medskip

In {\bf CPC6}-CSBLUB, we vary the six SM parameters independently 
under the constraint of $C_t^S=C_c^S=C_u^S$ and we find
that the gauge-Higgs couplings are consistent with the SM with the $1\sigma$ errors of
about 3\%-4\%. The central value of $C_W$ ($C_Z$) is slight above (below) the SM value of 1.
The up-quark and tau-lepton Yukawa couplings are about $2\sigma$ below the SM
with the $1\sigma$ errors of about 4\%.
The down-quark  Yukawa coupling is about $1\sigma$ below the SM
with the $1\sigma$ error of about 7\%.
We find that the minima for the positive and negative values of
$C_{\mu}^S$ are degenerate and, at the positive minimum,
the muon Yukawa coupling  is consistent with the SM 
with the $1\sigma$ error of about 13\%.
The best-fitted values of
the gauge-Higgs couplings $C_{W,Z}$ and  the Yukawa couplings of $C^S_{u,d,\tau}$
are the same at the two generate minima.
We observe that,
around the SM values of $C_\tau^S=C_\mu^S=1$,
the $1\sigma$ regions of the normalized couplings of
$\left(C_\tau^S\right)_{1\sigma}=[0.876\,,0.954]$ 
and $\left(C_\mu^S\right)_{1\sigma}=[0.930\,,1.179]$ marginally overlap
with no violation of lepton universality.
Comparing the CL regions shown in the left (right) panel of Fig.~\ref{fig:CPC6_CLregions} with
those in the upper (middle and lower) frames of the right panel of Fig.~\ref{fig:CPC5_CLregions},
we observe that the CL regions {\bf CPC6}-CSBLUB are very similar to those of {\bf CPC5}-LUB.

\medskip

Before moving to {\bf CPV}, we provide the following brief summary 
for the SM parameters obtained from 
{\bf CPC3}, {\bf CPC4}, {\bf CPC5}, and {\bf CPC6} fits:
\footnote{
For correlations among the fitting parameters in 
{\bf CPC2}, {\bf CPC3}, and {\bf CPC4},
see Appendix~\ref{sec:appendix_D}.}
\begin{itemize}
\item{$C_V$, $C_W$, $C_Z$}: consistent with the SM with the $1\sigma$ error of 2\%-3\%
\item{$C_u^S$, $C_{ud}^S$, $C_{u\ell}^S$}: 
about $2\sigma$ below the SM with the $1\sigma$ error of 3\%-4\%
\item{$C_{d\ell}^S$, $C_\ell^S$, $C_{\tau}^S$}: 
about $2\sigma$ below the SM with the $1\sigma$ error of 4\%-5\%
\item{$C_d^S$}: 
about $1\sigma$ below the SM with the $1\sigma$ error of 7\%-8\%
\item{$|C_\mu^S|$}: 
consistent with the SM with the $1\sigma$ error of 12-15\%
\end{itemize}
We further note that the BSM models predicting the same 
normalized Yukawa couplings to
the up- and down-type quarks and charged leptons are preferred.

\subsection{CP-violating fits}
%
%
\begin{table}[!t]
\centering
\caption{\label{tab:CPVN}
Varying parameters in the {\bf CPVn} fits and their subfits considered in this work.
The parameters not mentioned are supposed to take
the SM value of either 0 or 1.
For the total 17 CPV parameters, see Eq.~(\ref{eq:cpv_parameters}).
}  \vspace{1mm}
\renewcommand{\arraystretch}{1.1}
\begin{adjustbox}{width=12cm}
\begin{tabular}{c|c|c|c|c|c|c|c|c|c|c|c|c|c}
\hline
%
%
&
\multicolumn{4}{c|}{\bf CPV2} &
\multicolumn{5}{c|}{\bf CPV3} &
\multicolumn{2}{c|}{\bf CPV4} &
{\bf CPV5} &
{\bf CPV7} \\ \cline{2-14}
&
U & D & L & HC &
U & D & L & F & IUHC & IUF &
HCC & IUHCC & A \\ \hline
non-SM & & & & & & & & & $\Delta\Gamma_{\rm tot}$& $\Delta\Gamma_{\rm tot}$&
& $\Delta\Gamma_{\rm tot}$&  \\
Parameters& & & & $\Delta S^\gamma$& & & & & $\Delta S^\gamma$& &
$\Delta S^\gamma$& $\Delta S^\gamma$& \\
Varied& & & & $\Delta P^\gamma$& & & & & $\Delta P^\gamma$& &
$\Delta P^\gamma$& $\Delta P^\gamma$&  \\
& & & & & & & & & & & $\Delta S^g$& $\Delta S^g$&  \\
& & & & & & & & & & & $\Delta P^g$& $\Delta P^g$&  \\ \hline
SM& & & & & $C_V$& $C_V$& $C_V$& $C_V$& & $C_V$& & & $C_V$ \\
Parameters& $C_u^S$& & & & $C_u^S$& & & $C_f^S$ & & $C_f^S$& & & $C_u^S$ \\
Varied& $C_u^P$& & & & $C_u^P$& & & $C_f^P$ & & $C_f^P$&  & & $C_u^P$ \\
 & & $C_d^S$& & & & $C_d^S$& & & & & &  & $C_d^S$ \\
 & & $C_d^P$& & & & $C_d^P$& & & & & &  & $C_d^P$ \\
  & & & $C_\ell^S$& & & & $C_\ell^S$& & & & & &   $C_\ell^S$ \\
  & & & $C_\ell^P$& & & & $C_\ell^P$& & & & & &   $C_\ell^P$ \\
%
%
\hline
\end{tabular}
\end{adjustbox}
\end{table}

We generically label the CPV fits as {\bf CPVn} with {\bf n} standing for the
number of fitting parameters like as in the CPC fits. 
Since there are 17 parameters to fit most generally,
it is more challenging to exhaust all the possibilities than in the CPC fits.
Noting that CP violation is signaled by the simultaneous existence of
the Higgs couplings to the scalar and pseudoscalar fermion bilinears,
\footnote{
We suppose that the contributions from the triangle loops in which
non-SM heavy charged and/or colored fermions are running
result in
the coexistence of the scalar and pseudoscalar form factors of
$\Delta S^{\gamma,g}$ and $\Delta P^{\gamma,g}$  
when the Higgs boson simultaneously couples to the scalar and pseudoscalar bilinears of
the non-SM fermions.}
we consider the following CPV fits in this work:
\begin{itemize}
\item{\bf CPV2}: in this fit, we consider the four subfits as follows:
\begin{itemize}
\item{U}: vary $\{C_u^S\,,C_u^P\}$ for the case in which CP violation resides 
in the up-type quark sector
\item{D}: vary $\{C_d^S\,,C_d^P\}$ for the case in which CP violation resides 
in the down-type quark sector
\item{L}: vary $\{C_\ell^S\,,C_\ell^P\}$ for the case in which CP violation resides 
in the charged-lepton sector
\item{HC}: vary $\{\Delta S^\gamma\,,\Delta P^\gamma\}$ for the case in which CP violation occurs 
due to heavy electrically charged non-SM fermions coupling to the Higgs boson
\end{itemize}
\item{\bf CPV3}: in this fit, we consider the five subfits as follows:
\begin{itemize}
\item{U}: vary $\{C_V\,,C_u^S\,,C_u^P\}$ for the up-quark sector CP violation
\item{D}: vary $\{C_V\,,C_d^S\,,C_d^P\}$ for the down-quark sector CP violation
\item{L}: vary $\{C_V\,,C_\ell^S\,,C_\ell^P\}$ for the charged-lepton sector CP violation
\item{F}: vary $\{C_V\,,C_f^S\,,C_f^P\}$ assuming the universal normalized CPV couplings to the SM
quarks and charged leptons
\item{IUHC}: vary $\{\Delta\Gamma_{\rm tot}\,,\Delta S^\gamma\,,\Delta P^\gamma\}$ 
for the case in which CP violation occurs 
due to the $H$ couplings to heavy electrically charged non-SM fermions 
in the presence of light non-SM particles into which $H$ could decay
\end{itemize}
\item{\bf CPV4}:  in this fit, we consider the following two subfits:
\begin{itemize}
\item{IUF}: vary $\{\Delta\Gamma_{\rm tot},C_V,C_f^S,C_f^P\}$ assuming the 
universal normalized CPV couplings to the SM  fermions
in the presence of light non-SM particles into which $H$ could decay
\item{HCC}: vary $\{\Delta S^\gamma\,,\Delta P^\gamma\,,\Delta S^g\,,\Delta P^g\}$ 
for the case in which CP violation occurs 
due to the $H$ couplings to heavy electrically charged {\it and} colored non-SM fermions 
\end{itemize}
\item{\bf CPV5}: in this fit, we consider the following scenario:
\begin{itemize}
\item{IUHCC}: vary $\{\Delta\Gamma_{\rm tot}\,,
\Delta S^\gamma\,,\Delta P^\gamma\,,\Delta S^g\,,\Delta P^g\}$ 
for the case in which CP violation occurs 
due to the $H$ couplings to heavy electrically charged {\it and} colored non-SM fermions 
in the presence of light non-SM particles into which $H$ could decay
\end{itemize}
\item{\bf CPV7}: in this fit, we consider the following scenario:
\begin{itemize}
\item{A}: vary $\{C_V,C_u^S,C_u^P,C_d^S,C_d^P,C_\ell^S,C_\ell^P\}$ 
with the $H$ couplings to the SM particles like as in CPV A2HDM
\end{itemize}
\end{itemize}
We provide Table~\ref{tab:CPVN} for the summary of the CPV fits considered in this work
which explicitly shows the parameters varied in each subfit of {\bf CPVn}.

\medskip

Since the signal strengths are CP-even quantities, they do not contain
CPV products such as $C_{u,d,\ell,f}^S\times C_{u,d,\ell,f}^P$ and 
$S^{\gamma,g}\times P^{\gamma,g}$.
Therefore, the CL regions appear as a circle or an ellipse or some overlapping of them in the 
$(C_{u,d,\ell,f}^S,C_{u,d,\ell,f}^P)$ and
$(\Delta S^{\gamma,g},\Delta P^{\gamma,g})$ planes.

\subsubsection{{\bf CPV2} and {\bf CPV3}}
%
\begin{table}[!b]
\caption{\it
\label{tab:CPV23}
{\bf CPV2} and {\bf CPV3}: The best-fitted values in the four {\bf CPV2} 
and five {\bf CPV3} subfits. 
Also shown are the corresponding minimal
chi-square per degree of freedom ($\chi^2_{\rm min}$/dof), 
goodness of fit (gof), and 
$p$-value against the SM for compatibility
with the SM hypothesis.
For the SM, we obtain $\chi^2_{\rm SM}/{\rm dof}=82.3480/76$
and gof $=0.2895$.
}  \vspace{1mm}
\renewcommand{\arraystretch}{1.1}
\begin{adjustbox}{width= \textwidth}
\begin{tabular}{c|c|c|c|c|c|c|c|c|c}
\hline
%
%
&
\multicolumn{4}{c|}{\bf CPV2} &
\multicolumn{5}{c}{\bf CPV3}  \\ \cline{2-10}
&
U & D & L & HC &
U & D & L & F & IUHC  \\ \hline
non-SM & & & & & & & & & $\Delta\Gamma_{\rm tot}/{\rm MeV}=0.090^{+0.168}_{-0.157}$\\
Parameters& & & & $\Delta S^\gamma=-0.313^{+14.00}_{-0.175}$& & & & & 
$\Delta S^\gamma=-0.369^{+14.14}_{-0.207}$\\
Varied& & & & $\Delta P^\gamma=0.0^{+7.075}_{-7.074}~~~~~~$& & & & & 
$\Delta P^\gamma~=~0.0^{+7.162}_{-7.159}~~~~~~$\\
%
%
\hline
SM& & & & & $C_V=1.040^{+0.018}_{-0.029}$& $C_V=1.048^{+0.026}_{-0.026}$& $C_V=1.026^{+0.017}_{-0.016}$& $C_V=1.015^{+0.017}_{-0.036}$&  \\
Parameters& $C_u^S=0.987^{+0.026}_{-0.322}$& & & & $C_u^S=0.959^{+0.028}_{-0.143}$& & & $C_f^S=0.930^{+0.031}_{-0.081}$ & \\
Varied& $C_u^P=~~~0.0^{+0.469}_{-0.469}$& & & & $C_u^P=~~~0.0^{+0.341}_{-0.341}$& & & $C_f^P~=~0.0^{+0.267}_{-0.268}$ & \\
& & $C_d^S=0.978^{+0.029}_{-0.482}$& & & & $C_d^S=1.045^{+0.048}_{-0.671}$& & & \\
& & $C_d^P=~~~0.0^{+0.845}_{-0.844}$& & & & $C_d^P=~~~0.0^{+0.986}_{-0.989}$& & &  \\
& & & $C_\ell^S=0.939^{+0.038}_{-1.893}$& & & & $C_\ell^S=0.946^{+0.037}_{-1.921}$& &  \\
& & & $C_\ell^P=~~~0.0^{+0.964}_{-0.964}$& & & & $C_\ell^P=~~~0.0^{+0.975}_{-0.975}$& &  \\
%
%
\hline
$\chi^2_{\rm min}$/dof & 82.1034/74 & 81.7649/74 & 79.7168/74 & 79.2183/74 & 77.1617/73 & 78.4013/73 & 77.2007/73 & 74.3664/73 & 78.8971/73 \\
goodness of fit (gof) & 0.2427 & 0.2509 & 0.3040 & 0.3178 & 0.3471 & 0.3116 & 0.3460 & 0.4335 & 0.2979 \\
$p$-value against the SM & 0.8849 & 0.7470 & 0.2683 & 0.2091 & 0.1587 & 0.2673 & 0.1613 & 0.0464 & 0.3272 \\
\hline
\end{tabular}
\end{adjustbox}
\end{table}
\begin{figure}[!htb]
\begin{center}
\includegraphics[width=9.1cm]{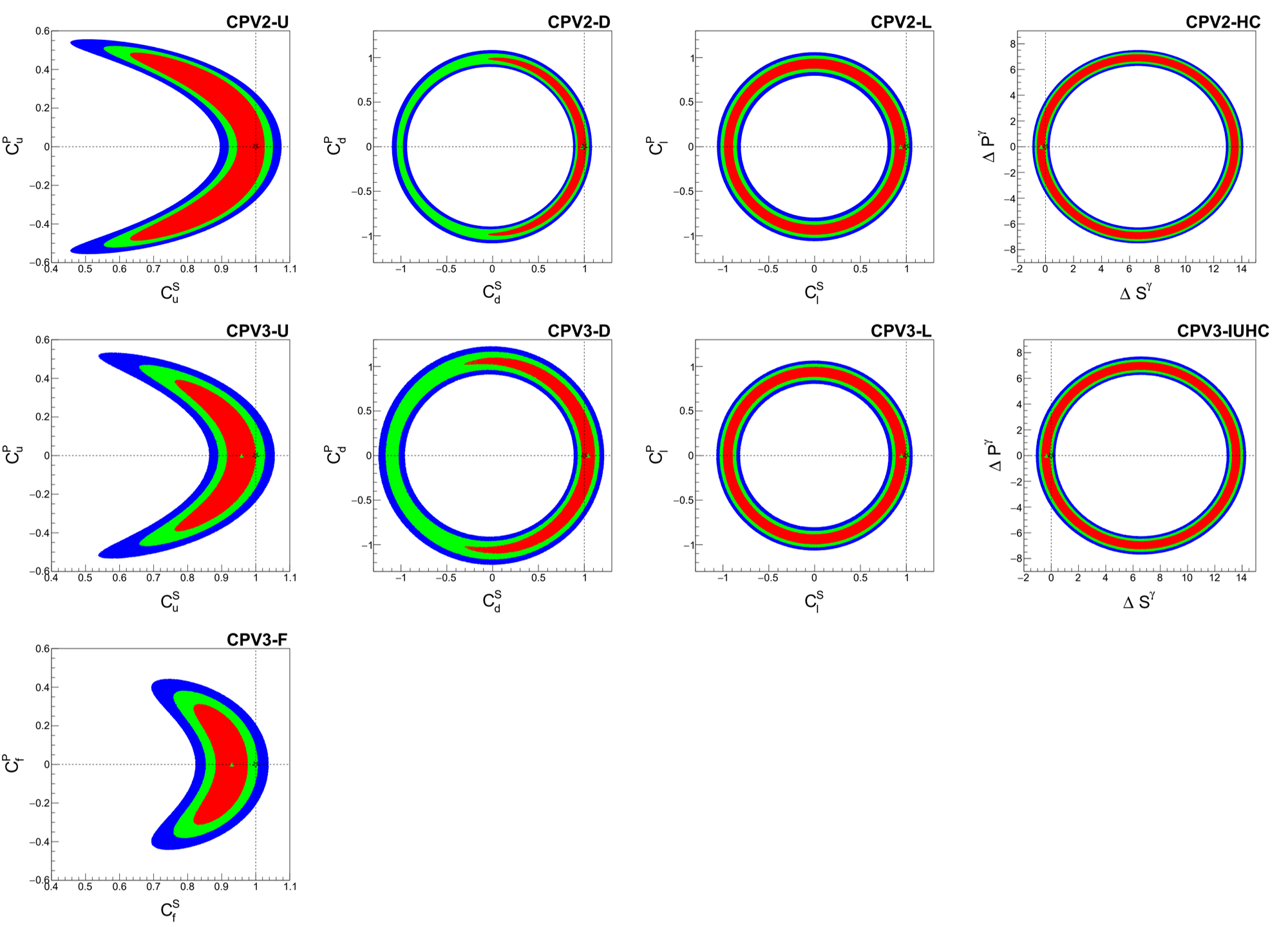}
\includegraphics[width=9.1cm]{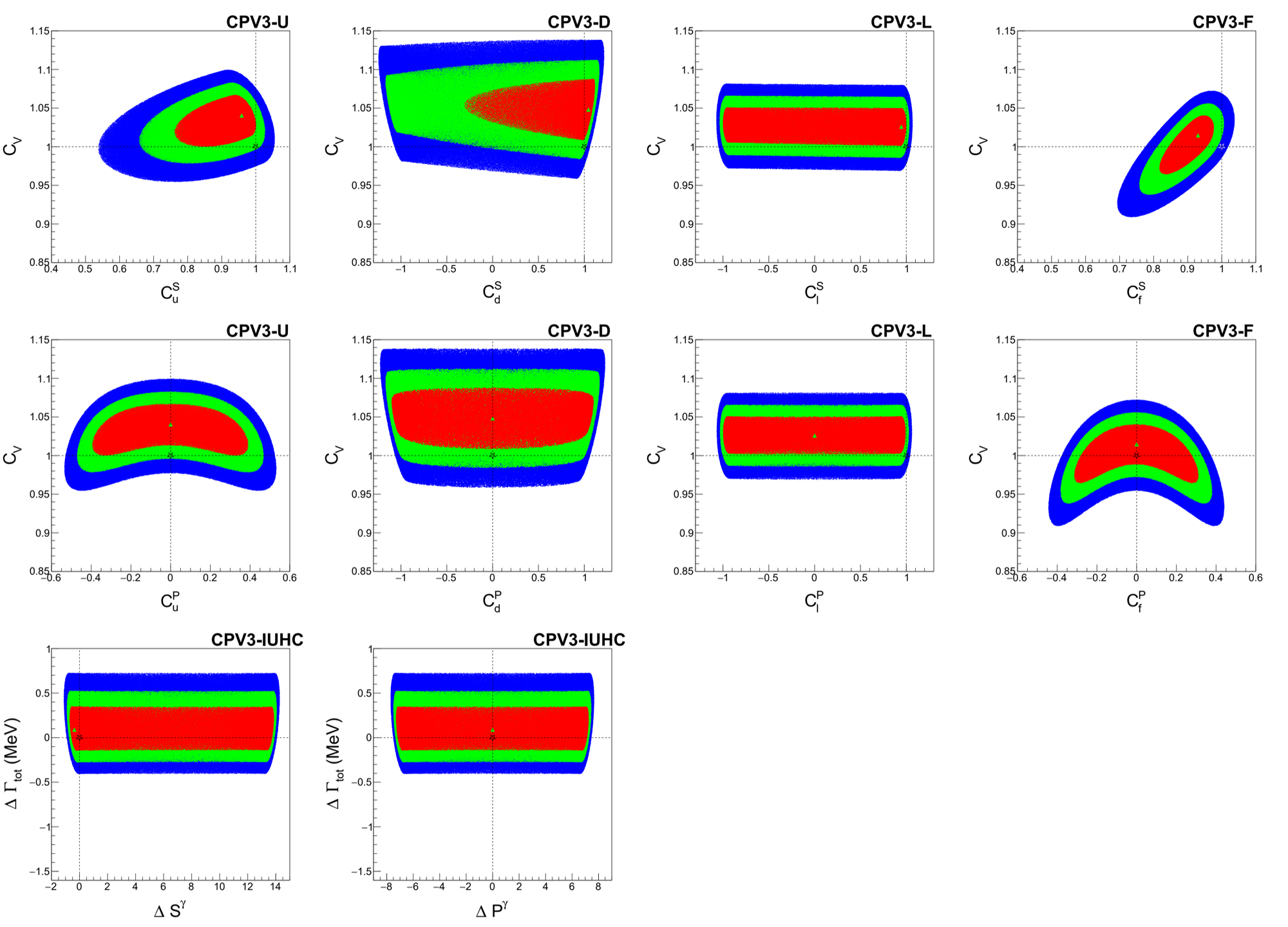}
\end{center}
\vspace{-0.5cm}
\caption{\it {\bf CPV2} and {\bf CPV3}:
[Left] 
The CL regions of the four {\bf CPV2} (upper) and 
five {\bf CPV3} (middle and lower)
subfits in the $(C_{u,d,\ell,f}^S,C_{u,d,\ell,f}^P)$ and
$(\Delta S^{\gamma},\Delta P^{\gamma})$ planes.
[Right]
The CL regions of the five {\bf CPV3} subfits in the 
$(C_{u,d,\ell,f}^{S}, C_V)$ (upper),
$(C_{u,d,\ell,f}^{P}, C_V)$ (middle),
$(\Delta S^\gamma,\Delta\Gamma_{\rm tot})$, and
$(\Delta P^\gamma,\Delta\Gamma_{\rm tot})$ (lower) planes.
The contour regions shown are for
$\Delta\chi^2\leq 2.3$ (red),
$\Delta\chi^2\leq 11.83$ (blue)
above the minimum, which correspond to
confidence levels of 68.27\%, 95\%, and 99.73\%, respectively.
In each frame, the vertical and horizontal lines locate the SM point denoted
by a star and the best-fit point is denoted by a triangle.
}
\label{fig:CPV23}
\end{figure}

We show the fitting results for the four {\bf CPV2} and five {\bf CPV3} subfits 
in Table~\ref{tab:CPV23}.
We have the largest gof value for {\bf CPV3}-F and note that
the $p$-values against the SM 
for compatibility with the SM hypothesis
are high in {\bf CPV2}-U and {\bf CPV2}-D
with $\chi^2_{\rm min} \sim \chi^2_{\rm SM}$. 
In the left panel of Fig.~\ref{fig:CPV23}, the CL regions 
are depicted in the 
$(C_{u,d,\ell,f}^S,C_{u,d,\ell,f}^P)$ and
$(\Delta S^{\gamma},\Delta P^{\gamma})$ planes for {\bf CPV2} and {\bf CPV3}.
The other {\bf CPV3} CL regions in the $(C_{u,d,\ell,f}^{S,P}, C_V)$,
$(\Delta S^\gamma,\Delta\Gamma_{\rm tot})$, and
$(\Delta P^\gamma,\Delta\Gamma_{\rm tot})$ planes are shown in
the right panel of Fig.~\ref{fig:CPV23}.

\medskip

In {\bf CPV2}-U,  we obtain the sickle-shaped CL region in the $(C_u^S, C_u^P)$ plane,
see the upper-left frame of the left panel of Fig.~\ref{fig:CPV23}.
This could be understood by observing 
that the top-Yukawa couplings are involved in the
ggF$+$bbH and ttH$\oplus$tH production processes and the $H\to\gamma\gamma$ decay mode.
From Eq.~(\ref{eq:ggF_coeff}) with the Run 2 decomposition coefficients in Table~\ref{tab:ggF_coeff},
Eq.~(\ref{eq:spaa_numeric}) with $S^\gamma_{\rm SM}=-6.542 + 0.046\,i$, and
Eq.~(\ref{eq:muP13}), we have
\begin{eqnarray}
\widehat\mu({\rm ggF+bbH})^{{\bf CPV2}-{\rm U}} &\simeq &
\left[1.04\,(C_u^S)^2-0.06\,C_u^S+0.02\right]  +  2.2\,(C_u^P)^2 \simeq
1.04\,\left(C_u^S-0.03\right)^2 + 2.2\,\left(C_u^P\right)^2 +0.02\,,
\nonumber \\[2mm]
\widehat\mu(\gamma\gamma)^{{\bf CPV2}-{\rm U}} &\simeq &
\left(-1.28+0.28\,C_u^S\right)^2 + 1.016\left(0.42\,C_u^P\right)^2\,,
\nonumber \\[2mm]
\widehat\mu({\rm ttH}\oplus{\rm tH})^{{\bf CPV2}-{\rm U}} &\simeq &
1.3\,\left[(C_u^S)^2+(C_u^P)^2\right]  -  0.8\,C_u^S  +  0.5 \simeq
1.3\,\left[\left(C_u^S-0.31\right)^2 + \left(C_u^P\right)^2\right] +0.38\,,
\end{eqnarray}
where the factor 1.016 in the second line for
$\widehat\mu(\gamma\gamma)$ takes account of the difference
in the QCD and electroweak corrections to the scalar and 
pseudoscalar parts~\cite{Choi:2021nql}.
Note that $\widehat\mu({\rm ggF+bbH})\simeq 1$ gives
an ellipse centered at $(C_u^S,C_u^P)\simeq (0,0)$
with  the lengths of the major ($C_u^S$) and minor ($C_u^P$) axes of 1 and $0.67$
while 
$\widehat\mu(\gamma\gamma)\simeq 1$ an ellipse centered at $(C_u^S,C_u^P)\simeq (4.6,0)$
with  the lengths of the major ($C_u^S$) and minor ($C_u^P$) axes of $3.6$ and $2.4$.
In addition, $\widehat\mu({\rm ttH}\oplus{\rm tH})\simeq 1$  gives
a circle centered at $(C_u^S,C_u^P)\simeq (0.3,0)$
with a radius of about $0.7$.
Both the ellipses and the circle
pass the SM point of $(C_u^S,C_u^P)=(1,0)$ as they should with
the ggF$+$bbH ellipse and the ttH$\oplus$tH circle  
extending to the negative $C_u^S$ direction 
from the SM point and
the $\gamma\gamma$ ellipse to the positive $C_u^S$ direction.
The overlapping of the two ellipses and a circle with some corresponding errors
explain the sickle-shaped CL region in the $(C_u^S, C_u^P)$ plane which
also appears in {\bf CPV3}-U and {\bf CPV3}-F, see the middle-left
and lower-left frames of the left panel of Fig.~\ref{fig:CPV23}.
We observe that the SM point lies outside the 68\% CL region in {\bf CPV3}-F
with $C_f^S=0.930^{+0.031}_{-0.081}$ which deviates from the SM point
more than $C_u^S$ in {\bf CPC2}-U and {\bf CPC3}-U with the smaller negative error.

\medskip

The circles in the $(C_d^S,C_d^P)$ planes for
{\bf CPV2}-D and {\bf CPV3}-D shown
in the two middle-left frames of the left panel of Fig.~\ref{fig:CPV23} are 
understood by noting that the signal strength of 
the $bb$ decay mode is given by
\begin{equation}
\widehat\mu(bb)^{{\bf CPV3}-{\rm D}} \simeq \frac{(C_d^S)^2+(C_d^P)^2}
{0.57\,[(C_d^S)^2+(C_d^P)^2]+0.25\,C_V^2+0.18}.
\end{equation}
The positive values of $C_d^S$ are preferred because of the 
interferences between the top- and bottom-quark contributions to ggF. We note that
$\Delta\chi^2$ above the minimum at $(C_d^S,C_d^P)\simeq (1,0)$
increases by the amount of about 5 while $C_d^S$ changes from $+1$ to $-1$.
When $C_d^P$ changes from $+1$ to $-1$,
$\Delta\chi^2$ above the minimum 
increases by the amount smaller than 2.

\medskip

The circles in the $(C_\ell^S,C_\ell^P)$ planes for
{\bf CPV2}-L and {\bf CPV3}-L shown
in the two middle-right frames of the left panel of Fig.~\ref{fig:CPV23} are
understood by noting that the signal strength of
the $\tau\tau$ and $\mu\mu$ decay modes is given by
$\widehat\mu(\tau\tau)\simeq \widehat\mu(\mu\mu) \simeq (C_\ell^S)^2+(C_\ell^P)^2$.
We note that $\Delta\chi^2$ above the minimum $(C_\ell^S,C_\ell^P)\simeq (0.94,0)$
increases by the amount less than 1 while $C_\ell^{S,P}$ changes from $+1$ to $-1$.
We note that the charged-lepton circles are smaller than
the down-type-quark circles.

\medskip

The circles in the $(\Delta S^\gamma,\Delta P^\gamma)$ planes for
{\bf CPV2}-HC and {\bf CPV3}-IUHC shown
in the two right frames of the left panel of Fig.~\ref{fig:CPV23} are 
understood by noting that 
$\widehat\mu(\gamma\gamma)=1$ gives a circle centered at 
$(\Delta S^\gamma,\Delta P^\gamma) \simeq (6.5,0)$ 
with the radius of about $6.5$ with
the signal strength of the $H\to\gamma\gamma$ decay mode given by
\begin{equation}
\widehat\mu(\gamma\gamma)^{{\bf CPV2}-{\rm HC}} \simeq 
\left(-1+\frac{\Delta S^\gamma}{|S^\gamma_{\rm SM}|}\right)^2 +
\left(\frac{\Delta P^\gamma}{|S^\gamma_{\rm SM}|}\right)^2\,.
\end{equation}

\medskip

From the ten frames of the right panel of Fig.~\ref{fig:CPV23},
we observe that the most of the SM points are outside of the 68\% CL regions
except {\bf CPV3}-F in the $(C_f^P,C_V)$ plane (middle-right) and 
{\bf CPV3}-IUHC (lower).
There are almost no correlations between
$C_V$ and $C_\ell^{S,P}$ (upper-middle-right and middle-middle-right) and
the correlation between
$C_V$ and $C_d^{P}$ in {\bf CPV3}-D (middle-middle-left) is 
weakly correlated.
We also see almost no correlations between 
$\Delta\Gamma_{\rm tot}$ and $\Delta S^\gamma$ (lower-left) and
$\Delta\Gamma_{\rm tot}$ and $\Delta P^\gamma$ (lower-middle) 
in {\bf CPV3}-IUHC.

\subsubsection{{\bf CPV4}, {\bf CPV5}, and {\bf CPV7}}
%
\begin{table}[!b]
\caption{\it
\label{tab:CPV457}
{\bf CPV4}, {\bf CPV5}, and {\bf CPV7}: The best-fitted values in the {\bf CPV4}, 
{\bf CPV5}, and {\bf CPV7} fits.
Also shown are the corresponding minimal
chi-square per degree of freedom ($\chi^2_{\rm min}$/dof), 
goodness of fit (gof), and 
$p$-value against the SM for compatibility
with the SM hypothesis.
For the SM, we obtain $\chi^2_{\rm SM}/{\rm dof}=82.3480/76$
and gof $=0.2895$.
}  \vspace{1mm}
\renewcommand{\arraystretch}{1.1}
\begin{adjustbox}{width=12cm}
\begin{tabular}{c|c|c|c|c}
\hline
%
%
&
\multicolumn{2}{c|}{{\bf CPV4}} &
{\bf CPV5} &
{\bf CPV7} \\ \cline{2-5}
&
IUF&HCC & IUHCC & A \\ \hline
non-SM &$\Delta\Gamma_{\rm tot}=0.0^{+0.155}$ & &$\Delta\Gamma_{\rm tot}=-0.029^{+0.211}_{-0.491}$&  \\
Parameters& & $\Delta S^\gamma=-0.400^{+14.21}_{-0.208}$&  $\Delta S^\gamma=-0.392^{+14.17}_{-0.213}$&
\\
Varied& &$\Delta P^\gamma=0.0^{+7.190}_{-7.183}~~~~~~$& 
$\Delta P^\gamma = 0.0^{+7.178}_{-7.176}~~~~~~$&  \\
&& $\Delta S^g=-0.032^{+0.031}_{-0.257}$& $\Delta S^g=-0.036^{+0.041}_{-0.609}$&  \\
&& $\Delta P^g=0.0^{+0.743}_{-0.745}~~~~~~$& $\Delta P^g=0.0^{+0.987}_{-0.988}~~~~~~$&  \\ \hline
SM& $C_V =1.0_{-0.020}~~~$& & & $C_V = 1.002^{+0.034}_{-0.034}$ \\
Parameters& $C_f^S=0.900^{+0.035}_{-0.050}$&& & $C_u^S = 0.927^{+0.037}_{-0.106}$ \\
Varied& $C_f^P=0.169^{+0.101}_{-0.440}\,,-0.169^{+0.440}_{-0.101}$&  
& & $C_u^P = 0.0^{+0.300}_{-0.294}~~~$ \\
&&  & &  $C_d^S = 0.902^{+0.075}_{-0.902}$ \\
&&  & &  $C_d^P = 0.0^{+0.951}_{-0.950}~~~$ \\
&&  & &   $C_\ell^S = 0.916^{+0.044}_{-0.916}$ \\
&&  & &   $C_\ell^P = 0.0^{+0.954}_{-0.952}~~~$ \\
%
%
\hline
$\chi^2_{\rm min}$/dof &74.3500/72&  78.1906/72 & 78.1707/71  &  74.1893/69 \\
goodness of fit (gof) & 0.4340 & 0.3175 & 0.2893  &  0.3129 \\
$p$-value against the SM  & 0.0461 & 0.2450  & 0.3825  &  0.3188 \\
\hline
\end{tabular}
\end{adjustbox}
\end{table}
%
%
\begin{figure}[!b]
\begin{center}
\includegraphics[width=9.1cm]{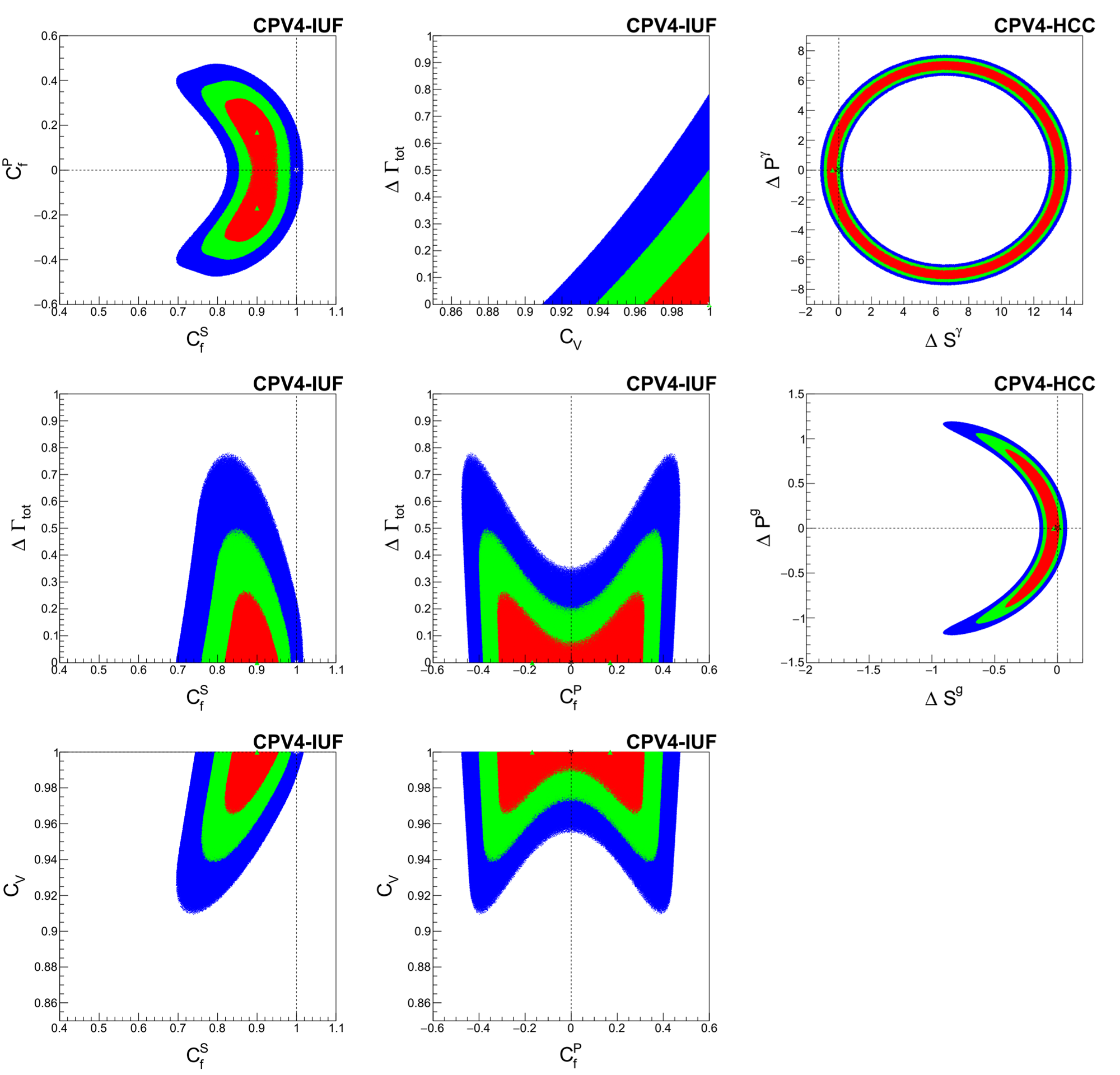}
\includegraphics[width=9.1cm]{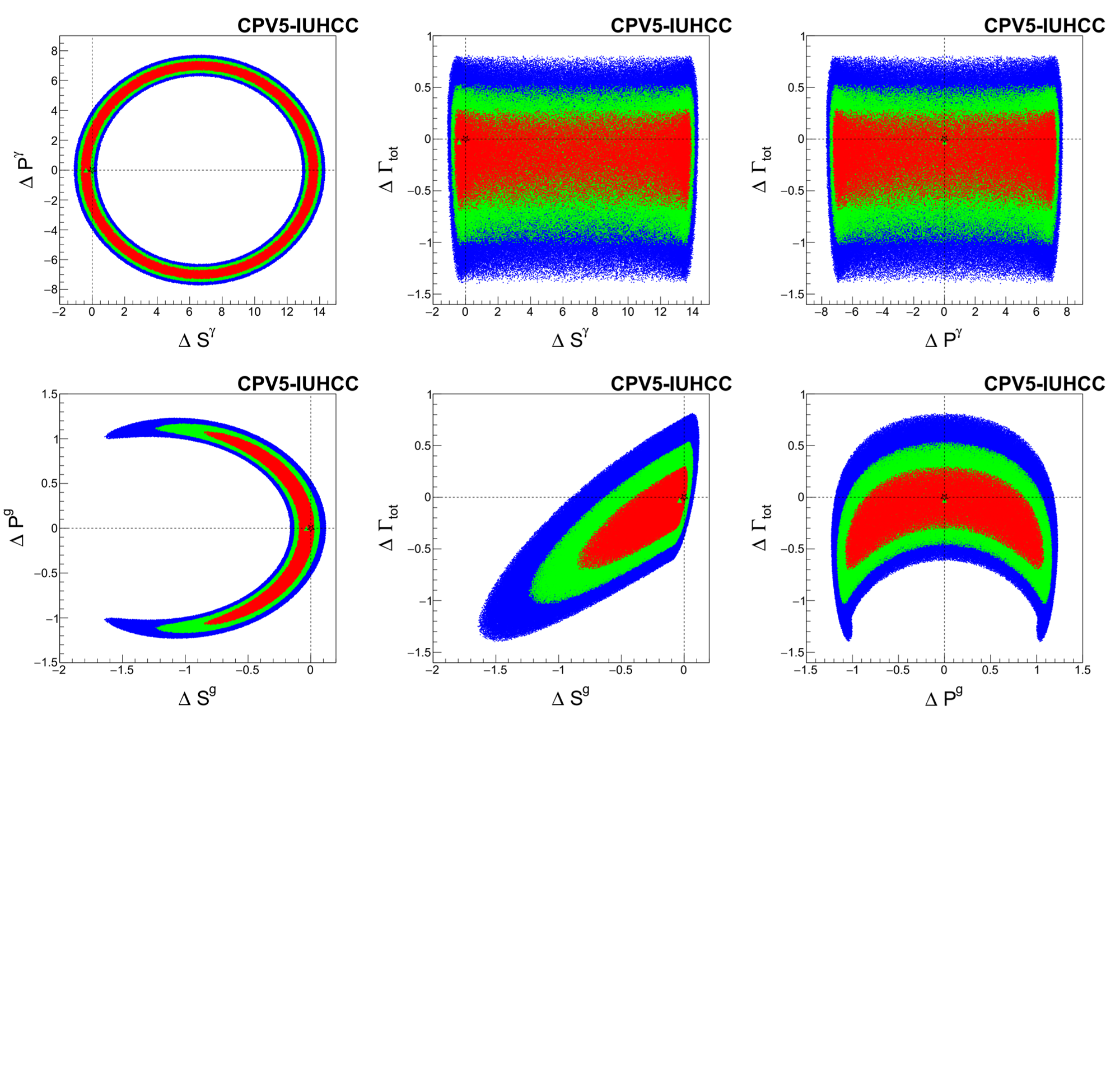}
\end{center}
\vspace{-0.5cm}
\caption{\it {\bf CPV4} and {\bf CPV5}:
[Left]
The CL regions of {\bf CPV4}-IUF (left and middle) and
{\bf CPV4}-HCC (right) in two-parameter planes.
[Right]
The CL regions of {\bf CPV5}-IUHCC in two-parameter planes.
The contour regions shown are for
$\Delta\chi^2\leq 2.3$ (red),
$\Delta\chi^2\leq 5.99$ (green),
$\Delta\chi^2\leq 11.83$ (blue)
above the minimum, which correspond to
confidence levels of 68.27\%, 95\%, and 99.73\%, respectively.
In each frame, the vertical and horizontal lines locate the SM point denoted
by a star and the best-fit point is denoted by a triangle.
}
\label{fig:CPV45}
\end{figure}
%
%
\begin{figure}[!t]
\begin{center}
\includegraphics[width=9.1cm]{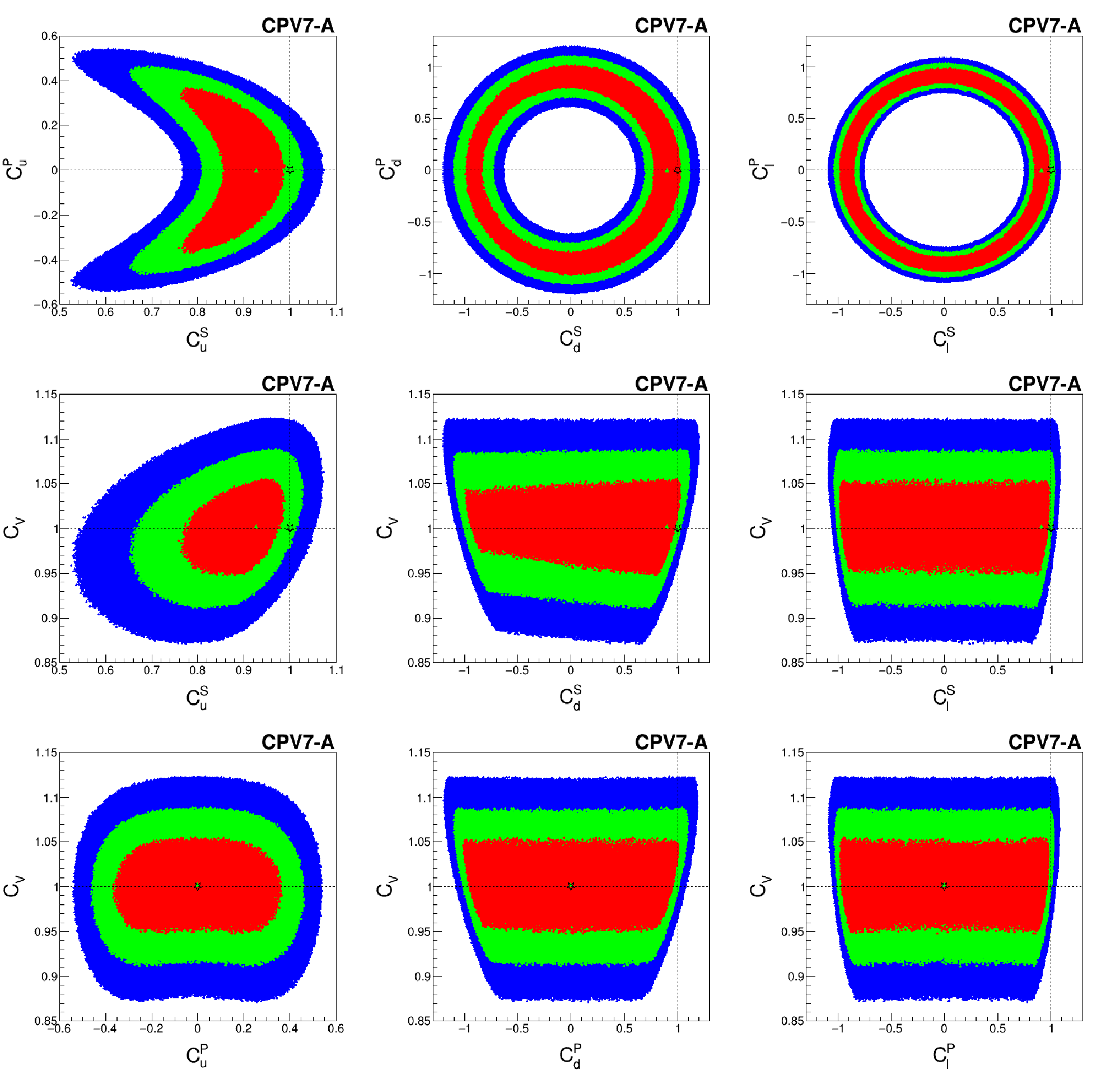}
\end{center}
\vspace{-0.5cm}
\caption{\it {\bf CPV7}:
The CL regions of {\bf CPV7}-A in two-parameter planes.
The contour regions shown are for
$\Delta\chi^2\leq 2.3$ (red),
$\Delta\chi^2\leq 5.99$ (green),
$\Delta\chi^2\leq 11.83$ (blue)
above the minimum, which correspond to
confidence levels of 68.27\%, 95\%, and 99.73\%, respectively.
In each frame, the vertical and horizontal lines locate the SM point denoted
by a star and the best-fit point is denoted by a triangle.
}
\label{fig:CPV7}
\end{figure}

We show the fitting results for {\bf CPV4}, {\bf CPV5}, and {\bf CPV7}
in Table~\ref{tab:CPV457}.
Note that, in {\bf CPV4}-IUF, the parameters are not bounded like as in the HP scenarios
and we implement the fit under the constraints of
$\Delta\Gamma_{\rm tot}\geq 0$ and $C_{V}\leq 1$.
We have the largest gof value for {\bf CPV4}-IUF which is slightly higher than
that of {\bf CPV3}-F.
The left panel of Fig.~\ref{fig:CPV45} is for {\bf CPV4}.
In the six left and middle frames, the CL regions in IUF
are depicted and, in the two upper- and middle-right frames,
those in HCC are shown in the $(\Delta S^\gamma,\Delta P^\gamma)$ and
$(\Delta S^g,\Delta P^g)$ planes.
The right panel of Fig.~\ref{fig:CPV45} is for {\bf CPV5}-IUHCC and
the CL regions in the $(\Delta S^\gamma,\Delta P^\gamma)$ and
$(\Delta S^g,\Delta P^g)$ planes are depicted in the two left frames.
In Fig.~\ref{fig:CPV7}, we show the CL regions of {\bf CPV7}
in the $(C_{u,d,\ell}^S,C_{u,d,\ell}^P)$
planes in the upper three frames and some others below them.

\medskip

In {\bf CPV4}-IUF, $\Delta\Gamma_{\rm tot}$ and $C_V$ are driven to the SM values under 
the constraints of $\Delta\Gamma_{\rm tot}\geq 0$ and $C_V\leq 1$ and,
being different from {\bf CPV2} and {\bf CPV3}, we have the 
two degenerate minima at $(C_f^S,C_f^P)\simeq (0.90,\pm 0.17)$.
The central value of $C_f^S$ is smaller than those in the {\bf CPC} fits 
by the amount of 2\%-3\% 
which is compensated by the relation of
$\sqrt{(C_f^S)^2+(C_f^P)^2} \simeq 0.92$ at the minima.
The CL regions are shown in the left and middle six frames 
of the left panel of Fig.~\ref{fig:CPV45}.
Both the two degenerate minima are in the 68\% CL region and the SM point is outside the
95\% CL region, see the left three frames for $C_f^P$, $\Delta\Gamma_{\rm tot}$, and
$C_V$ versus $C_f^S$.

\medskip
%
In {\bf CPV4}-HCC, we have the $\widehat\mu(\gamma\gamma)$ circle
centered at $(\Delta S^\gamma,\Delta P^\gamma) \simeq (7,0)$ 
with the radius of about $7$, see the upper-right frame of the
left panel of Fig.~\ref{fig:CPV45}.
There is no visible
change in $\Delta\chi^2$ above the minimum along the circle
passing the center of the 68\% CL region.
On the other hand, we obtain the sickle-shaped CL region in 
the $(\Delta S^g, \Delta P^g)$ plane, 
see the middle-right frame of the left panel of Fig.~\ref{fig:CPV45}.
This is understood by the overlapping of the 
$\widehat\mu({\rm ggF}+{\rm bbH})$ and $\widehat\mu({\cal D}\neq \gamma\gamma,gg)$ circles:
\begin{eqnarray}
\widehat\mu({\rm ggF}+{\rm bbH})^{{\bf CPV4}-{\rm HCC}} &= &
\left[1+1.778\,\Delta S^g+0.758\, (\Delta S^g)^2\right] + 0.779\, (\Delta P^g)^2 \simeq
0.76\,(\Delta S^g +1.15)^2 + 0.78\,(\Delta P^g)^2\,,
\nonumber \\[2mm]
\widehat\mu({\cal D}\neq \gamma\gamma,gg)^{{\bf CPV4}-{\rm HCC}} &\simeq  &
\frac{1}{0.92+0.08\left[
\left(1+{\Delta S^g}/{|S^g_{\rm SM}|}\right)^2 +
0.96\left({\Delta P^g}/{|S^g_{\rm SM}|}\right)^2\right]}\,.
\end{eqnarray}
For $\widehat\mu({\rm ggF}+{\rm bbH})$, we use Eq.~(\ref{eq:ggF_coeff}) 
with the Run 2 decomposition coefficients in Table~\ref{tab:ggF_coeff}.
We note that $\widehat\mu({\rm ggF}+{\rm bbH})=1$
gives a circle centered at 
$(\Delta S^g,\Delta P^g) \simeq (-1.15,0)$ 
with the radius of about $1.15$.
In the second line for $\widehat\mu({\cal D}\neq \gamma\gamma,gg)$,
the factor 0.96 takes account of the difference
in the QCD and electroweak corrections to the scalar and 
pseudoscalar parts~\cite{Choi:2021nql}.
With $S^g_{\rm SM}=0.636 + 0.071\,i$ and $P^g_{\rm SM}=0$, we note that
$\widehat\mu({\cal D}\neq \gamma\gamma,gg)=1$ gives a circle centered at 
$(\Delta S^g,\Delta P^g) \simeq (-0.64,0)$ 
with the radius of about $0.64$ which is smaller than the
$\widehat\mu({\rm ggF}+{\rm bbH})$ circle.
Note that we obtain the sickle-shaped CL region in 
the $(\Delta S^g, \Delta P^g)$ plane because we consider the
ggF$+$bbH production process beyond LO in QCD.

\medskip

In {\bf CPV5}-IUHCC, compared to {\bf CPV4}-HCC, we vary $\Delta\Gamma_{\rm tot}$ additionally.
The best-fitted values are similar to those in {\bf CPV4}-HCC
with a bit larger errors for $\Delta S^g$ and $\Delta P^g$. We observe that
the sickle-shaped CL region in the $(\Delta S^g,\Delta P^g)$ plane extends
at the cost of negative $\Delta\Gamma_{\rm tot}$, see the middle three
frames in the right panel of Fig.~\ref{fig:CPV45}.
Note that $\Delta\Gamma_{\rm tot}$ is almost insensitive to
$\Delta S^\gamma$ and $\Delta P^\gamma$ in the allowed regions.

\medskip

In {\bf CPV7}-A, $C_V$ is consistent with the SM with the $1\sigma$ error less than 4\%.
The scalar couplings $C_{u,d,\ell}^S$ below the SM with the positive 
$1\sigma$ errors of 4-8\%.  The negative $1\sigma$ errors are larger
or much larger: about 10\% for $C_u^S$ and almost 100\% for $C_{d,\ell}^S$.
Comparing with the best-fitted values in {\bf CPC4}-A, we find that the 
central values and the positive errors of $C_{u,d,\ell}^S$ are similar while the
negative errors extend to the negative direction and the positive and
negative regions are connected for $C_{d,\ell}^S$: compare
the CL regions shown in
the upper frames of the left panel of Fig~\ref{fig:CPC4_CLregions} ({\bf CPC4}-A)
and
those in the middle frames of Fig~\ref{fig:CPV7} ({\bf CPV7}-A).
For the pseudoscalar couplings, $C_u^P$ is constrained around its SM value
of 0 with the $1\sigma$ error of 30\%.
For the other pseudoscalar couplings of $C_d^P$ and and $C_\ell^P$,
we have the $1\sigma$ errors of about 100\%.
We clearly see the SM points outside of the 68\% CL in the 
$(C_u^S,C_u^P)$  and $(C_u^S,C_V)$ 
planes, see the upper- and middle-left frames of Fig.~\ref{fig:CPV7}.

\subsection{Predictions for $H\to Z\gamma$}
\label{subsec:Za}
Recently, the ATLAS and CMS collaborations report the first evidence
for the Higgs boson decay to a $Z$ boson and a photon 
with a statistical significance of 3.4 standard deviations
based on the Run 2 data with 140/fb luminosity for 
each experiment~\cite{ATLAS:2023yqk}.
The combined analysis gives the measured signal yield of $2.2\pm 0.7$ 
times the SM prediction which corresponds
to $B(H\to Z\gamma)=(3.4\pm 1.1)\times 10^{-3}$ assuming SM Higgs boson production
cross sections.
\footnote{The SM prediction for  $B(H\to Z\gamma)$ is $1.58\times 10^{-3}$
\cite{Choi:2021nql}.}

\medskip

The loop-induced Higgs couplings to a $Z$ boson and a photon
are similarly described as 
those to two photons and two gluons by using the scalar and pseudoscalar form factors
of $S^{Z\gamma}$ and $P^{Z\gamma}$.
For the detailed description and analytic structure of them,
we refer to Ref.~\cite{Choi:2021nql}.
Taking $M_H=125$ GeV, we have
\begin{eqnarray}
\label{eq:spza_numeric}
S^{Z\gamma} &=&
	-12.3401  \, g_{_{HWW}}
	+  0.6891  \, g_{H\bar{t}t}^S
	+ (-0.0186 + 0.0111 \, i ) \, g_{H\bar{b}b}^S
	+ (-0.0005 + 0.0002 \, i) \,g_{H\tau\tau}^S + \Delta S^{Z\gamma}  \,,
\nonumber \\
P^{Z\gamma} &=&
	1.0459  \, g_{H\bar{t}t}^P
	+ (-0.0219 + 0.0112 \, i ) \, g_{H\bar{b}b}^P
	+ (-0.0006 + 0.0002 \, i) \,g_{H\tau\tau}^P + \Delta P^{Z\gamma} \,,
\end{eqnarray}
retaining only the dominant contributions from third-generation SM 
fermions and the charged gauge bosons $W^\pm$
and introducing $\Delta S^{Z\gamma}$ and $\Delta P^{Z\gamma}$ to
parametrize contributions from the triangle loops in which
non-SM charged particles are running.
In the SM limit, $S^{Z\gamma}_{\rm SM} =-11.6701 + 0.0114 \, i$ and 
$P^{Z\gamma}_{\rm SM} =0$.

%
%
\begin{figure}[!b]
\begin{center}
\includegraphics[width=9.1cm]{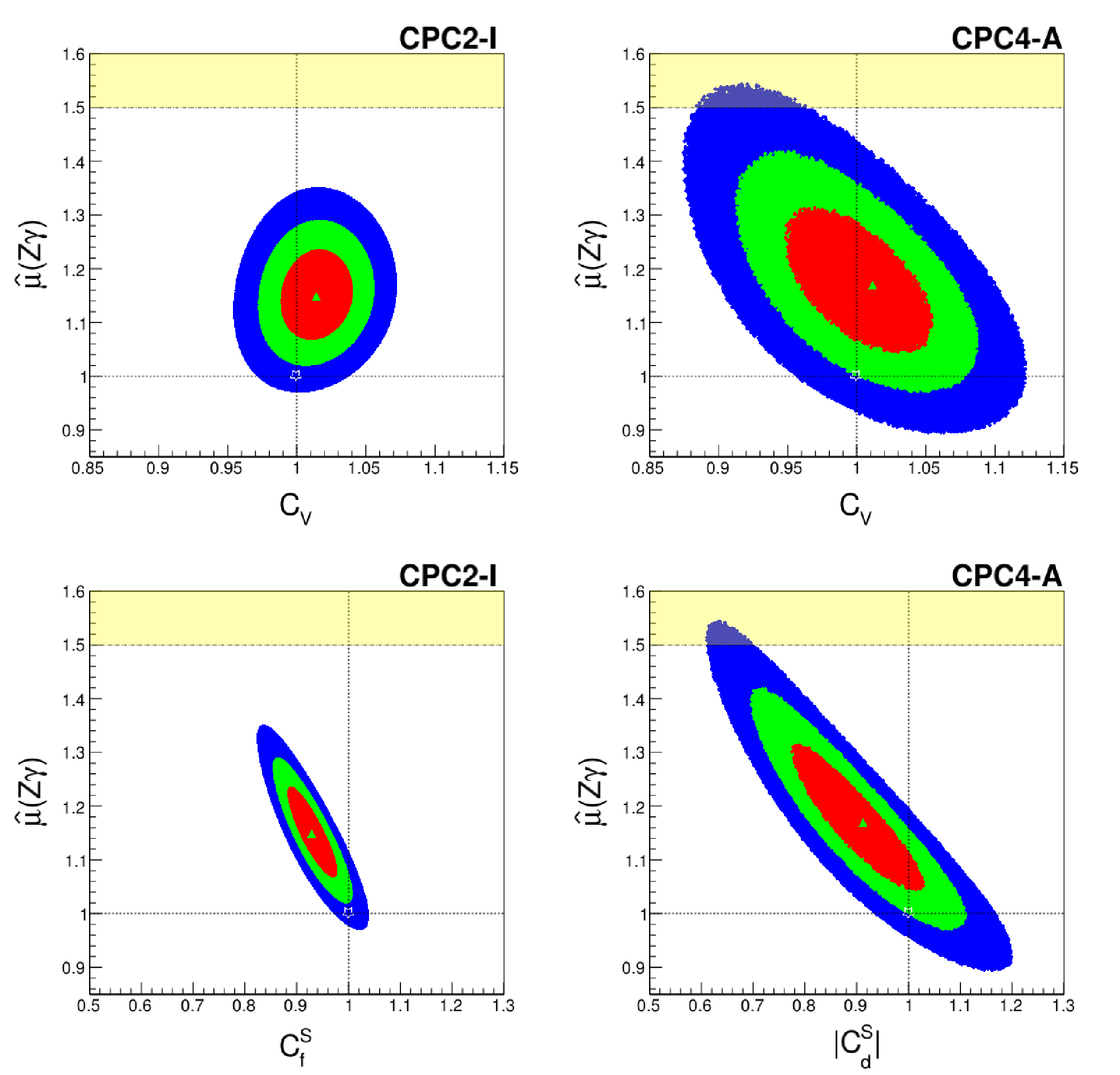}
\includegraphics[width=9.1cm]{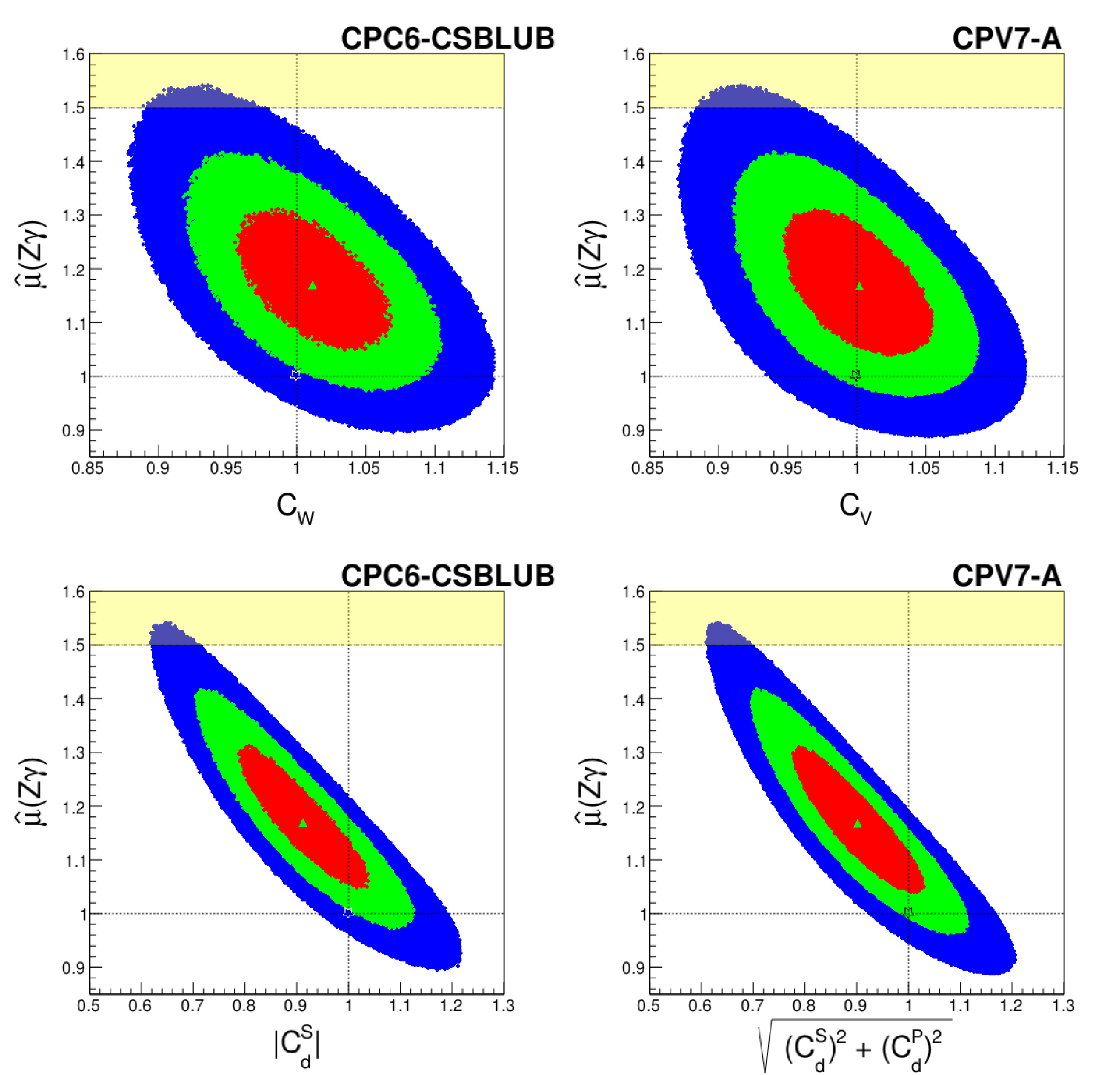}
\end{center}
\vspace{-0.5cm}
\caption{\it
Predictions for $\widehat\mu({Z\gamma})$ in {\bf CPC2}-I, {\bf CPC4}-A,
{\bf CPC6}-CSBLUB, and {\bf CPV7}-A.
The upper frames are versus $C_V$, $C_V$, $C_W$, and $C_V$ and
the lower ones versus $C_f^S$, $|C_d^S|$, $|C_d^S|$ and
$\sqrt{\left(C_d^S\right)^2+\left(C_d^P\right)^2}$ from 
{\bf CPC2}-I to {\bf CPV7}-A.
In each frame, the horizontal line at $\widehat\mu({Z\gamma})=1.5$
denotes the lower boundary of the shaded $1\sigma$ region of the measured 
$H\to Z\gamma$ signal strength of $2.2\pm 0.7$~\cite{ATLAS:2023yqk}.
The contour regions shown are for
$\Delta\chi^2\leq 2.3$ (red), 
$\Delta\chi^2\leq 5.99$ (green),
$\Delta\chi^2\leq 11.83$ (blue) 
above the minimum, which correspond to
confidence levels of 68.27\%, 95\%, and 99.73\%, respectively.
In each frame, the vertical and horizontal lines locate the SM point denoted
by a star and the best-fit point is denoted by a triangle.
}
\label{fig:HZa}
\end{figure}

\medskip

We first examine how large $\widehat\mu (Z\gamma)$ can be in {\bf CPC2}-I in which
$C_V$ and $C_f^S$ are varied in the absence of non-SM particles contributing
to $\Delta S^{Z\gamma}$.  In this scenario, we have
\begin{eqnarray}
\widehat\mu (Z\gamma)^{\rm{\bf CPC2}-I} 
= \frac{\Gamma_{\rm tot}(H_{\rm SM})}{\Gamma_{\rm tot}(H)}\,
\frac{\left|S^{Z\gamma}\right|^2}{\left|S^{Z\gamma}_{\rm SM}\right|^2}
\simeq 4\,\frac{\left|1.06\,C_V-0.06\,C^S_f\right|^2}{C_V^2+3\,(C^S_f)^2}\,,
\end{eqnarray}
where we use Eq.~(\ref{eq:CPC2I_GoG})  for
$\Gamma_{\rm tot}(H_{\rm SM})/\Gamma_{\rm tot}(H)$.
Using the best-fitted values of $C_V=1.015\pm 0.017$ and 
$C_f^S=0.930\pm 0.031$ in {\bf CPC2}-I, see Table~\ref{tab:CPC2}, we have
\begin{equation}
\widehat\mu(Z\gamma)^{\rm{\bf CPC2}-I}\simeq 1.15^{+0.14}_{-0.13} \ (95\%~{\rm CL})\,,
\end{equation}
leading to the enhanced Higgs decay into $Z\gamma$ by the amount of 30\%
at the upper boundary of the 95\% CL region
which is in the right direction to be consistent with the the measured signal strength
of $2.2\pm 0.7$. We note that $C_f^S$ fitted below the SM value of 1 
increases $\widehat\mu(Z\gamma)$, see the two {\bf CPC2}-I
frames of Fig.~\ref{fig:HZa}.
In {\bf CPC4}-A and {\bf CPC6}-CSBLUB where we have the more fitting parameters
of the gauge-Higgs and Yukawa couplings but still with $\Delta S^{Z\gamma}=0$,
we find that
\begin{equation}
\widehat\mu(Z\gamma)^{\rm{\bf CPC4}-A} \simeq 
\widehat\mu(Z\gamma)^{\rm{\bf CPC6}-CSBLUB} \simeq 
1.17^{+0.25}_{-0.19}  \ (95\%~{\rm CL})\,,
\end{equation}
which leads to the enhanced Higgs decay into $Z\gamma$ by the amount of 40\%
at the upper boundary of the 95\% CL region
with the larger errors compared to {\bf CPC2}-I,
see the four {\bf CPC4}-A and {\bf CPC6}-CSBLUB frames of Fig.~\ref{fig:HZa}.
We observe that CP violation does not alter the situation,
see the two {\bf CPV7}-A frames of Fig.~\ref{fig:HZa}.
These observations indicate that
one might need nonvanishing $\Delta S^{Z\gamma}\sim -5$ 
to accommodate the measured $H\to Z\gamma$ signal strength
of $2.2\pm 0.7$ comfortably.
For global fits including the $H\to Z\gamma$ data, see
Appendix~\ref{sec:appendix_E}.

%
%

\section{Conclusions}
\label{sec:conclusions}
We perform global fits of the Higgs boson couplings to
the full Higgs datasets collected at the LHC
with the integrated luminosities per experiment of approximately
5/fb at 7 TeV, 20/fb at 8 TeV, and up to 139/fb at 13 TeV.
To enhance the sensitivity of our global analysis,
we combine the LHC Run 1 dataset with the two Run 2 datasets
separately given by the ATLAS and CMS collaborations
ignoring correlations among them.
We have carefully chosen the 76 
production-times-decay signal strengths and, based on them,
we consistently reproduce the global and 
individual (production and decay) signal strengths in the literature.
We further demonstrate that our combined analysis
based on the 76 experimental signal strengths and
the theoretical ones elaborated in this work reliably reproduce
the fitting results presented in Ref.~\cite{ATLAS:2016neq} (Run 1) and 
Refs.~\cite{ATLAS:2022vkf,CMS:2022dwd} (Run 2) 
within $0.5$ standard deviations.
Note that we have included the production signal strength 
for the tH process to accommodate the new feature of the LHC Run 2 data 
and considered the ggF production process beyond leading order in QCD
to match the level of precision of the LHC Run 2 data.

\medskip

\begin{figure}[!t]
\begin{center}
\includegraphics[width=19.1cm]{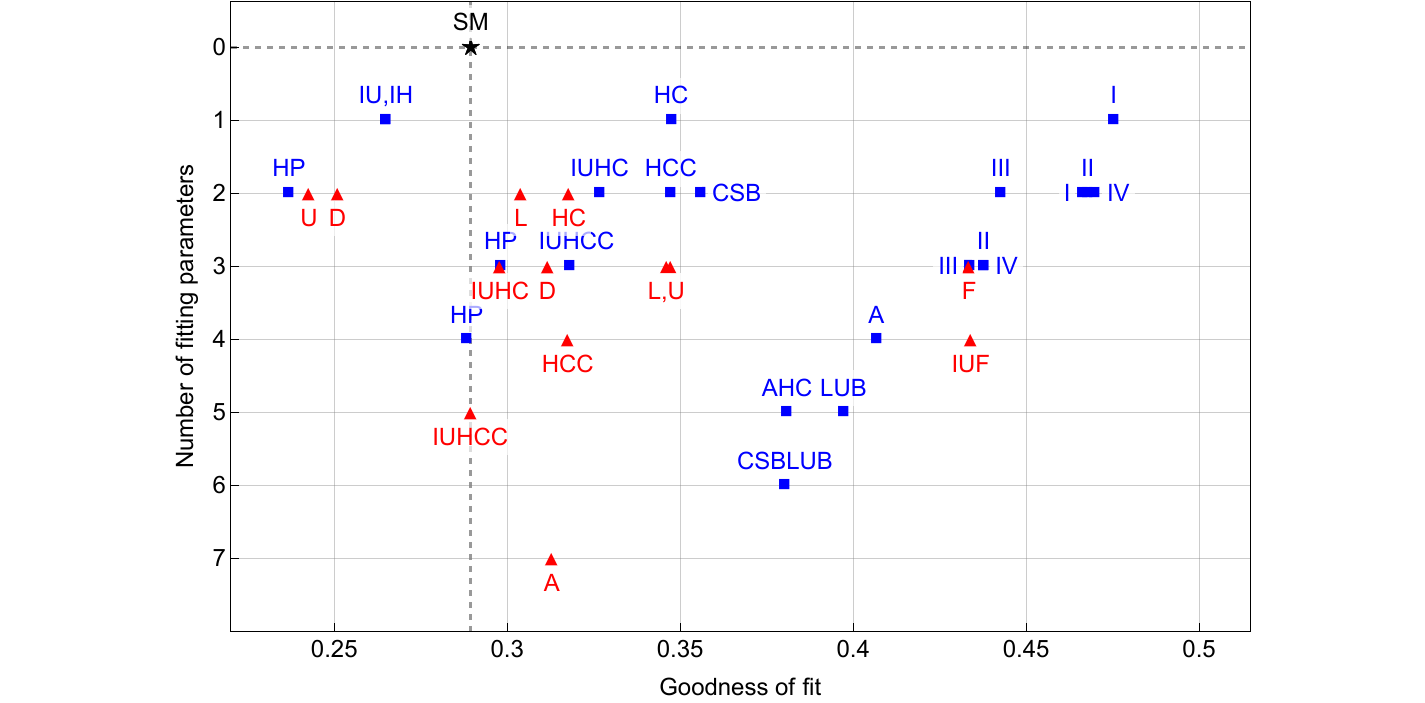}
\end{center}
\vspace{-0.5cm}
\caption{\it
Goodness of fit of the {\bf CPCn} (blue boxes) and {\bf CPVn} (red triangles)
subfits considered in this work.  The SM point is denoted by a star.
}
\label{fig:GOF}
\end{figure}

We have implemented the 22 CPC subfits from {\bf CPC1} to  {\bf CPC6} in
Table~\ref{tab:CPCN} and the 13 CPV subfits from {\bf CPV2} to  {\bf CPV7} in 
Table~\ref{tab:CPVN} taking account of various scenarios 
found in  several well-motivated BSM models.
Our extensive and comprehensive analysis
reveals that the LHC Higgs precision data are no longer  
best described by the SM Higgs boson.
\footnote{We remind that the SM value of goodness of fit is only $0.29$ with
$\chi^2_{\rm SM}/{\rm dof}=82/76$.}
For example, in {\bf CPC2}-I for which we obtain the higher 
gof value of $0.47$ than in the SM and the low $p$-value of $0.02$ 
for compatibility with the SM, we find the following
best-fitted values of
$$
C_V^{\rm{\bf CPC2}-I} = 1.015 \pm 0.017\,; \ \ \
\left(C_f^S\right)^{\rm{\bf CPC2}-I} = 0.930 \pm 0.031\,,
$$
with $C_V$ being consistent with the SM with the $1\sigma$ error of 2\% and
$C_f^S$ {\it below} the SM by more than 2 standard deviations
with the $1\sigma$ error of 3\%. We show that this could be understood
by looking into the individual decay signal strengths presented in
Table~\ref{tab:pdss76}: $\gamma\gamma$ gives the relation
$C_f^S \sim 3\,C_V-2.1$ around the SM point under which
$WW^*$ and $ZZ^*$ drive $C_V$ near to the SM value of 1 while
the Yukawa couplings are driven smaller to match the signal strengths 
of about $0.9$ for $pp\to H\to bb$ and $pp\to H\to \tau\tau$.
In {\bf CPC3}, {\bf CPC4}, {\bf CPC5}, and {\bf CPC6}  where 
we have the more fitting parameters of the gauge-Higgs and Yukawa couplings,
we find that these features remain the same 
but with a bit larger  $1\sigma$ errors. Explicitly, we observe
the following behavior of the gauge-Higgs and Yukawa couplings
to the SM particles:
\begin{itemize}
\item{$C_V$, $C_W$, $C_Z$}: consistent with the SM with the $1\sigma$ error of 2\%-3\%
\item{$C_u^S$, $C_{ud}^S$, $C_{u\ell}^S$}: 
about $2\sigma$ {\it below} the SM with the $1\sigma$ error of 3\%-4\%
\item{$C_{d\ell}^S$, $C_\ell^S$, $C_{\tau}^S$}: 
about $2\sigma$ {\it below} the SM with the $1\sigma$ error of 4\%-5\%
\item{$C_d^S$}: 
about $1\sigma$ {\it below} the SM with the $1\sigma$ error of 7\%-8\%
\item{$\left|C_\mu^S\right|$}: 
consistent with the SM with the $1\sigma$ error of 12-15\%
\end{itemize}
Incidentally, in  many of the two-parameter planes,
the SM points locate outside the 68\% CL region easily and 
even the 95\% CL region sometimes.
In Fig.~\ref{fig:GOF}, we compare the gof values of
all the {\bf CPCn} and {\bf CPVn} subfits considered in 
this work. We indeed observe that the most of them have the better 
goodness of fit than the SM.
Incidentally, we note that
CP violation is largely unconstrained by the LHC Higgs data
with the CL regions appearing as a circle or an ellipse or some 
overlapping of them in the CP-violating two-parameter planes.
We explain the details of how the ellipses and circles emerge in several
subfits of {\bf CPVn}.
Especially, in {\bf CPV4}-HCC and {\bf CPV5}-IUHCC,
we note that the sickle-shaped CL regions in 
the $(\Delta S^g, \Delta P^g)$ plane are obtained since we consider
the ggF production beyond LO in QCD,

\medskip

Interestingly, we find 
that the BSM models predicting the same normalized Yukawa couplings to
the up- and down-type quarks and charged leptons are preferred. 
For example,
among the four types of 2HDMs classified according to 
the Glashow-Weinberg condition to avoid FCNCs,
this could be achieved only in the type-I 2HDM.
Last but not least, we note that
the reduced Yukawa couplings help to explain 
the combined $H\to Z\gamma$ signal strength of $2.2\pm 0.7$ recently 
reported by the ATLAS and CMS collaborations~\cite{ATLAS:2023yqk}.
But one might need nonvanishing $\Delta S^{Z\gamma}\sim -5$ 
to comfortably accommodate the large central value of $2.2$.

%

%

%
%
\section*{Acknowledgment}
We thank Kingman Cheung for the informative comments 
regarding $H\to Z\gamma$ and statistics.
Our thanks go to 
Seong Youl Choi,
Pyungwon Ko,
Chan Beom Park,
Seodong Shin,
Minho Son,
Jeonghyeon Song, and
Hwidong Yoo for ample
discussions at 2024 LSSU Theory Meeting for Future Colliders.
This work was supported by the National Research Foundation (NRF) of Korea
Grant No. NRF-2021R1A2B5B02087078 (Y.H., D.-W.J., J.S.L.).
The work of D.-W.J. was also supported in part by
the NRF of Korea Grant No. NRF-2019R1A2C1089334,
No. NRF-2021R1A2C2011003, and No. RS-2023-00246268
and in part by IBS under the project code IBS-R018-D3.
The work of J.S.L. was also supported in part by
the NRF of Korea Grant No. NRF-2022R1A5A1030700.

\section*{Appendices}

\def\theequation{\Alph{section}.\arabic{equation}}
\begin{appendix}

\setcounter{equation}{0}
\section{$\widehat\mu({\rm ggF})$ beyond LO in QCD}
\label{sec:appendix_A}
The ggF productions signal strength is given by
\begin{equation}
\widehat\mu({\rm ggF)}=\frac{\sigma_{\rm ggF}}{\sigma^{\rm SM}_{\rm ggF}}\,.
\end{equation}
At LO, $\widehat\mu({\rm ggF)}$ is given by the following ratio in terms of the
absolute squares of the relevant scalar and pseudoscalar form factors:
\begin{equation}
\widehat\mu({\rm ggF})^{\rm LO} =
\frac{\left|S^g(M_H)\right|^2+\left|P^g(M_H)\right|^2}
{\left|S^g_{\rm SM}(M_H)\right|^2}\,,
\end{equation}
numerator of 
which depends on the model-independent Yukawa couplings of
$g^{S,P}_{H\bar t t}$,
$g^{S,P}_{H\bar b b}$, and
$g^{S,P}_{H\bar c c}$, and the non-SM parameters 
of $\Delta S^g$ and $\Delta P^g$
denoting the contributions from the triangle loops in which
non-SM colored particles are running.
The LO ggF production signal strength 
should be reliable only if some higher order corrections to 
the non-SM cross section $\sigma_{\rm ggF}$ 
and those to the SM cross section $\sigma^{\rm SM}_{\rm ggF}$
are largely canceled out in the ratio. 
It turns out this is not the case with the QCD corrections 
\cite{Harlander:2013qxa} 
and, in this work,
we consider the production signal strength beyond LO in QCD.
%
%
Using the numerical expressions for the form factors
given in Eq.~(\ref{eq:spgg_numeric}) for $M_H=125$ GeV, we have
the following LO ggF production signal strength:
\begin{eqnarray}
\label{eq:mu_ggF_lo}
\widehat\mu({\rm ggF})^{\rm LO} &\simeq &
 1.158\,(g^S_{H\bar t t})^2
+0.014\,(g^S_{H\bar b b})^2
-0.145\,(g^S_{H\bar t t}g^S_{H\bar b b})
-0.032\,(g^S_{H\bar t t}g^S_{H\bar c c})
+3.364\,(g^S_{H\bar t t}\Delta S^g)
+2.444\,(\Delta S^g)^2
\nonumber \\[2mm]
&+&
 2.680\,(g^P_{H\bar t t})^2
+0.016\,(g^P_{H\bar b b})^2
-0.254\,(g^P_{H\bar t t}g^P_{H\bar b b})
-0.051\,(g^P_{H\bar t t}g^P_{H\bar c c})
+5.118\,(g^P_{H\bar t t}\Delta P^g)
+2.444\,(\Delta P^g)^2\,, \nonumber \\
\end{eqnarray}
assuming that $\Delta S^g$ and $\Delta P^g$ are real and the 
interferences terms proportional to the products of
$g^{S,P}_{H\bar b b}\times g^{S,P}_{H\bar c c}$,
$g^{S}_{H\bar b b\,,H\bar c c}\times \Delta S^g$, 
$g^{P}_{H\bar b b\,,H\bar c c}\times \Delta P^g$ and
the diagonal terms $(g^{S,P}_{H\bar c c})^2$
have been neglected.

\medskip

To go beyond LO in QCD, to begin with, we consider the contributions
from top-, bottom-, and charm-quark loops taking
$\Delta S^g=\Delta P^g=0$.
In this case, the ggF production cross section of 
a CP-mixed Higgs boson $H$ might be organized as follow:
\begin{eqnarray}
\sigma_{\rm ggF} &=&
\sigma^S_{\rm ggF}(g^S_{H\bar t t},g^S_{H\bar b b},g^S_{H\bar c c}) +
\sigma^P_{\rm ggF}(g^P_{H\bar t t},g^P_{H\bar b b},g^P_{H\bar c c}) \,,
\end{eqnarray}
where
\begin{eqnarray}
\sigma^S_{\rm ggF}(g^S_{H\bar t t},g^S_{H\bar b b},g^S_{H\bar c c}) &\equiv &
(g^S_{H\bar t t})^2\,\sigma^S_{tt}+
(g^S_{H\bar b b})^2\,\sigma^S_{bb}+
(g^S_{H\bar t t}g^S_{H\bar b b})\,\sigma^S_{tb}+
(g^S_{H\bar t t}g^S_{H\bar c c})\,\sigma^S_{tc}+
{\cal O}(\sigma^S_{cc},\sigma^S_{bc})\,,
\nonumber \\[2mm]
\sigma^P_{\rm ggF}(g^P_{H\bar t t},g^P_{H\bar b b},g^P_{H\bar c c}) &\equiv &
(g^P_{H\bar t t})^2\,\sigma^P_{tt}+
(g^P_{H\bar b b})^2\,\sigma^P_{bb}+
(g^P_{H\bar t t}g^P_{H\bar b b})\,\sigma^P_{tb}+
(g^P_{H\bar t t}g^P_{H\bar c c})\,\sigma^P_{tc}+
{\cal O}(\sigma^P_{cc},\sigma^P_{bc})\,.
\end{eqnarray}
%
\begin{table}[t]
\centering
\caption{\it
\label{tab:ggF_cx}
$\sigma^S_{\rm ggF}$ at N$^3$LO and
$\sigma^P_{\rm ggF}$ at NNLO
for several combinations of the relevant Yukawa couplings
obtained by
using {\tt SusHi-1.7.0}~\cite{Harlander:2012pb,Harlander:2016hcx}
with {\bf PDF4LHC15}~\cite{Butterworth:2015oua}.
For each combination of the Yukawa couplings,
$\sigma_{bbH}$ at NNLO is also shown.
We consider three values of $\sqrt{s}=7$ TeV, 8 TeV, and 13 TeV 
and $M_H=125$ GeV has been taken.
The renormalization and factorization scales are chosen
$\mu_R=\mu_F=M_H/2$ for $\sigma_{\rm ggF}$ and
$\mu_R=4 \mu_F=M_H$ for $\sigma_{bbH}$.
When $g^{S,P}_{H\bar t t}=1$ and
$g^{S,P}_{H\bar b b}=g^{S,P}_{H\bar c c}=0$,
the LO ggF cross sections are also shown in parentheses.
%
\\}
\begin{adjustbox}{width= \textwidth}
\begin{tabular}{c|c|c|c|c|c|c|c|c||c|c|c|c|c|c|c|c|c}
\hline
\multicolumn{3}{c|}{Couplings} &
\multicolumn{3}{c|}{$\sigma^S_{\rm ggF}$ (pb)} & 
\multicolumn{3}{c||}{$\sigma_{bbH}$ (pb)} &
\multicolumn{3}{c|}{Couplings} &
\multicolumn{3}{c|}{$\sigma^P_{\rm ggF}$ (pb)} & 
\multicolumn{3}{c}{$\sigma_{bbH}$ (pb)} \\ \hline
$g_{H\bar{t}t}^S$ & $g_{H\bar{b}b}^S$ & $g_{H\bar{c}c}^S$ &
7 TeV & 8 TeV & 13 TeV & 7 TeV & 8 TeV & 13 TeV &
$g_{H\bar{t}t}^P$ & $g_{H\bar{b}b}^P$ & $g_{H\bar{c}c}^P$ &
7 TeV & 8 TeV & 13 TeV & 7 TeV & 8 TeV & 13 TeV  \\ \hline
$1$ & $1$ & $1$ &
$16.66$ & $21.21$ & $48.29$ & $0.18$ & $0.23$ & $0.55$ &
$1$ & $1$ & $1$ &
$35.85$ & $45.57$ & $103.23$ & $0.18$ & $0.23$ & $0.55$ \\ \hline
$1$ & $0$ & $0$ &
$17.70$ & $22.50$ & $50.95$ & $0$ & $0$ & 0 &
$1$ & $0$ & $0$ &
$37.87$ & $48.09$ & $108.52$ & $0$ & $0$ & 0 \\ 
 &  &  &
$(5.91)$ & $(7.46)$  & ($16.44$) & & &  &
 &  &  &
$(13.67)$ & $(17.26)$ & ($38.06$) & & &   \\
$0$ & $1$ & $0$ &
$0.12$ & $0.15$ & $0.32$ & $0.18$ & $0.23$ & $0.55$ &
$0$ & $1$ & $0$ &
$0.13$ & $0.16$ & $0.34$ & $0.18$ & $0.23$ & $0.55$ \\ 
$0$ & $0$ & $1$ & 
$0.004$ & $0.005$ & $0.01$ & $0$ & $0$ & 0 &
$0$ & $0$ & $1$ &
$0.004$ & $0.005$ & $0.01$ & $0$ & $0$ & 0 \\ \hline
$1$ & $1$ & $0$ &
$16.84$ & $21.43$ & $48.75$ & $0.18$ & $0.23$ & $0.55$ &
$1$ & $1$ & $0$ &
$36.17$ & $45.98$ & $104.08$ & $0.18$ & $0.23$ & $0.55$ \\ 
$1$ & $0$ & $1$ &
$17.48$ & $22.22$ & $50.39$ & $0$ & $0$ & 0 &
$1$ & $0$ & $1$ &
$37.49$ & $47.63$ & $107.55$ & $0$ & $0$ & 0 \\ 
$0$ & $1$ & $1$ &
$0.16$ & $0.20$ & $0.44$ & $0.18$ & $0.23$ & $0.55$ &
$0$ & $1$ & $1$ &
$0.17$ & $0.22$ & $0.47$ & $0.18$ & $0.23$ & $0.55$ \\ 
\hline
\end{tabular}
\end{adjustbox}
\end{table}
%


In Table~\ref{tab:ggF_cx}, we present 
various ggF and bbH cross sections 
obtained by using {\tt SusHi-1.7.0} 
\cite{Harlander:2012pb,Harlander:2016hcx}
for several combinations of the $g^{S,P}_{H\bar q q}$  couplings
at $\sqrt{s}=7$ TeV, 8 TeV, and 13 TeV.
Neglecting
$\sigma_{cc}^{S,P}\lsim 0.01$ pb  and
$\sigma_{bc}^{S,P}\lsim 0.1$ pb, 
at each value of $\sqrt{s}$,
one may derive the interference cross sections as follows:
\begin{eqnarray}
\sigma_{tb}^{S,P} &=& \sigma_{\rm ggF}^{S,P}(1,1,1)
-\sigma_{\rm ggF}^{S,P}(1,0,1)-\sigma_{\rm ggF}^{S,P}(0,1,0)\,,
\nonumber \\[2mm]
\sigma_{tc}^{S,P} &=& \sigma_{\rm ggF}^{S,P}(1,1,1)
-\sigma_{\rm ggF}^{S,P}(1,1,0)\,,
\end{eqnarray}
together with the diagonal ones
$\sigma_{tt}^{S,P}=\sigma_{\rm ggF}^{S,P}(1,0,0)$ and
$\sigma_{bb}^{S,P}=\sigma_{\rm ggF}^{S,P}(0,1,0)$.
Explicitly, at $\sqrt{s}=13$ TeV, we have
$\sigma_{tt}^{S,P}$,
$\sigma_{bb}^{S,P}$,
$\sigma_{tb}^{S,P}$, and
$\sigma_{tc}^{S,P}$ in pb:
\begin{eqnarray}
\label{eq:sig_qq}
&&
\sigma^S_{tt} = \,\,\,50.95\,, \ \
\sigma^S_{bb} = 0.32\,, \ \
\sigma^S_{tb} =-2.42\,, \ \
\sigma^S_{tc} =-0.46\,; \nonumber \\[2mm]
&&
\sigma^P_{tt} = 108.52\,, \ \
\sigma^P_{bb} = 0.34\,, \ \
\sigma^P_{tb} =-4.67\,, \ \
\sigma^P_{tc} =-0.85\,.
\end{eqnarray}
Then, one might obtain
\begin{eqnarray}
\label{eq:mu_ggF_tbc_13}
\left.\widehat\mu({\rm ggF})\right|_{\rm 13\,TeV}^{tbc} =
\left.\frac{\sigma_{\rm ggF}}{\sigma^{\rm SM}_{\rm ggF}}\right|_{\rm 13\,TeV}^{tbc}
&=& \left.
\frac{\sum_{X=S,P} \left[
(g^X_{H\bar t t})^2\,\sigma^X_{tt}+
(g^X_{H\bar b b})^2\,\sigma^X_{bb}+
(g^X_{H\bar t t}g^X_{H\bar b b})\,\sigma^X_{tb}+
(g^X_{H\bar t t}g^X_{H\bar c c})\,\sigma^X_{tc}
\right]
}{\sigma^S_{\rm ggF}(1,1,1)}\right|_{\rm 13\,TeV}\nonumber \\[2mm]
&=&
 1.055\,(g^S_{H\bar t t})^2
+0.007\,(g^S_{H\bar b b})^2
-0.050\,(g^S_{H\bar t t}g^S_{H\bar b b})
-0.010\,(g^S_{H\bar t t}g^S_{H\bar c c})
\nonumber \\[2mm]
&+&
 2.248\,(g^P_{H\bar t t})^2
+0.007\,(g^P_{H\bar b b})^2
-0.097\,(g^P_{H\bar t t}g^P_{H\bar b b})
-0.018\,(g^P_{H\bar t t}g^P_{H\bar c c})\,.
\end{eqnarray}
Similarly, using the cross sections at $\sqrt{s}=7$ TeV and 8 TeV 
shown in Table~\ref{tab:ggF_cx}, we also obtain
\begin{eqnarray}
\label{eq:mu_ggF_tbc_78}
\left.\widehat\mu({\rm ggF})\right|_{\rm 7\oplus 8\,TeV}^{tbc} =
\left.\frac{\sigma_{\rm ggF}}{\sigma^{\rm SM}_{\rm ggF}}\right|_{\rm 7\oplus 8\,TeV}^{tbc}
&=& \left.
\frac{\sum_{X=S,P} \left[
(g^X_{H\bar t t})^2\,\sigma^X_{tt}+
(g^X_{H\bar b b})^2\,\sigma^X_{bb}+
(g^X_{H\bar t t}g^X_{H\bar b b})\,\sigma^X_{tb}+
(g^X_{H\bar t t}g^X_{H\bar c c})\,\sigma^X_{tc}
\right]
}{\sigma^S_{\rm ggF}(1,1,1)}\right|_{\rm 7\oplus 8\,TeV}\nonumber \\[2mm]
&=&
 1.061\,(g^S_{H\bar t t})^2
+0.007\,(g^S_{H\bar b b})^2
-0.055\,(g^S_{H\bar t t}g^S_{H\bar b b})
-0.011\,(g^S_{H\bar t t}g^S_{H\bar c c})
\nonumber \\[2mm]
&+&
 2.268\,(g^P_{H\bar t t})^2
+0.007\,(g^P_{H\bar b b})^2
-0.105\,(g^P_{H\bar t t}g^P_{H\bar b b})
-0.019\,(g^P_{H\bar t t}g^P_{H\bar c c})\,,
\end{eqnarray}
where we use the luminosity-weighted cross sections for the Run 1 data
at $\sqrt{s}=7\oplus 8$ TeV:
\begin{eqnarray}
\left.
\sigma^{S,P}_{qq^(\prime)}\right|_{7\oplus 8\,{\rm TeV}}
&=&\frac{5.1~{\rm fb}^{-1} \times
\left.\sigma^{S,P}_{qq^(\prime)}\right|_{7\,{\rm TeV}} +
19.6~{\rm fb}^{-1} \times
\left.\sigma^{S,P}_{qq^(\prime)}\right|_{8\,{\rm TeV}}}
{5.1~{\rm fb}^{-1}+19.6~{\rm fb}^{-1}}\,,
\end{eqnarray}
for $qq^{(\prime)}=tt,bb,tb,tc$ and similarly for 
$\left.\sigma^{S}_{\rm ggF}(1,1,1)\right|_{7\oplus 8\,{\rm TeV}}$.
We note that the decomposition coefficients are almost independent
of $\sqrt{s}$.
\footnote{We have also checked that the variation of the coefficients 
due to the change of $M_H$ in the range between 125 GeV and $125.5$ GeV
are also negligible.}
On the other hand,
comparing with the LO result Eq.~(\ref{eq:mu_ggF_lo}),
we observe that the coefficients proportional to
$(g^S_{H\bar t t})^2$ and $(g^P_{H\bar t t})^2$ decrease
by the factors of $0.92$ and $0.85$, respectively, while
the other coefficients by factors of about 2 to 3.

\medskip
Next, neglecting the interference terms proportional to the products of
$g^{S}_{H\bar b b\,,H\bar c c}\times \Delta S^g$
and assuming $H$ is CP even with $P^g=0$,
we address the case with $\Delta S^g\neq 0$.
In this case, including the QCD corrections to the top-quark loops,
the form factor  $S^g$ might be written as
\begin{equation}
S^g = (1+\epsilon^S_{tt})\, S^g_{tt} + \Delta S^g\,,
\end{equation}
where $\epsilon^S_{tt}$ denotes the QCD corrections with $\epsilon^S_{tt}=0$ at LO and
$S^g_{tt} =2/3\,g^S_{H \bar t t}$ in the limit of $M_t\to \infty$.
The cross section is proportional to $|S^g|^2$ and it might be given by
\begin{equation}
\sigma^S_{\rm ggF} ={\cal A}^S\, |S^g|^2
= {\cal A}^S\, \left| (1+\epsilon^S_{tt}) S^g_{tt} + \Delta S^g \right|^2
= {\cal A}^S\,\left\{
|1+\epsilon^S_{tt}|^2 \left|S^g_{tt} \right|^2
+2\real\left[(1+\epsilon^S_{tt})S^g_{tt} (\Delta S^g)^* \right]
+ \left|\Delta S^g \right|^2
\right\}\,.
\end{equation}
When $\Delta S^g$ is real, one may reorganize it as follow
\begin{equation}
\sigma^S_{\rm ggF} = (g^S_{H\bar t t})^2\,\sigma^S_{tt}
+ (g^S_{H\bar t t})(\Delta S^g)\,\sigma^{S}_{t\Delta}
+ (\Delta S^g)^2\,\sigma^S_{\Delta\Delta}
\end{equation}
by identifying
\begin{eqnarray}
{\cal A}^S\,|1+\epsilon^S_{tt}|^2 \left|S^g_{tt}\right|^2 &=&
(g^S_{H\bar t t})^2\,\sigma^S_{tt} \equiv
(g^S_{H\bar t t})^2\,
\left(\sigma_{tt}^S\right)^{\rm LO}\,|1+\epsilon^S_{tt}|^2\,, \nonumber \\[2mm]
2{\cal A}^S\,\real\left[(1+\epsilon^S_{tt})S^g_{tt} (\Delta S^g)^*\right] &=&
(g^S_{H\bar t t})(\Delta S^g)\,\sigma^{S}_{t\Delta}\,, \nonumber \\[2mm]
{\cal A}^S\,\left|\Delta S^g \right|^2 &=&
(\Delta S^g)^2\,\sigma^S_{\Delta\Delta}\,.
\end{eqnarray}
Regarding $\sigma^S_{tt}$ and $\left(\sigma_{tt}^S\right)^{\rm LO}$ as inputs
and taking $S^g_{tt} =2/3\,g^S_{H \bar t t}$, we have
\begin{eqnarray}
|1+\epsilon^S_{tt}| = \sqrt{
{\sigma^S_{tt}}/{\left(\sigma_{tt}^S\right)^{\rm LO}} }\,, \ \ \
{\cal A}^S =\frac{9}{4}\,\left(\sigma_{tt}^S\right)^{\rm LO}\,,
\end{eqnarray}
which lead to
\begin{eqnarray}
\label{eq:sig_S_tD}
\sigma^{S}_{t\Delta} &=& \frac{4}{3}\, {\cal A}^S\,
\real\left(1+\epsilon^S_{tt}\right) \leq
3\,\sqrt{\sigma^S_{tt}\,\left(\sigma_{tt}^S\right)^{\rm LO}} \,, \nonumber \\[2mm]
\sigma^{S}_{\Delta\Delta} &=& {\cal A}^S
= \frac{9}{4}\,\left(\sigma_{tt}^S\right)^{\rm LO}\,,
\end{eqnarray}
by noting that $\real\left(1+\epsilon^S_{tt}\right)\leq |1+\epsilon^S_{tt}|$.
Similarly, starting from
\begin{equation}
P^g = (1+\epsilon^P_{tt})\, P^g_{tt} + \Delta P^g \ \ {\rm and} \ \
\sigma^P_{\rm ggF} ={\cal A}^P\, |P^g|^2\,,
\end{equation}
one might have
\begin{equation}
\sigma^P_{\rm ggF} = (g^P_{H\bar t t})^2\,\sigma^P_{tt}
+ (g^P_{H\bar t t})(\Delta P^g)\,\sigma^{P}_{t\Delta}
+ (\Delta P^g)^2\,\sigma^P_{\Delta\Delta}\,,
\end{equation}
where
\begin{eqnarray}
\label{eq:sig_P_tD}
\sigma^{P}_{t\Delta} &=& 2\, {\cal A}^P\,
\real\left(1+\epsilon^P_{tt}\right) \leq
2\,\sqrt{\sigma^P_{tt}\,\left(\sigma_{tt}^P\right)^{\rm LO}} \,, \nonumber \\[2mm]
\sigma^{P}_{\Delta\Delta} &=& {\cal A}^P
= \left(\sigma_{tt}^P\right)^{\rm LO}\,,
\end{eqnarray}
assuming that $\Delta P^g$ is real, taking
$P^g_{tt}=g^P_{H\bar t t}$ in the infinite $M_t$ limit, and using
$\real\left(1+\epsilon^P_{tt}\right)\leq |1+\epsilon^P_{tt}|
=\sqrt{ {\sigma^P_{tt}}/{\left(\sigma_{tt}^P\right)^{\rm LO}} }$.

\begin{table}[t!]
\caption{\it
\label{tab:sig_tD}
$\sigma^S_{t\Delta}$, $\sigma^S_{\Delta\Delta}$,
$\sigma^P_{t\Delta}$, and $\sigma^P_{\Delta\Delta}$ in pb
at $\sqrt{s}=7$ TeV, 8 TeV, and 13 TeV
obtained by using the relations given by
Eqs.~(\ref{eq:sig_S_tD}) and (\ref{eq:sig_P_tD})
and the cross sections
$\sigma^{S,P}_{tt}$ and $\left(\sigma^{S,P}_{tt}\right)^{\rm LO}$
given in Table~\ref{tab:ggF_cx}
for $(g^S_{H\bar t t},g^S_{H\bar b b},g^S_{H\bar c c})=(1,0,0)$.
The estimation has been done
in the limit of $M_t\to\infty$
taking $\real(1+\epsilon_{tt}^{S,P}) \approx
\left|1+\epsilon_{tt}^{S,P}\right|$
under the assumption that
$\Delta S^g$ and $\Delta P^g$ are real.
\\}
\begin{center}
\begin{tabular}{c||c|c||c|c}
\hline
$\sqrt{s}$ (TeV) &
$\sigma^S_{t\Delta}$ & $\sigma^S_{\Delta\Delta}$ &
$\sigma^P_{t\Delta}$ & $\sigma^P_{\Delta\Delta}$ \\
\hline
7 &
$30.67$ & $13.29$ & $45.50$ & $13.67$ \\
8 &
$38.86$ & $16.78$ & $57.62$ & $17.26$ \\
13 &
$86.84$ & $37.00$ & $128.54$ & $38.06$ \\
\hline
\end{tabular}
\end{center}
\end{table}


In Table~\ref{tab:sig_tD}, we show the cross sections
$\sigma^{S,P}_{t\Delta}$ and $\sigma^{S,P}_{\Delta\Delta}$ in pb
at $\sqrt{s}=7$ TeV, 8 TeV, and 13 TeV assuming $\Delta S^g$ and 
$\Delta P^g$ are real.
We use the values of the cross sections
$\sigma^{S,P}_{tt}=\sigma^{S,P}_{\rm ggF}(1,0,0)$ and 
$(\sigma^{S,P}_{tt})^{\rm LO}$
in Table~\ref{tab:ggF_cx} together with 
the relations given by
Eqs.~(\ref{eq:sig_S_tD} and (\ref{eq:sig_P_tD}).
For $\sigma^{S,P}_{t\Delta}$, we take the approximation
$\real(1+\epsilon_{tt}^{S,P}) \approx
\left|1+\epsilon_{tt}^{S,P}\right|$
and, for the contributions from the triangle top-quark loops,
we take the $M_t\to\infty$ limit.
Then, with the cross sections
$\sigma^{S,P}_{t\Delta}$ and $\sigma^{S,P}_{\Delta\Delta}$ given,
we derive 
\begin{eqnarray}
\label{eq:mu_ggF_tD_13}
\left.\widehat\mu({\rm ggF})\right|_{\rm 13\,TeV}^{t\Delta} =
\left.\frac{\sigma_{\rm ggF}}{\sigma^{\rm SM}_{\rm ggF}}\right|_{\rm 13\,TeV}^{t\Delta}
&=& \left.
\frac{\sum_{X=S,P} \left[
(g^X_{H\bar t t})^2\,\sigma^X_{tt}+
(g^X_{H\bar t t})(\Delta X^g)\,\sigma^X_{t\Delta}+
(\Delta X^g)^2\,\sigma^X_{\Delta\Delta}
\right]
}{\sigma^S_{\rm ggF}(1,1,1)}\right|_{\rm 13\,TeV}\nonumber \\[2mm]
&=&
 1.055\,(g^S_{H\bar t t})^2
+1.799\,(g^S_{H\bar t t})(\Delta S^g)
+0.766\,(\Delta S^g)^2
\nonumber \\[2mm]
&+&
 2.248\,(g^P_{H\bar t t})^2
+2.662\,(g^P_{H\bar t t})(\Delta P^g)
+0.788\,(\Delta P^g)^2\,,
\end{eqnarray}
and, similarly as before,
\begin{eqnarray}
\label{eq:mu_ggF_tD_78}
\left.\widehat\mu({\rm ggF})\right|_{\rm 7\oplus 8\,TeV}^{t\Delta} &=&
 1.061\,(g^S_{H\bar t t})^2
+1.834\,(g^S_{H\bar t t})(\Delta S^g)
+0.792\,(\Delta S^g)^2
\nonumber \\[2mm]
&+&
 2.268\,(g^P_{H\bar t t})^2
+2.719\,(g^P_{H\bar t t})(\Delta P^g)
+0.815\,(\Delta P^g)^2\,.
\end{eqnarray}
We again note that the decomposition coefficients are almost independent
of $\sqrt{s}$ and,
comparing with the LO result Eq.~(\ref{eq:mu_ggF_lo}),
we observe that the coefficients 
proportional to the products of 
$(g^S_{H\bar t t})(\Delta S^g)$ and
$(g^P_{H\bar t t})(\Delta P^g)$
and the squares of $(\Delta S^g)^2$ and $(\Delta P^g)^2$
decrease by factors of about 2 to 3.

\medskip

Finally, combining
$\left.\widehat\mu({\rm ggF})\right|^{tbc}$ given by 
Eqs.~(\ref{eq:mu_ggF_tbc_13}) and (\ref{eq:mu_ggF_tbc_78}) and
$\left.\widehat\mu({\rm ggF})\right|^{t\Delta}$ given by
Eqs.~(\ref{eq:mu_ggF_tD_13}) and (\ref{eq:mu_ggF_tD_78}),
we have arrived at
\begin{eqnarray}
\label{eq:mu_ggF_All}
\left.\widehat\mu({\rm ggF})\right|_{\rm 13\,TeV} &= &
 1.055\,(g^S_{H\bar t t})^2
+0.007\,(g^S_{H\bar b b})^2
-0.050\,(g^S_{H\bar t t}g^S_{H\bar b b})
-0.010\,(g^S_{H\bar t t}g^S_{H\bar c c})
+1.799\,(g^S_{H\bar t t}\Delta S^g)
+0.766\,(\Delta S^g)^2
\nonumber \\[2mm]
&+&
 2.248\,(g^P_{H\bar t t})^2
+0.007\,(g^P_{H\bar b b})^2
-0.097\,(g^P_{H\bar t t}g^P_{H\bar b b})
-0.018\,(g^P_{H\bar t t}g^P_{H\bar c c})
+2.662\,(g^P_{H\bar t t}\Delta P^g)
+0.788\,(\Delta P^g)^2\,; \nonumber \\[2mm]
\left.\widehat\mu({\rm ggF})\right|_{\rm 7\oplus 8\,TeV} &= &
 1.061\,(g^S_{H\bar t t})^2
+0.007\,(g^S_{H\bar b b})^2
-0.055\,(g^S_{H\bar t t}g^S_{H\bar b b})
-0.011\,(g^S_{H\bar t t}g^S_{H\bar c c})
+1.834\,(g^S_{H\bar t t}\Delta S^g)
+0.792\,(\Delta S^g)^2
\nonumber \\[2mm]
&+&
 2.268\,(g^P_{H\bar t t})^2
+0.007\,(g^P_{H\bar b b})^2
-0.105\,(g^P_{H\bar t t}g^P_{H\bar b b})
-0.019\,(g^P_{H\bar t t}g^P_{H\bar c c})
+2.719\,(g^P_{H\bar t t}\Delta P^g)
+0.815\,(\Delta P^g)^2\,. \nonumber \\
\end{eqnarray}

\setcounter{equation}{0}
\section{Single variable behavior}
\label{sec:appendix_B}
In each box of Table~\ref{tab:SVB1} and Table~\ref{tab:SVB2},
we show the best-fitted value, goodness of fit, and 
$p$-value against the SM for compatibility
with the SM hypothesis
when only a single non-SM or SM parameter is varied
while all the other ones are taking the SM value of either 0 or 1.
Also shown is $\Delta\chi^2$ above the minimum versus the single parameter varied.
Note that four of them are the same as in {\bf CPC1} and, otherwise,
this is to check and see the chi-square behavior rather than
to address some physics cases.
%

\begin{table}[!htb]
\caption{\it
\label{tab:SVB1}
The best-fitted value, goodness of fit (gof), and
$p$-value against the SM for compatibility with the SM hypothesis
when only a single parameter is varied.
Also shown is $\Delta\chi^2$ above the minimum versus the single parameter varied.
For the SM, we obtain $\chi^2_{\rm SM}/{\rm dof}=82.3480/76$
and gof $=0.2895$.
}  \vspace{1mm}
\renewcommand{\arraystretch}{0.8}
\begin{adjustbox}{width= \textwidth}
\begin{tabular}{|c|c|c|c|c|} \hline
Varying& $C_{Vf}=1.005^{+0.017}_{-0.017}$ & 
$C_{V}=1.028^{+0.016}_{-0.016}$ & 
$C_{W}=1.031^{+0.018}_{-0.018}$ &
 $C_{Z}=0.992^{+0.030}_{-0.030}$  \\ 
Parameter & & & &  \\ \hline
gof & 0.2649 & 0.3449 & 0.3867 & 0.2641  \\ \hline
$p$-value & 0.7590 & 0.0810 & 0.0349 & 0.8014  \\ \hline
& \includegraphics[width=4cm]{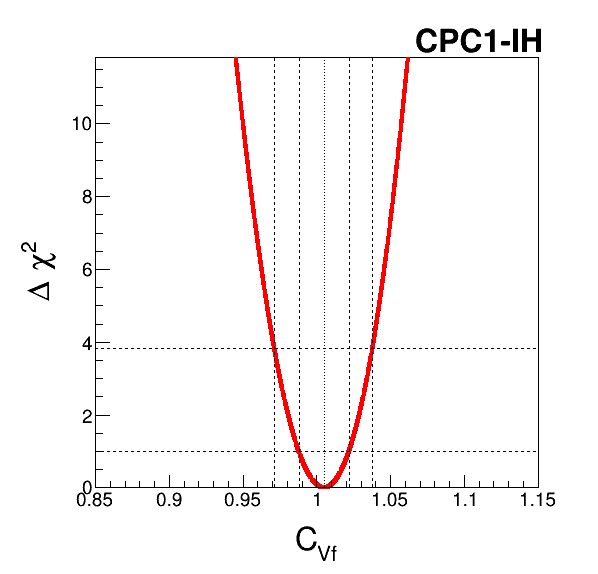}
& \includegraphics[width=4cm]{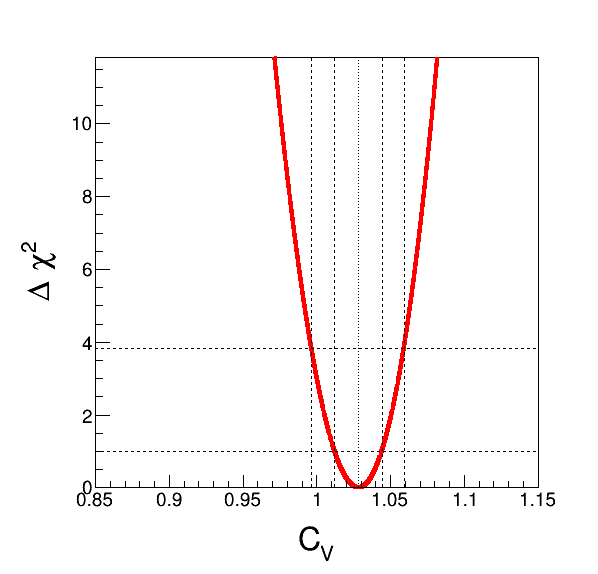}
& \includegraphics[width=4cm]{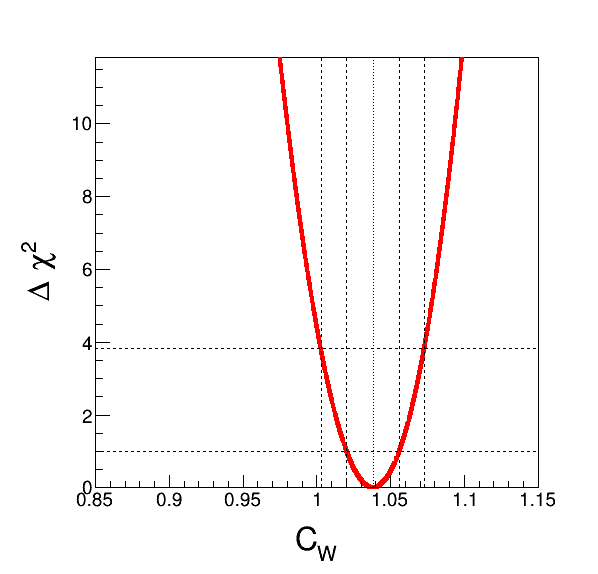}
& \includegraphics[width=4cm]{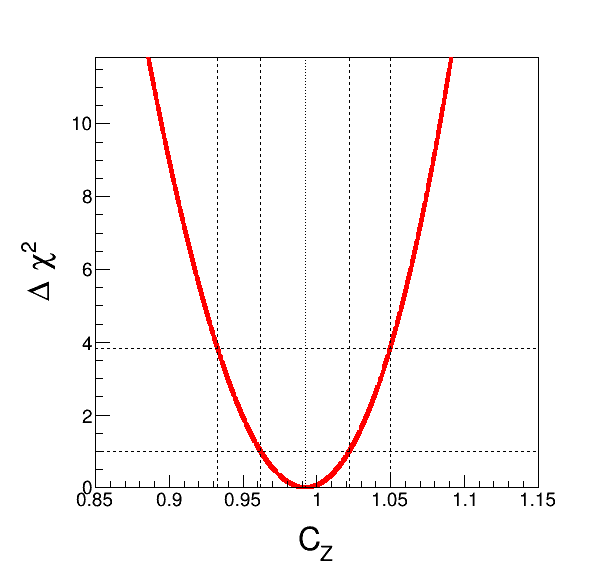}
\\ \hline \hline
Varying & $C_f^S=0.920^{+0.029}_{-0.029}$ & 
$C_u^S=0.987^{+0.026}_{-0.026}$ & 
$C_t^S=0.988^{+0.025}_{-0.025}$ & 
$C_c^S=0.847^{+0.513}_{-2.221}$  \\ 
Parameter & & & &  \\ \hline
gof & 0.4753 & 0.2687 & 0.2680 & 0.2643  \\ \hline
$p$-value & 0.0071 & 0.6209 & 0.6421 & 0.7948  \\ \hline
& \includegraphics[width=4cm]{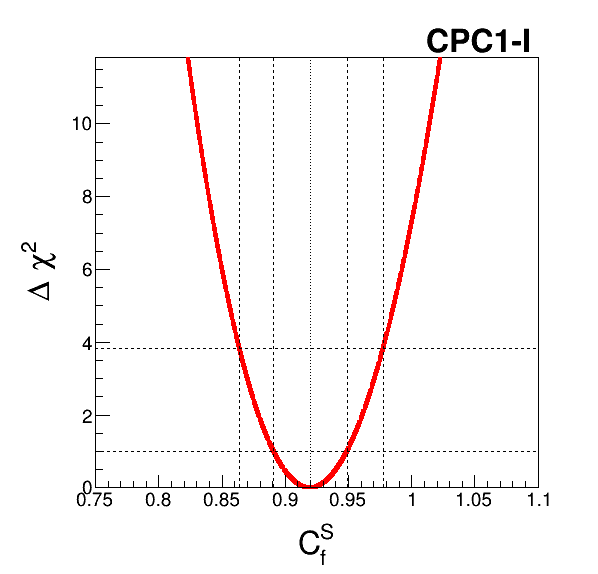}
& \includegraphics[width=4cm]{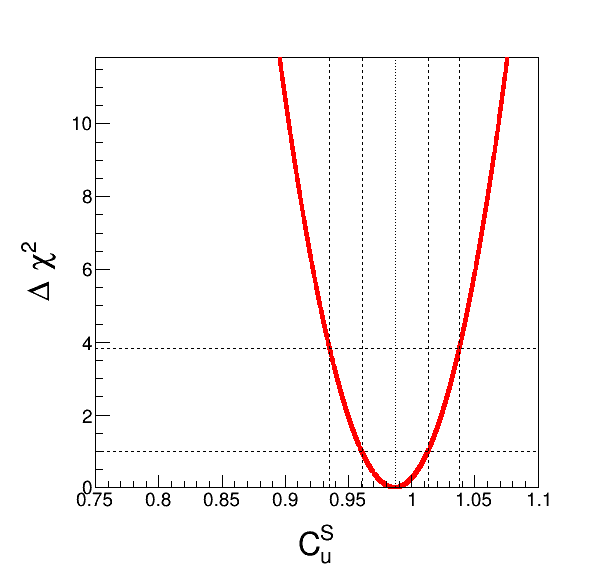}
& \includegraphics[width=4cm]{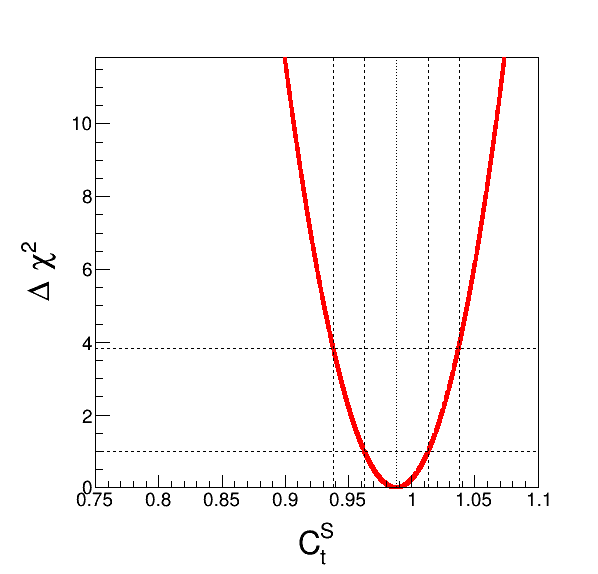}
& \includegraphics[width=4cm]{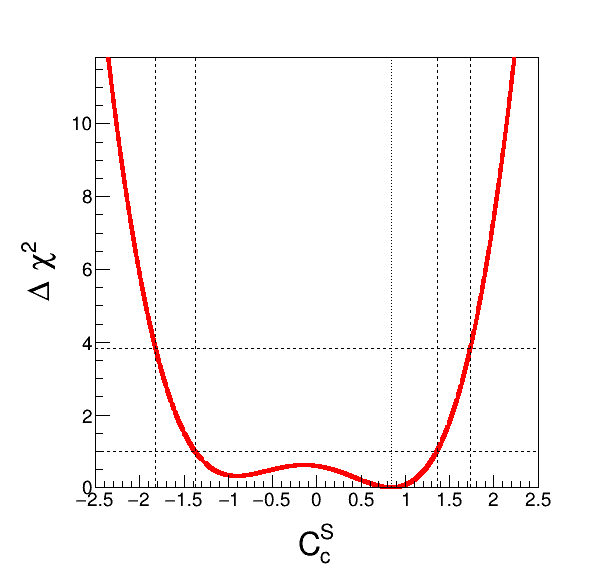}
\\ \hline \hline
Varying & 
$C_d^S=0.978^{+0.029}_{-0.028}$ & 
$C_\ell^S=0.939^{+0.038}_{-0.039}$ & 
$C_\tau^S=0.933^{+0.038}_{-0.039}$ & 
$C_\mu^S=+1.078^{+0.154}_{-0.180}$,  \\ 
Parameter & & & & $~~~~~~\, -1.078_{-0.154}^{+0.180}$ \\ \hline
gof & 0.2774 & 0.3330 & 0.3451 & 0.2678  \\ \hline
$p$-value & 0.4450 & 0.1048 & 0.0806 & 0.6473  \\ \hline
& \includegraphics[width=4cm]{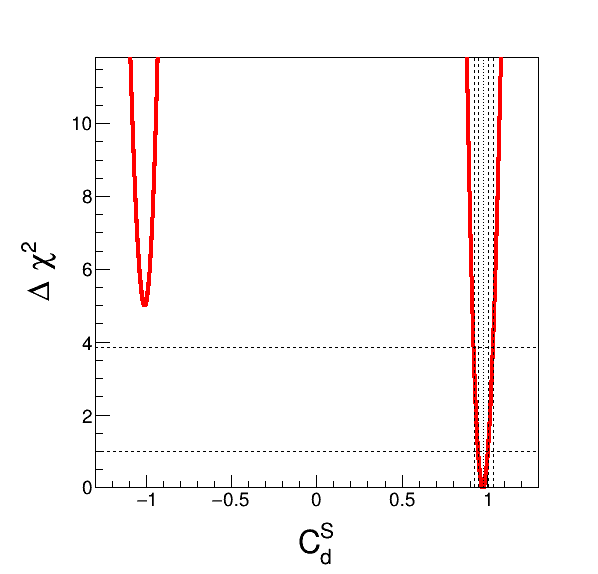}
& \includegraphics[width=4cm]{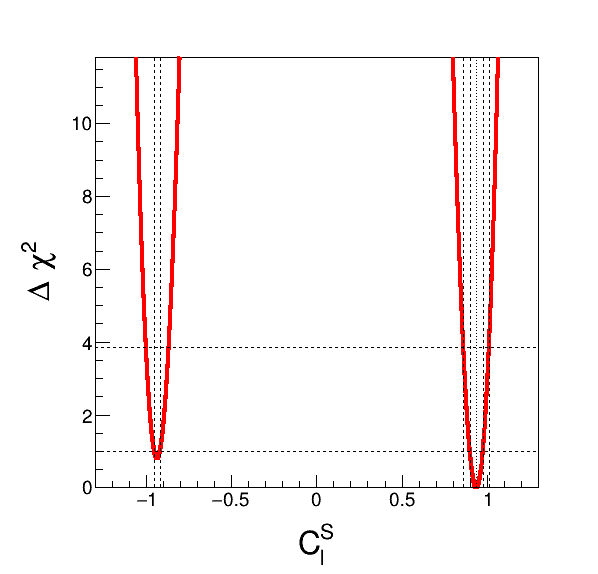}
& \includegraphics[width=4cm]{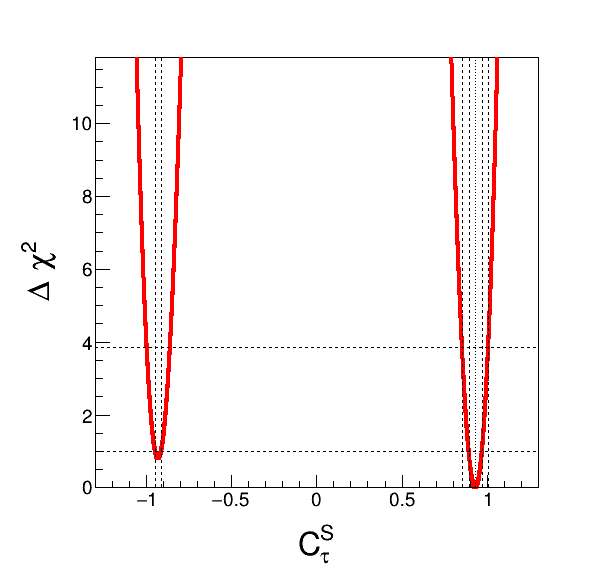}
& \includegraphics[width=4cm]{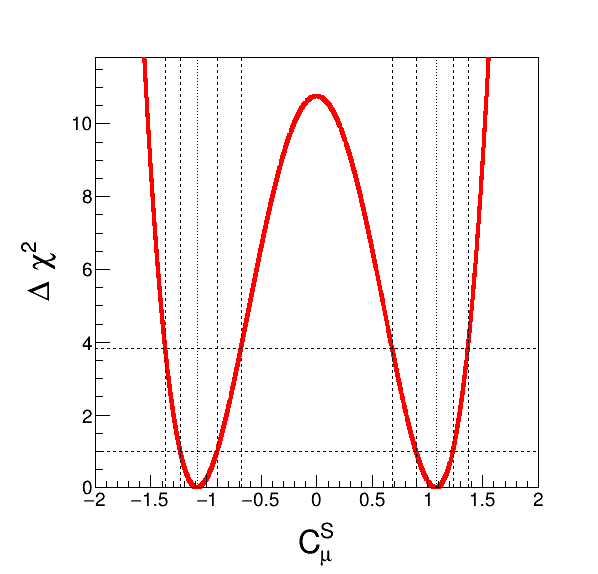}
\\ \hline \hline
Varying & 
$\Delta S^\gamma=-0.313^{+0.176}_{-0.176}$ & 
$\Delta S^g=-0.006^{+0.029}_{-0.030}$ & 
$\Delta \Gamma_{\rm tot}=-0.042^{+0.142}_{-0.132}$ MeV & 
\multirow{3}{*}{} \\ 
Parameter & & & &  \\ \cline{1-4}
gof & 0.3474 & 0.2636 & 0.2649 &   \\ \cline{1-4}
$p$-value & 0.0769 & 0.8425 & 0.7590 &   \\ 
\hline
& \includegraphics[width=4cm]{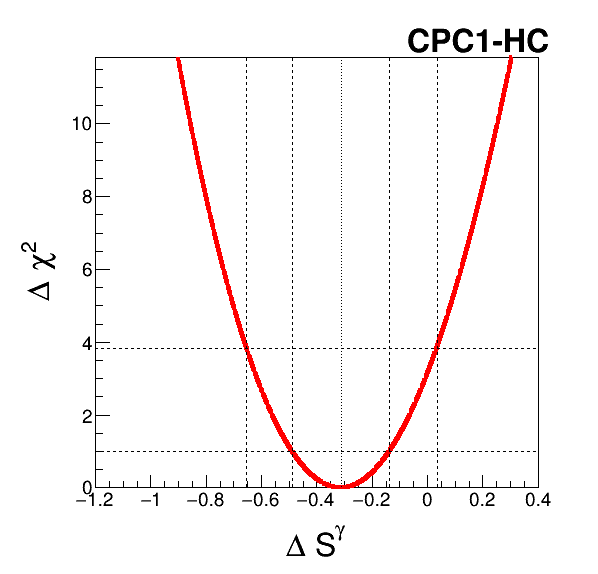}
& \includegraphics[width=4cm]{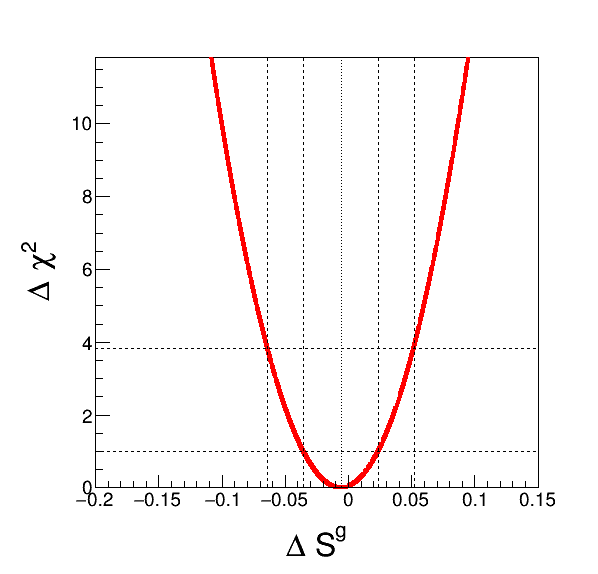}
& \includegraphics[width=4cm]{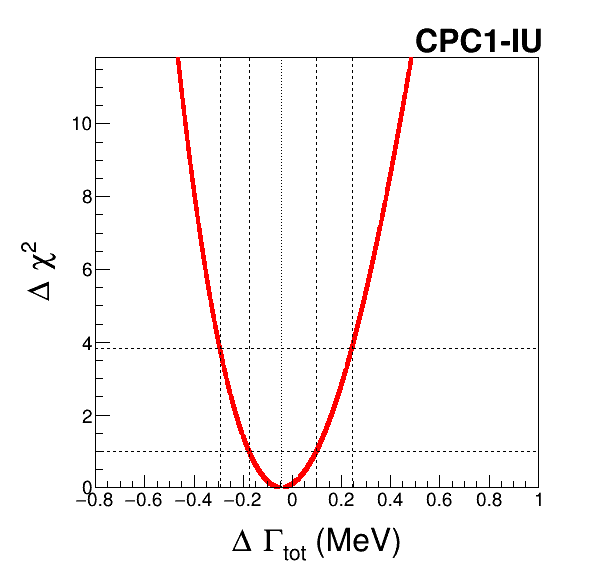}
&
\\ \hline 
\end{tabular}
\end{adjustbox}
\end{table}
%
%
\begin{table}[!htb]
\caption{\it
\label{tab:SVB2}
Continued from Table \ref{tab:SVB1}.
}  \vspace{1mm}
\begin{adjustbox}{width= \textwidth}
\begin{tabular}{|c|c|c|c|c|} \hline
Varying& 
$C_f^P=0.0^{+0.112}_{-0.112}$ & 
$C_u^P=0.0^{+0.152}_{-0.152}$ & 
$C_t^P=+0.031^{+0.119}_{-0.031}$, & 
$C_c^P=0.0^{+0.951}_{-0.951}$  \\ 
Parameter& &  & $~~~~~~\, -0.031_{-0.119}^{+0.031}$ &  \\ \hline
gof & 0.2626 & 0.2626 & 0.2626 & 0.2626\\ \hline
$p$-value & 1.0 & 1.0 & 0.9635 & 1.0  \\ \hline
& \includegraphics[width=4cm]{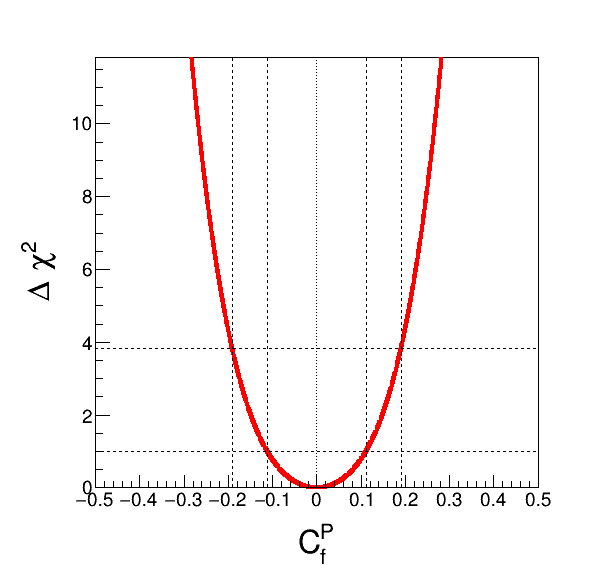}
& \includegraphics[width=4cm]{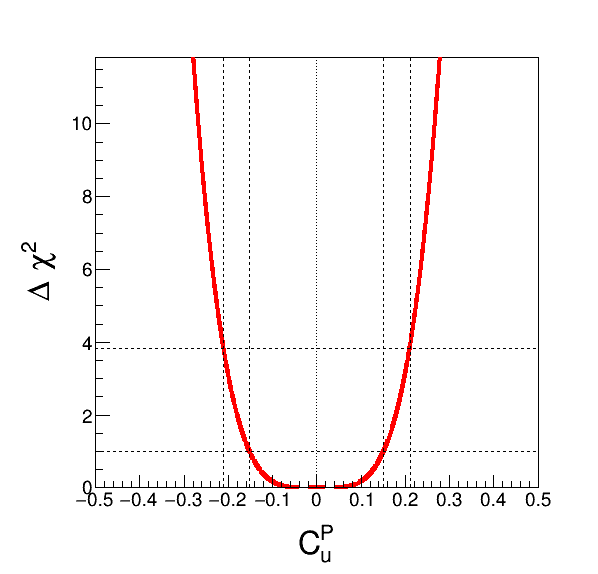}
& \includegraphics[width=4cm]{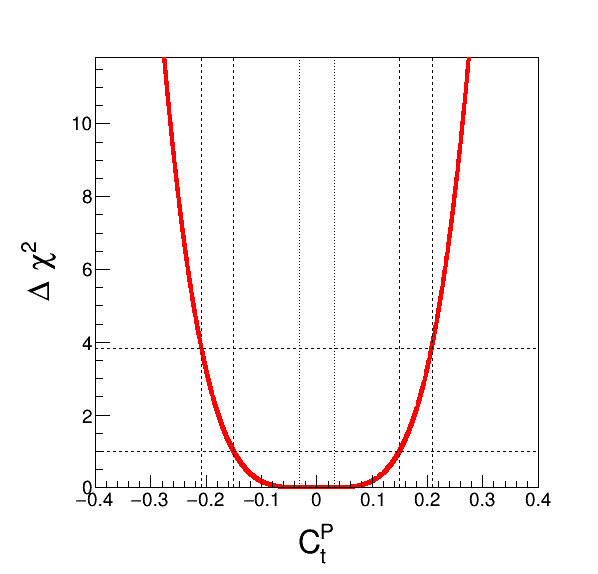}
& \includegraphics[width=4cm]{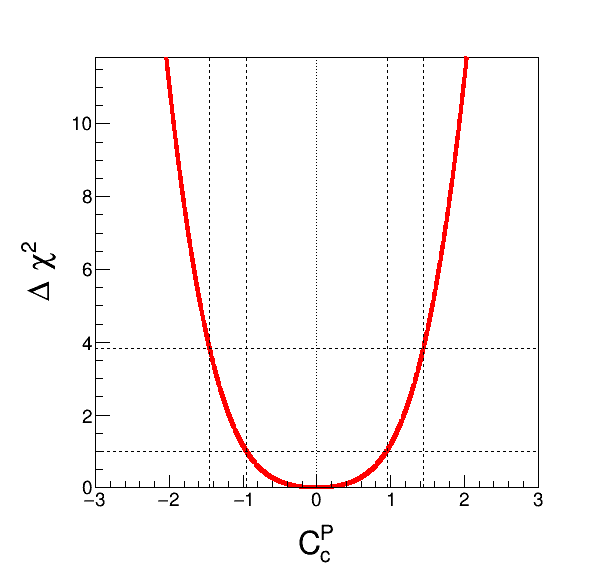}
\\ \hline \hline
Varying& 
$C_d^P = 0.0^{+0.172}_{-0.172}$ & 
$C_\ell^P = 0.0^{+0.145}_{-0.145}$ & 
$C_\tau^P = 0.0^{+0.141}_{-0.141}$ & 
$C_\mu^P = +0.397^{+0.312}_{-1.106}$,  \\ 
Parameter& &  & & $~~~~~~\, -0.397_{-0.312}^{+1.106}$  \\ \hline
\,\,gof & 0.2626 & 0.2626 & 0.2626 & 0.2678  \\ \hline
$p$-value & 1.0 & 1.0 & 1.0 & 0.6473  \\ \hline
& \includegraphics[width=4cm]{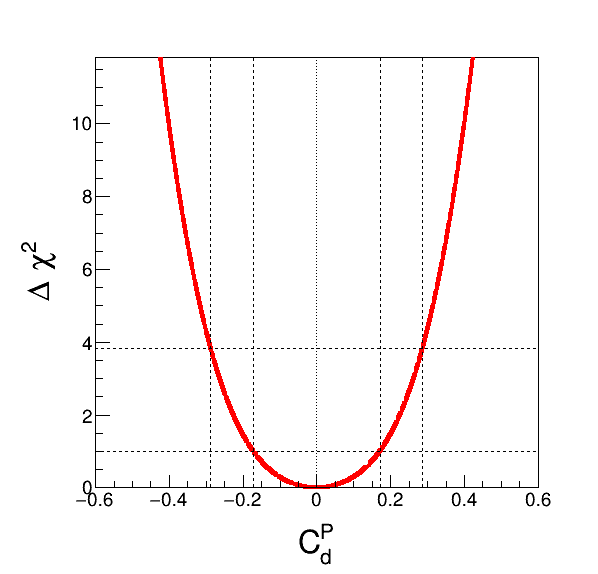}
& \includegraphics[width=4cm]{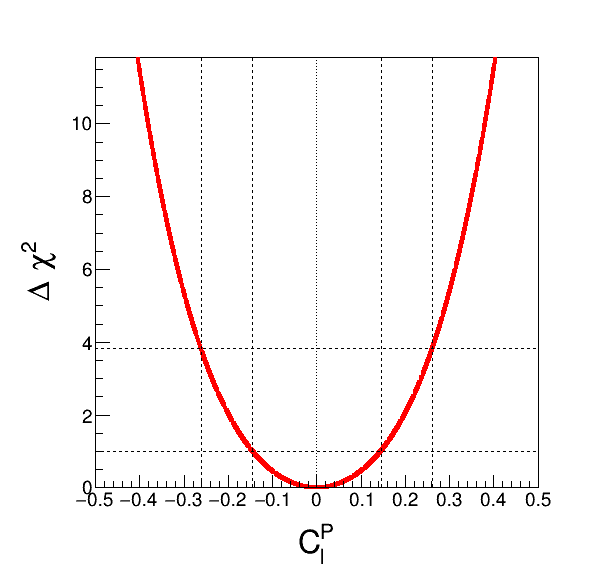}
& \includegraphics[width=4cm]{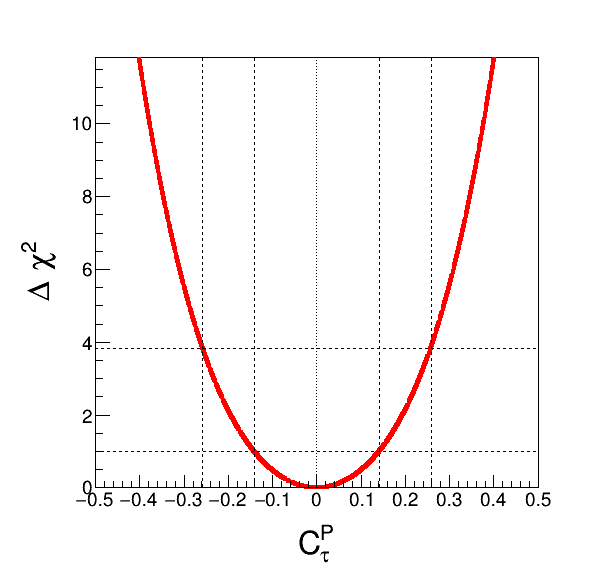}
& \includegraphics[width=4cm]{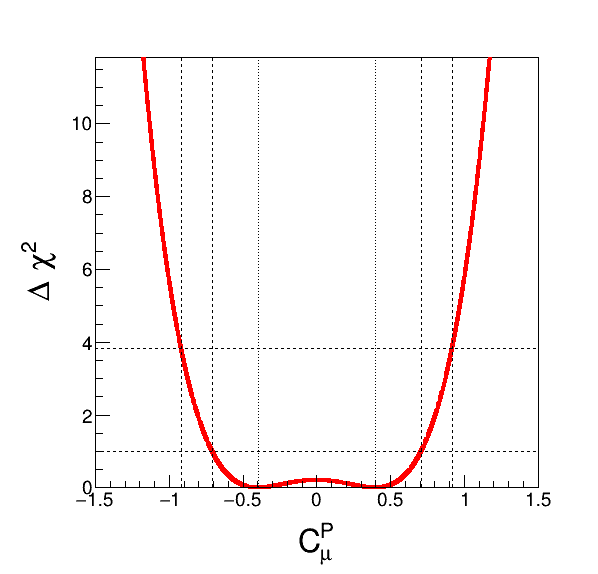}
\\ \hline \hline
Varying & 
$\Delta P^\gamma = +2.053^{+0.529}_{-0.703}$, & 
$\Delta P^g = 0.0^{+0.242}_{-0.242}$ &  & \\
Parameter & $~~~~~~~~\, -2.053_{-0.529}^{+0.703}$ & & &  \\ \cline{1-3}
gof & 0.3473 & 0.2626 & &  \\ \cline{1-3}
$p$-value & 0.0770 & 1.0 & & \\  
\hline
& \includegraphics[width=4cm]{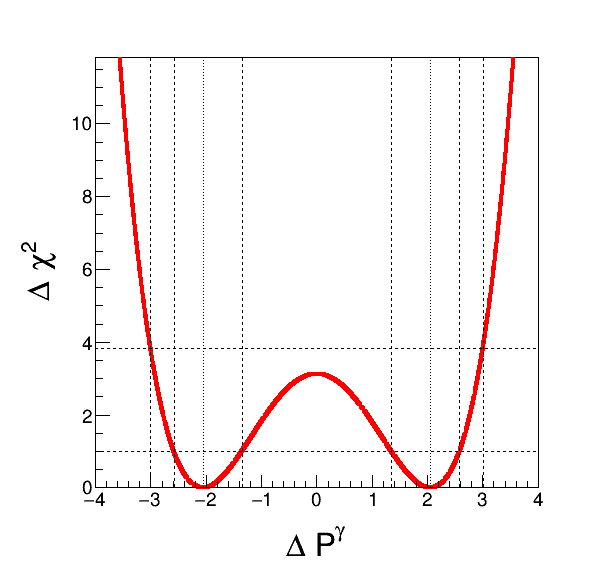}
& \includegraphics[width=4cm]{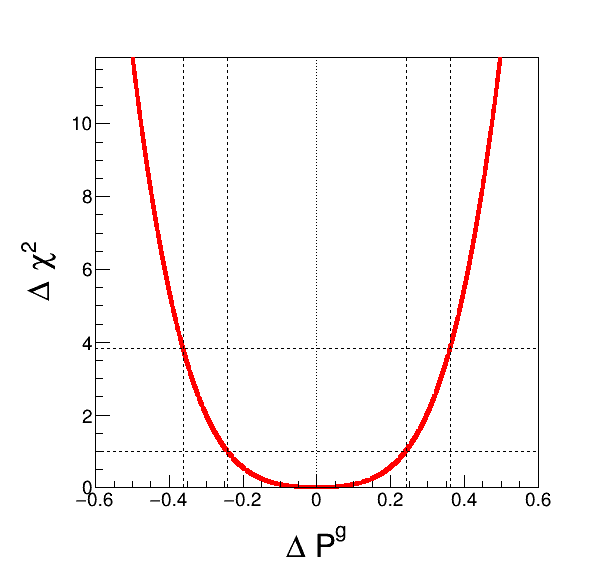} & & \\ \hline 
\end{tabular}
\end{adjustbox}
\end{table}

\newpage
\setcounter{equation}{0}
\section{Parametric dependence of $\widehat\mu(\gamma\gamma)$ in {\bf CPC2}-I}
\label{sec:appendix_C}
In {\bf CPC2}-I in which $C_V$ and $C_f^S$ are varied, we have
the following decay signal strength for $H\to\gamma\gamma$:
\begin{eqnarray}
\widehat\mu (\gamma\gamma)^{\rm{\bf CPC2}-I} 
= \frac{\Gamma_{\rm tot}(H_{\rm SM})}{\Gamma_{\rm tot}(H)}\,
\frac{\left|S^\gamma\right|^2}{\left|S^\gamma_{\rm SM}\right|^2}
\simeq 4\,\frac{\left|1.27\,C_V-0.27\,C^S_f\right|^2}{C_V^2+3\,(C^S_f)^2}
\end{eqnarray}
where we use Eq.~(\ref{eq:decay_signal_strength}),
Eq.~(\ref{eq:spaa_numeric}) together
with $\Delta S^\gamma=0$ and $S^\gamma_{\rm SM}=-6.542 + 0.046\,i$, 
and
\begin{eqnarray}
\label{eq:CPC2I_GoG}
\frac{\Gamma_{\rm tot}(H)}{\Gamma_{\rm tot}(H_{\rm SM})} &\simeq &
\left[\Gamma(H\to bb)+
\Gamma(H\to WW^*)+
\Gamma(H\to gg)+
\Gamma(H\to \tau\tau)+
\Gamma(H\to cc)+
\Gamma(H\to ZZ^*)\right]/\Gamma_{\rm tot}(H_{\rm SM}) \nonumber \\[2mm]
&=&
C_V^2\,\left[B(H\to WW^*)+B(H\to ZZ^*)\right]_{\rm SM} +
(C_f^S)^2\,\left[B(H\to bb)+B(H\to gg)+B(H\to \tau\tau)+
B(H\to cc)\right]_{\rm SM} \nonumber \\[2mm]
&\simeq&
\left[C_V^2 + 3\,(C_f^S)^2\right]/4 \,. \nonumber
\end{eqnarray}
Introducing
$C_V\equiv 1+\delta_V$ and
$C_f^S\equiv 1+\delta_f^S$, we have
\begin{equation}
\widehat\mu(\gamma\gamma) \simeq 1 \ + \ 2\,(\delta_V-\delta_f^S)\,.
\end{equation}
Incorporating the six production processes of ggF, VBF, WH, ZH, ttH, and tH with
\begin{equation}
\widehat\mu({\rm ggF}+{\rm bbH})=
\widehat\mu({\rm ttH})= (C_f^S)^2\,; \ \ \
\widehat\mu({\rm VBF})=\widehat\mu({\rm WH})
=\widehat\mu({\rm ZH})= C_V^2\,; \ \ \
\widehat\mu({\rm tH})= 3(C_f^S)^2+3.4\,C_V^2-5.4\,C_f^SC_V\,,
\end{equation}
we find that the global behavior of the theoretical
$pp\to H\to\gamma\gamma$ signal strength around the SM point
could be described by the following relation:
\begin{equation}
\label{eq:ss_aa}
\mu(\sum{\cal P},\gamma\gamma) \sim
1 \ + \ 3\,\delta_V-\delta_f^S \ \simeq \ 1.1 \ \pm \ 0.07\,,
\end{equation}
where we use $\mu(\sum{\cal P},\gamma\gamma)=1.10\pm 0.07$,
see Table~\ref{tab:pdss76}.
Note that the above relation is equivalent to
$C_f^S \sim 3\,C_V-2.1$ for the central value of $1.1$
which is quoted below Eq.~(\ref{eq:CPC2I_best}).

\begin{table}[!htb]
\caption{\it
\label{tab:Correlations}
Correlations $\rho_{xy}=\rho\,(x,y)$
between the two fitting parameters  of $x$ and $y$
in {\bf CPC2}, {\bf CPC3}, and {\bf CPC4}.}
\setlength{\tabcolsep}{2.5ex}
\renewcommand{\arraystretch}{1.1}
\centering
\begin{tabular}{|l|l||.{3.4}|.{3.4}|.{3.4}|.{3.4}|}
\hline
\multicolumn{1}{|c|}{Fit} & \multicolumn{1}{c||}{$\rho\,(x,y)$} & \multicolumn{1}{c|}{$1\sigma$} &
\multicolumn{1}{c|}{68.27\% CL} & \multicolumn{1}{c|}{95\% CL} & \multicolumn{1}{c|}
{ 99.73\% CL} \\ \hline \hline
\textbf{CPC2}-IUHC & $\rho\,(\Delta S^\gamma,\Delta \Gamma_{\rm tot})$ & -0.493 & -0.495 & -0.503
& -0.485  \\ \hline
\textbf{CPC2}-HCC & $\rho\,(\Delta S^g,\Delta S^\gamma)$ & 0.424 & 0.420 & 0.428 & 0.414  \\
\hline
\textbf{CPC2}-CSB & $\rho\,(C_W, C_Z)$ & 0.113 & 0.092 & 0.091 & 0.092  \\ \hline
\textbf{CPC2}-I & $\rho\,(C_V, C^S_f)$ & 0.386 & 0.383 & 0.380 & 0.379  \\ \hline
\textbf{CPC2}-II & $\rho\,(C^S_u, C^S_{d\ell})$ & 0.632 & 0.630 & 0.630 & 0.630  \\ \hline
\textbf{CPC2}-III & $\rho\,(C^S_{ud}, C^S_{\ell})$ & 0.228 & 0.225 & 0.202 & 0.202  \\ \hline
\textbf{CPC2}-IV & $\rho\,(C^S_{u\ell}, C^S_{d})$ & 0.710 & 0.710 & 0.715 & 0.706  \\ \hline
\hline
\multirow{3}{*}{\textbf{CPC3}-IUHCC}
        & $\rho\,(\Delta S^g, \Delta S^\gamma)$
        & 0.122 & 0.123 & 0.114 & 0.116 \\ \cline{2-6}
        & $\rho\,(\Delta S^g,\Delta \Gamma_{\rm tot})$
        & 0.641 & 0.638 & 0.639 & 0.665 \\ \cline{2-6}
        & $\rho\,(\Delta S^\gamma,\Delta \Gamma_{\rm tot})$
        & -0.309 & -0.295 & -0.303 & -0.291 \\ \hline
\multirow{3}{*}{\textbf{CPC3}-II}
        & $\rho\,(C_V, C^S_u)$
        & 0.101 & 0.108 & 0.104 & 0.099 \\ \cline{2-6}
        & $\rho\,(C_V, C^S_{d\ell})$
        & 0.742 & 0.741 & 0.740 & 0.737 \\ \cline{2-6}
        & $\rho\,(C^S_u, C^S_{d\ell})$
        & 0.494 & 0.492 & 0.488 & 0.484 \\ \hline
\multirow{3}{*}{\textbf{CPC3}-III}
        & $\rho\,(C_V, C^S_{ud})$
        & 0.404 & 0.389 & 0.397 & 0.391 \\ \cline{2-6}
        & $\rho\,(C_V, C^S_{\ell})$
        & 0.211 & 0.212 & 0.210 & 0.210 \\ \cline{2-6}
        & $\rho\,(C^S_{ud}, C^S_{\ell})$
        & 0.288 & 0.282 & 0.278 & 0.279 \\ \hline
\multirow{3}{*}{\textbf{CPC3}-IV}
        & $\rho\,(C_V, C^S_{u\ell})$
        & 0.659 & 0.653 & 0.650 & 0.652 \\ \cline{2-6}
        & $\rho\,(C_V, C^S_{d})$
        & 0.884 & 0.883 & 0.879 & 0.878 \\ \cline{2-6}
        & $\rho\,(C^S_{u\ell},C^S_{d})$
        & 0.830 & 0.826 & 0.827 & 0.825 \\ \hline
\hline
\multirow{6}{*}{\textbf{CPC4}-A}
        & $\rho\,(C_V,C^S_u)$
        & 0.515 & 0.508 & 0.481 & 0.499 \\ \cline{2-6}
        & $\rho\,(C_V,C^S_d)$
        & 0.884 & 0.873 & 0.861 & 0.855 \\ \cline{2-6}
        & $\rho\,(C^S_{u\ell},C^S_d)$
        & 0.779 & 0.755 & 0.741 & 0.729 \\ \cline{2-6}
        & $\rho\,(C_V,C^S_\ell)$
        & 0.647 & 0.616 & 0.604 & 0.595 \\ \cline{2-6}
        & $\rho\,(C^S_u,C^S_\ell)$
        & 0.504 & 0.447 & 0.417 & 0.410 \\ \cline{2-6}
        & $\rho\,(C^S_d,C^S_\ell)$
        & 0.679 & 0.647 & 0.622 & 0.595 \\ \hline
\end{tabular}
\end{table}
\setcounter{equation}{0}
\section{Correlations in {\bf CPC2}, {\bf CPC3}, and {\bf CPC4}}
\label{sec:appendix_D}
In Table~\ref{tab:Correlations}, we present
correlations among the fitting parameters in 
{\bf CPC2}, {\bf CPC3}, and {\bf CPC4} obtained by
fitting the CL contours in the $x$-$y$ plane
to the ellipses given by
\begin{equation}
\frac{(x-\widehat x)^2}{\sigma_x^2}+
\frac{(y-\widehat y)^2}{\sigma_y^2}-
2\rho_{xy}\frac{(x-\widehat x)}{\sigma_x} \frac{(y-\widehat y)}{\sigma_y}=
R^2(1-\rho_{xy}^2)
\end{equation}
with
$R^2=\Delta\chi^2= 1$ ($1\sigma$), $2.3$ (68.27\% CL), $5.99$ (95\% CL), and
$11.83$ (99.73\% CL).
Note that we consider the CL regions around the best-fit point 
of $(x,y)=(\widehat x,\widehat y)$ 
and drop the HP scenarios in which the CL contours  are not closed.
The best-fitted values of $\widehat x$ and $\widehat y$
and the $1\sigma$ errors of $\sigma_{x,y}$ are taken from
Table~\ref{tab:CPC2} and Table~\ref{tab:CPC34} and,
when the upper and lower $1\sigma$ errors are different from each other, 
we take the average of them.
In Table~\ref{tab:Correlations}, 
the dependence of $\rho$ on $\Delta\chi^2$ could
be considered as deviations from Gaussianity and 
the correlations are to be understood as approximations especially
when the values of $\rho$ are fluctuating noticeably.
For the fits with the higher number of fitting parameters with $n\geq 5$
and the more precise and detailed information even when $n\leq 4$, 
the grids for the boundaries of various
CL regions are available upon request to the authors.

\setcounter{equation}{0}
\section{Fits including $H\to Z\gamma$ data with
$\widehat\mu(Z\gamma)=2.2\pm 0.7$ : 
{\bf CPC2}-${\rm HC}$ and {\bf CPC6}-${\rm AHC}$}
\label{sec:appendix_E}
%
In this Appendix, we perform global fits of the Higgs boson couplings to
the extended Higgs datasets by including the Higgs boson decay to a 
$Z$ boson and a photon~\cite{ATLAS:2023yqk} which has been recently reported 
after the appearance of Refs.~\cite{ATLAS:2022vkf,CMS:2022dwd}.
Therefore,
including the 76 experimental signal strengths shown in 
Tables~\ref{tab:tev}, \ref{tab:78all}, \ref{tab:ATLAS13}, and \ref{tab:CMS13},
we consider 77 signal strengths in total in this Appendix.
More precisely, 
since the combined analysis in Ref.~\cite{ATLAS:2023yqk}
is based on the measured branching ratio of
$B(H\to Z\gamma)=(3.4\pm 1.1)\times 10^{-3}$ assuming 
the SM Higgs boson production cross sections,
we consider the following 77th experimental signal strength
\begin{equation}
\mu^{\rm EXP}({\cal Q},Z\gamma) \simeq 
\widehat\mu({\cal Q})\,\widehat\mu(Z\gamma)
=\sum_{{\cal P}_i\subset {\cal Q}}\,
\widehat\mu({\cal P}_i)\,\widehat\mu(Z\gamma)
=\widehat\mu(Z\gamma)=2.2\pm 0.7\,. 
\end{equation}
For the SM, we obtain $\chi^2_{\rm SM}/{\rm dof}=85.2868/77$
and ${\rm gof}=0.2424$.
Note that, compared to the case without including the $H\to Z\gamma$ data, 
\footnote{We recall that, without
including the $H\to Z\gamma$ data,  $\chi^2_{\rm SM}/{\rm dof}=82.3480/76$
and gof $=0.2895$.}
$\chi_{\rm SM}^2$ increases by the amount of 
$(1-2.2)^2/0.7^2=2.9388$ 
while gof decreases by $0.0471$.

\medskip

Since one might need nonvanishing 
$\Delta S^{Z\gamma}$ to resolve the tension in the 
measured $H\to Z\gamma$ signal strength
as discussed in subsection~\ref{subsec:Za}, 
we consider HC scenarios in which there exist heavy electrically 
charged non-SM particles leading to nonvanishing
$\Delta S^{\gamma}$ and $\Delta S^{Z\gamma}$ simultaneously.
To be specific, we consider the following two CPC fits:
\begin{itemize}
\item{{\bf CPC2}-HC}:  vary $\{\Delta S^\gamma\,,\Delta S^{Z\gamma}\}$ 
with the gauge-Higgs and Yukawa couplings the same as in the SM
\item{{\bf CPC6}-AHC}: vary $\{\Delta S^\gamma\,,\Delta S^{Z\gamma}\,,
C_V\,,C_u^S,C_d^S,C_\ell^S\}$ 
with the gauge-Higgs and Yukawa couplings like as in A2HDM
\end{itemize}
Note that we have promoted {\bf CPC1}-HC and {\bf CPC5}-AHC by employing 
$\Delta S^{Z\gamma}$ as an additional varying parameter.

\medskip

In {\bf CPC2}-HC, we obtain the following best-fitted values and $1\sigma$ errors:
\footnote{We have obtained the same best-fitted value for 
$\Delta S^\gamma$ at the both degenerate minima for the positive and
negative values of $\Delta S^{Z\gamma}$.}
\begin{equation}
\Delta S^\gamma = -0.318^{+0.176}_{-0.176}\,; \ \ \
\Delta S^{Z\gamma} = -5.698^{+3.013}_{-2.571}\,, \ 
+29.038^{+2.568}_{-3.015}\,,
\end{equation}
with
$\chi^2_{\rm min}/{\rm dof}=79.1649/75$,
gof $=0.3489$, and
$p$-value against the SM $=0.0468$.
We observe that $\Delta S^\gamma$ is fitted similarly as in {\bf CPC1}-HC,
see Table~\ref{tab:CPC1}, and there are two degenerate minima 
for the negative ($<0$) and positive ($>0$) values of $\Delta S^{Z\gamma}$ 
as dictated by the relation
$\left|S^{Z\gamma}_{\rm SM}+\left.\Delta S^{Z\gamma}\right|_{<0}\right|^2 =
\left|S^{Z\gamma}_{\rm SM}+\left.\Delta S^{Z\gamma}\right|_{>0}\right|^2$
with $S^{Z\gamma}_{\rm SM} =-11.6701 + 0.0114 \, i$, 
see Eq.~(\ref{eq:spza_numeric}).
Note that $\Delta S^\gamma$ shows a $1.8\sigma$ deviation from the SM
while $\left.\Delta S^{Z\gamma}\right|_{<0}$ deviates from the SM
by $1.9\sigma$.

\medskip

\begin{figure}[h!]
\begin{center}
\includegraphics[width=6cm]{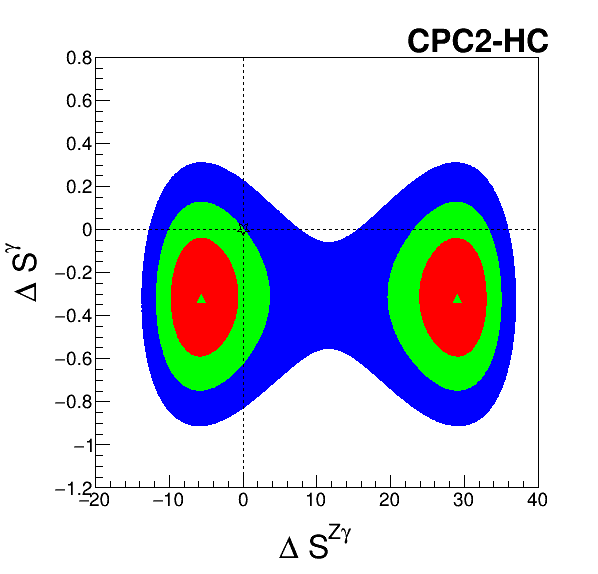}
\includegraphics[width=6cm]{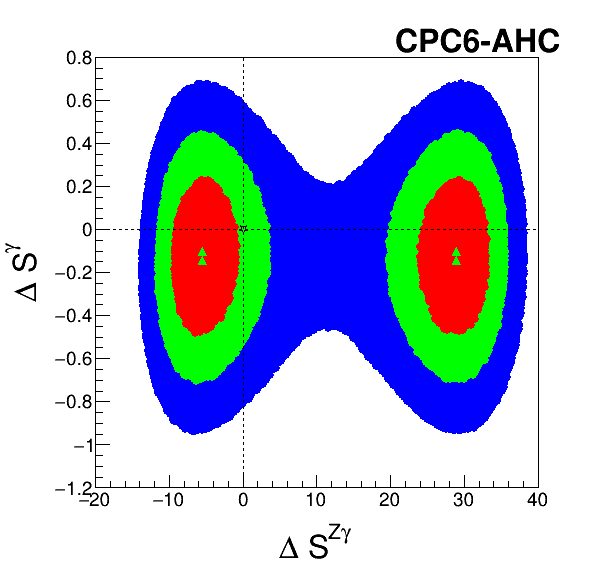}
\end{center}
\vspace{-0.5cm}
\caption{\it
The CL regions of {\bf CPC2}-{\rm HC} (left) and {\bf CPC6}-{\rm AHC} (right)
in the $(\Delta S^{Z\gamma},\Delta S^{\gamma})$ plane.
The contour regions shown are for
$\Delta\chi^2\leq 2.3$ (red),
$\Delta\chi^2\leq 5.99$ (green),
$\Delta\chi^2\leq 11.83$ (blue)
above the minimum, which correspond to
confidence levels of 68.27\%, 95\%, and 99.73\%, respectively.
In each frame, the vertical and horizontal lines locate the SM point denoted
by a star and the best-fit points are denoted by triangles.
}
\label{fig:CPC2HC_CPC6AHC}
\end{figure}

In {\bf CPC6}-AHC, there are four degenerate minima 
depending on the signs of $C_\ell^S$ and $\Delta S^{Z\gamma}$.
See Table~\ref{tab:CPC6-AHC} for the best-fitted values of the six fitting parameters
at each minimum.
We note that the best-fitted values 
of $C_V$ and $C_{u,d}^S$ at the four degenerate minima are 
almost the same like as in {\bf CPC5}-AHC.
For the statistical measures, we have
gof $=0.3810$ and
$p$-value against the SM $=0.0795$
with $\chi^2_{\rm min}/{\rm dof}=73.9864/71$.
Note that $\Delta S^\gamma$ is consistent with the SM
while $\left.\Delta S^{Z\gamma}\right|_{<0}$ deviates from the SM
by $2.0\sigma$.
In Fig.~\ref{fig:CPC2HC_CPC6AHC},
the CL regions of {\bf CPC2}-{\rm HC} (left) and {\bf CPC6}-{\rm AHC} (right)
are shown
in the $(\Delta S^{Z\gamma},\Delta S^{\gamma})$ plane.
We observe that the SM points locate outside the 68\% CL regions.
%
\begin{table}[!t]
\caption{\it
\label{tab:CPC6-AHC}
{\bf CPC6}-AHC: The best-fitted values at the four degenrate minima
in {\bf CPC6}-AHC.
We have
$\chi^2_{\rm min}/{\rm dof}=73.9864/71$,
gof $=0.3810$, and $p$-value against the SM $=0.0795$.
For the SM, in contrast, we obtain $\chi^2_{\rm SM}/{\rm dof}=85.2868/77$
and gof $=0.2424$.}
\setlength{\tabcolsep}{2.5ex}
\renewcommand{\arraystretch}{1.1}
\centering
\begin{tabular}{|c||r|r|r|r|}
\hline
Fitting parameter&\multicolumn{4}{c|}{Best-fitted values} \\ \hline
        $C^S_\ell $ & $+0.921^{+0.047}_{-0.046}$ &$+0.921^{+0.044}_{-0.045}$&
$-0.921^{+0.044}_{-0.045}$ & $-0.921^{+0.045}_{-0.046}$ \\ \hline 
        $\Delta S^{Z\gamma}$ & $-5.548^{+2.768}_{-2.367}$ & $+28.956^{+2.742}_{-2.840}$ &
$-5.548^{+2.745}_{-2.429}$& $+28.956^{+2.679}_{-2.828}$ \\ \hline  \hline
        $\Delta S^{\gamma}$ & $-0.102^{+0.205}_{-0.211}$ &  $-0.102^{+0.217}_{-0.208}$ &
$-0.146^{+0.207}_{-0.203}$ & $-0.146^{+0.204}_{-0.200}$ \\ \hline 
        $C_V$ & $0.999^{+0.033}_{-0.033}$ &$0.999^{+0.035}_{-0.033}$&$0.999^{+0.032}_{-0.033}$
&$0.999^{+0.034}_{-0.035}$ \\ \hline
        $C^S_u$ & $0.930^{+0.039}_{-0.040}$ &$0.930^{+0.041}_{-0.039}$& $0.930^{+0.039}_{-0.039}$ &
$0.930^{+0.040}_{-0.037}$ \\ \hline
        $C^S_d$ & $0.908^{+0.078}_{-0.077}$ &$0.908^{+0.082}_{-0.077}$& $0.908^{+0.078}_{-0.074}$ &
$0.908^{+0.078}_{-0.080}$ \\ \hline
\end{tabular}
\end{table}
%

\end{appendix}


\end{document}